\documentclass[11pt]{article}
\usepackage{a4}
\usepackage{epsfig}
\renewcommand{\caption}[2]
{\refstepcounter{figure}
\label{#1}
\begin{description}
\item[Fig.\ \ref{#1}]{#2}
\end{description}}

\newcommand{\fcaption}[2]
{\refstepcounter{table}
\label{#1}
\begin{description}
\item[Table \ref{#1}]{#2}
\end{description}}

\begin {document}
\begin{title}
{
\hfill{\small {\bf MKPH-T-98-3}}\\
{\bf Electroweak Processes in Few-Nucleon Systems}
\footnote{Lectures held at the Int.\ Workshop on Few-Body Problems in 
Nuclear Physics and Related Fields, ECT$^\ast$, Trento, Italy, September 
8-27, 1997}
\footnote{Supported by the Deutsche 
Forschungsgemeinschaft (SFB 201)}}
\end{title}
\bigskip
\author{Hartmuth Arenh\"ovel \\
Institut f\"ur Kernphysik\\ 
Johannes Gutenberg-Universit\"at\\
D-55099 Mainz, Germany}
\maketitle
\begin{abstract}
After a brief introduction into the basic ingredients of electroweak theory 
as a spontaneously broken local, non-Abelian gauge symmetry, the general 
properties of the electromagnetic current and two-photon operators are 
discussed. In particular, the consequences of gauge invariance and the 
resulting low energy theorems are reviewed. The multipole decomposition of 
the current operators and the general Siegert theorem are presented. 
The specific forms of vector and axial one-nucleon currents are given, 
together with lowest order $\pi$ meson exchange and isobar currents as well 
as meson production currents. A brief overview is given on the most important 
one- and two-boson processes. Electron scattering in the 
one-boson-approximation is then considered in greater detail. 
Formal expressions of the cross section for inclusive and exclusive 
processes are given, including parity violating contributions from 
$\gamma$-$Z$ interference 
as well as from parity violating components in the hadronic wave function. 
Specific electromagnetic reactions on the deuteron are then discussed  
with respect to the influence of meson exchange currents, isobar 
configurations in the deuteron groundstate, relativistic contributions and 
the role of $\pi$-meson retardation. Furthermore, recent results on 
coherent and incoherent $\pi$- and $\eta$-photoproduction are presented 
as well as a discussion of the Gerasimov-Drell-Hearn sum rule and
the effect of a parity violating deuteron component on
inclusive electron scattering off the deuteron for quasifree kinematics.
The review closes with a summary and a brief outlook. 
\end{abstract}

\newpage
\tableofcontents
\newpage
\setcounter{section}{0}
\setcounter{equation}{0}
\setcounter{figure}{0}
\setcounter{table}{0}

\section{Introduction }
\label{sec1}

One of the major goals of present day research in the field of medium 
energy physics is to 
clarify the role of effective subnuclear degrees of freedom (d.o.f.) 
in terms of nucleon, 
meson and isobar d.o.f., and their relationship to the underlying more 
fundamental d.o.f.\ of quantum chromodynamics (QCD). In other words, 
the question is, what is the connection of such effective d.o.f.\ to 
the presumably underlying quark-gluon dynamics of QCD. A large fraction of 
our present knowledge about the internal constitution of nuclei or more 
general of hadrons has been obtained from electromagnetic reactions. 
Indeed, throughout the history of nuclear and elementary particle physics, 
starting with the discovery of the nucleus by Coulomb scattering of 
$\alpha$-particles off a gold foil in 1909 up to the discovery of 
point-like objects as internal constituents in the proton in deep 
inelastic lepton scattering in 1967, 
the electromagnetic probe has provided us with a wealth of important 
information on the internal structure of hadronic systems. 

The salient features of the electromagnetic (e.m.) interaction, which make 
it such a valuable tool, are well known and may be summarized as follows: 
\begin{itemize}
\item
The e.m.\ interaction is already known from classical physics (Maxwell 
theory).
\item
The e.m.\ interaction fulfills the basic requirements of a fundamental 
interaction, i.e.,  it incorporates the principles of special relativity 
and represents the simplest case of a gauge theory, namely an Abelian gauge 
theory.
\item
The e.m.\ interaction is  weak enough, characterized by the small fine 
structure constant $\alpha=1/137$, so that in most 
cases lowest order perturbation theory is sufficient, allowing a simple 
interpretation of observables in terms of charge and current density matrix 
elements. 
\end{itemize}

Another important source of information is provided by reactions involving 
the weak interaction as, for example, mani\-fest in beta decay. 
Originally, the weak interaction was considered 
as an independent fundamental force, and it was only much later that 
its unification with electromagnetism was established, although already 
Fermi intuitively had formulated his famous four-fermion theory of beta 
decay along the scheme of the e.m.\ interaction. Indeed, already 
in 1935, Yukawa \cite{Yuk35} had hypothesized a heavy boson as mediator of 
the weak interaction, which at low energy would effectively give the point 
coupling of the Fermi theory. The various stages of the historical 
development of the theory of the weak interaction may be summarized 
as follows:
\begin{itemize}
\item
Discovery of radioactivity by Becquerel in 1896, and subsequent 
identification of beta rays as electrons. 
\item
Postulate of the  neutrino by Pauli in 1930 in order to save the 
energy conservation law in beta decay. 
\item
First formulation of a theory for beta decay by Fermi in 1934 as a 
 current-current interaction, modelled in analogy to the e.m.\ case. 
\item
Discovery of parity violation in beta decay in 1957 by Wu, 
Ambler, Hayward, Hoppes, Hudson \cite{WuA57} after 
a critical analysis of the status of experimental evidence for parity 
conservation in weak processes by Lee and Yang in 1956 
\cite{LeY56}. 
\item
Formulation of the  $(V-A)$-theory by Gell-Mann and 
Feynman \cite{FeG58} and independently by Sudarshan 
and Marshak \cite{SuM58} in 1958. Postulate of universality of the 
current-current interaction \cite{FeG58}, and renewed hypothesis of an 
intermediate vector boson as mediator of the weak interaction. 
\end{itemize}

Finally, about 30 years ago, e.m.\ and weak interactions were unified. 
The essential steps towards this unification were 
\begin{itemize}
\item
Study  of a local non-Abelian gauge symmetry by Yang and Mills 
\cite{YaM54} in 1954 for the case of an isospin $SU(2)$ symmetry. It is the 
simplest case of what now is called in general a Yang-Mills theory. 
\item
Introduction of spontaneous breaking of a local gauge symmetry 
as mass generating mechanism for vector bosons by Higgs, Kibble and others 
in 1964-66 (massive Yang-Mills theory). 
\item
Unification of e.m.\ and weak interactions via a spontaneously broken 
$SU(2)_L\times U(1)$ gauge symmetry by Glashow, 
Weinberg and Salam in 1967.
\item
Proof of  renormalizability of electroweak theory by 't Hooft in 1971.
\end{itemize}

I will begin these lectures by reviewing first the essential ideas of the 
unified electroweak theory as a spontaneously broken non-Abelian gauge theory 
leaving out all the finer details which may be found in appropriate 
textbooks, e.g.\ in \cite{BeB83,Hua92,Nac86,Wei95}. In Sect.\ \ref{sec3}, the 
properties of the e.m.\ interaction operators, gauge conditions, low energy 
theorems, multipole decomposition and generalized Siegert operators will be 
briefly discussed. Explicit expressions for the current operators will be 
collected in Sect.\ \ref{sec4}. Then in Sect.\ \ref{sec5}, an overview on 
basic electroweak processes is given. A variety of specific reactions 
on the deuteron will be considered in Sect.\ \ref{sec6} 
like, e.g., photo- and electrodisintegration, coherent and 
incoherent meson photoproduction, the Gerasimov-Drell-Hearn sum rule and 
parity violation in electron scattering. I will close with some conclusions 
and an outlook. 

    % introduction

%\newpage
\setcounter{equation}{0}
\setcounter{figure}{0}
\setcounter{table}{0}

\section{The Electroweak Interaction as a Non-Abelian Gauge Theory} 
\label{sec2}

As already mentioned, the two basic ingredients for the unified 
description of e.m.\ and weak processes are (i) a non-Abelian gauge 
symmetry, and (ii) spontaneous symmetry breaking. 

\subsection{Abelian and Non-Abelian Gauge Symmetry}
In order to illustrate the fact how the postulation of invariance under a 
local gauge symmetry leads to the existence of an interacting massless 
vector field, the gauge field, let me consider first the simplest case of an  
Abelian gauge symmetry for a free complex scalar field  $\Phi(x)$ 
having a Lagrangian 
\begin{eqnarray}
{\cal L}_0(\Phi,\,\partial_\mu\Phi)= \partial^\mu\Phi^\ast \partial_\mu\Phi
-V(\Phi^\ast \Phi)\,.\label{scalarlag}
\end{eqnarray}
This Lagrangian is evidently invariant under a global 
 $U(1)$ gauge transformation (phase transformation)
\begin{eqnarray}
\Phi\,& \Longrightarrow& \,  U(\theta)\, \Phi\,,\quad 
U(\theta)= e^{-i\theta}\,,\\
\partial_\mu\Phi\,&\Longrightarrow&\, 
 U(\theta)\, \partial_\mu\Phi\,,
\end{eqnarray}
where  $\theta$ is a real constant parameter. 
According to the Noether theorem, this invariance then leads to a conserved 
vector current
\begin{eqnarray}
\partial^\mu j_\mu=0
\quad \mbox{with}\quad
j_\mu = i \Phi^\ast \stackrel{\leftrightarrow}{\partial_\mu}\Phi\,.
\end{eqnarray}

It is very useful for the understanding of a local gauge symmetry to give a 
geometrical interpretation of this phase transformation \cite{BeB83}. 
The fact, that the 
field is complex, may be visualized by representing it as a vector in a 
complex plane, an internal two-dimensional space, which is attached to 
each space-time point. In principle, one could imagine to assign different 
frames to these internal spaces at different space-time points. 
In this case, however, 
one could not compare vectors at different points anymore. In other words 
the affine connection would be lost. In fact, in writing a complex 
field amplitude, the tacit assumption is made, that at different 
points the internal frames can be identified establishing thus the affine 
connection, which is necessary for taking the derivative. 
Therefore, a global gauge transformation can be viewed as a 
global rotation of the internal frame at each space-time point 
by the same angle  $\theta$ so that the affine connection is not lost 
(see Fig.\ \ref{figglobtrafo}). 

\begin{figure}[b]
\centerline {\psfig{file=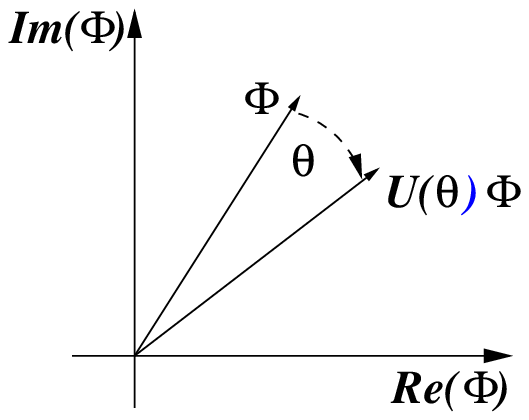,width=6cm,angle=0}}
\caption{figglobtrafo}
{Visualization of a phase transformation as rotation in an 
internal space.}
\end{figure}

This interpretation is important for the understanding of a local gauge 
transformation which is defined by 
\begin{eqnarray}
\Phi(x)\,\Longrightarrow \, U(\theta(x))\, \Phi(x)\,,
\end{eqnarray}
where  $\theta(x)$ is a real arbitrary parameter function of $x$. The 
motivation for considering such local transformations of the internal 
frames is, that the physics should not depend on the local choice of the 
internal frame \cite{Wey29}. Under such a transformation, the Lagrangian 
is no longer invariant, because it does not commute anymore with the 
derivative, instead one has 
\begin{eqnarray}
\partial_\mu\Phi\, \Longrightarrow \, 
U(\theta(x))\,\Big(
\partial_\mu \Phi(x)-i \partial_\mu\theta(x)\Phi(x)\Big)\,.
\end{eqnarray}
This fact can be understood easily in the geometric 
interpretation of the gauge transformation, because the internal coordinate 
systems are now rotated by an arbitrary angle $\theta(x)$, i.e., 
differently at different points which means that the 
affine connection is lost. In other words, 
the field vectors at different points cannot be compared to each other anymore 
and thus the derivative is no longer well defined. 

At this point the gauge field comes into play, because 
this problem can be cured by introducing a new vector field
 $A_\mu(x)$, by which one defines a parallel displacement
of internal vectors from one point to another so that internal vectors 
at different points can be compared. For this reason, Weyl has coined the 
name ``gauge field'' for $A_\mu(x)$. It is also subject to a certain 
gauge transformation as is specified below. 
The parallel displacement of a field 
vector at point $x$ to the point $x+dx$ via the gauge field is defined by 
\begin{eqnarray}
^{x+dx}\Phi(x):=(1-ieA_\mu(x)dx^\mu)\,{^x\Phi(x)}\,,
\end{eqnarray}
where  ${^x\Phi(x)}$ denotes the representation of $\Phi(x)$ in the internal 
basis at  point  $x$ and correspondingly  $^{x+dx}\Phi(x)$ 
its representation in the internal basis at  point  $x+dx$, 
i.e., the  same internal vector in different internal
coordinate systems. Here, $e$ is a coupling constant which will characterize 
the strength of the interaction of the gauge field with the given scalar field. 
\begin{figure}[h]
\centerline{\psfig{file=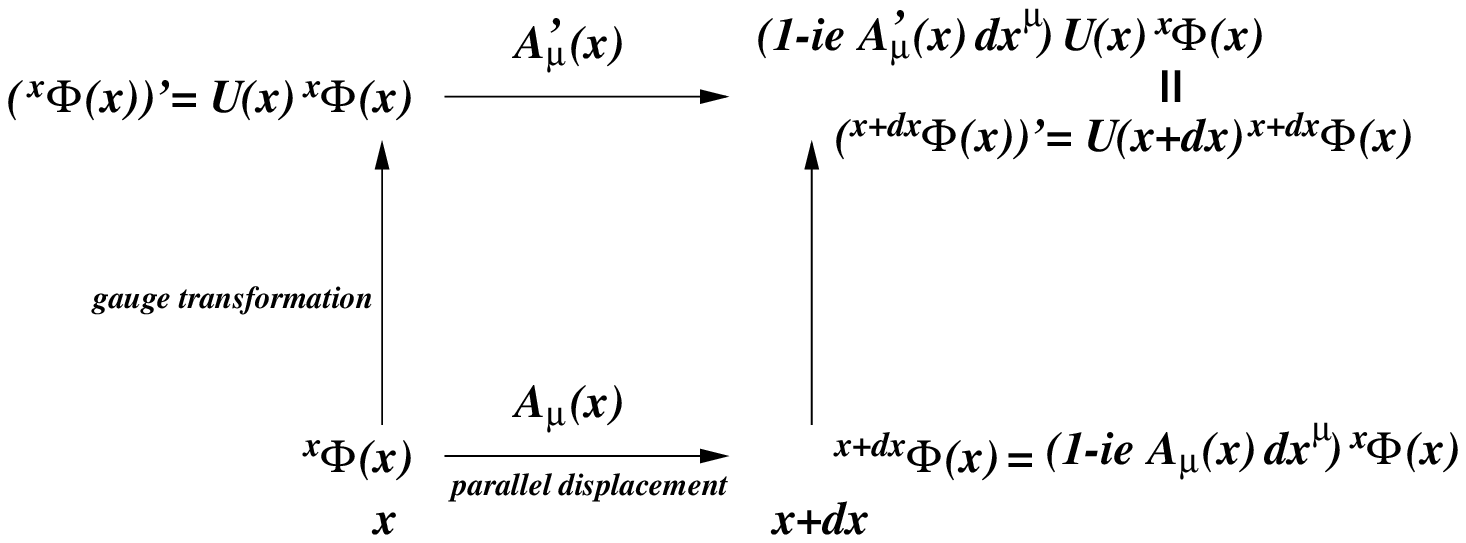,width=14cm,angle=0}}
\caption{figlocaltrafo} 
{Parallel displacement and local gauge transformation.
}
\end{figure}

In order to have a meaningful definition of this parallel displacement, 
i.e., independent from any local gauge transformation, one has to require 
that one gets the same result if one first makes a gauge transformation 
and then a parallel displacement or vice versa (see Fig.\ \ref{figlocaltrafo}).
This condition then leads to the well known transformation law for the 
gauge field  $A_\mu(x)$ 
\begin{eqnarray}
A_\mu(x)\, &  \Longrightarrow & \, 
A_\mu^{\prime}(x)=A_\mu(x)+\frac{1}{e}\partial_\mu \theta(x)\,.
\label{fieldgt}
\end{eqnarray}
With the help of the gauge field, one then introduces the 
covariant derivative 
\begin{eqnarray}
D_\mu:=\partial_\mu + ie A_\mu(x)\,,
\end{eqnarray}
which transforms under a local gauge transformation according to
\begin{eqnarray}
D_\mu\phi(x)\, & \Longrightarrow&\, 
 U(\theta(x))\,D_\mu\phi(x)\,,
\end{eqnarray}
i.e., $D_\mu$ commutes with the gauge transformation. 
Replacing then in (\ref{scalarlag}) the ordinary derivatives by the 
covariant derivatives, yielding the Lagrangian 
\begin{eqnarray}
{\cal L}_0(\phi,\,D_\mu\phi)= D^\mu\phi^\ast D_\mu\phi
-V(\phi^\ast \phi)\,,
\end{eqnarray}
one has achieved invariance of  ${\cal L}_0$ under a local gauge 
transformation. 
\begin{figure}[h]
\centerline{\psfig{file=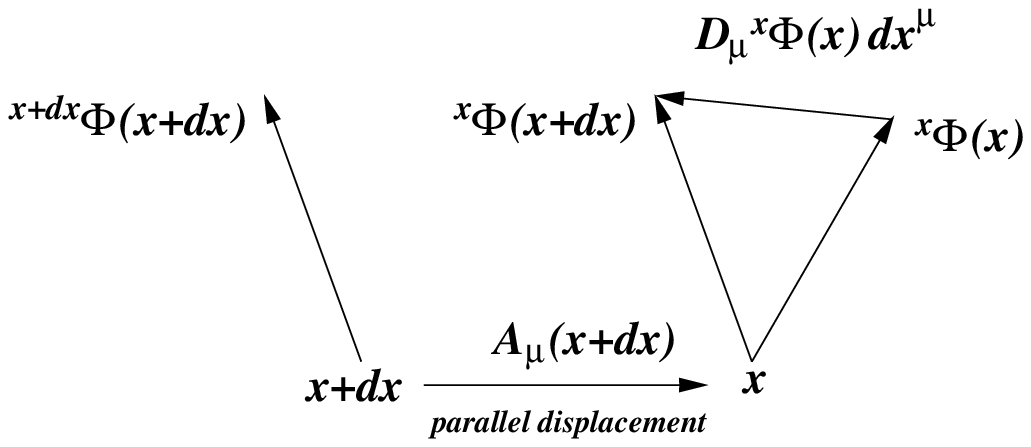,width=12cm,angle=0}}
\caption{figcovdif}
{Geometrical interpretation of the covariant derivative.
}
\end{figure}

The  geometrical meaning of the covariant derivative becomes clear if one 
compares the field at two neighbouring points $x$ and $x+dx$ with the help of 
the parallel displacement (Fig.\ \ref{figcovdif}). 
One readily notices that the covariant derivative mediates 
the connection of the field at two neighbouring points. In other words, 
the difference of the field vectors at the points $x+dx$ and $x$ evaluated 
at the point $x$ is just given by the covariant derivative. 
\begin{eqnarray}
^{x}\Phi(x+dx)-{^x\Phi(x)}=D_\mu{^x\Phi(x)}\,dx^\mu\,.
\end{eqnarray}
For finite distances, the  connection depends on the 
path, because the components of the covariant derivative  do not 
commute, instead one has 
\begin{eqnarray}
\Big[D_\mu,\,D_\nu\Big]&=&
ie \Big(\partial_\mu A_\nu - 
\partial_\nu A_\mu\Big)\nonumber\\
&=& ie F_{\mu\nu}\,,
\end{eqnarray}
where the  field strength tensor $F_{\mu\nu}$ is defined by the last line. 
Obviously, the  field strength tensor is invariant under the local gauge 
transformations of (\ref{fieldgt}).

Until now, the gauge field was introduced as an external field. Thus, 
the final step is to introduce the gauge field dynamics by a gauge invariant 
Lagrangian for the free gauge field
\begin{eqnarray}
{\cal L}_{g.f.} = -\frac{1}{4} F_{\mu\nu} F^{\mu\nu}\,,
\end{eqnarray}
which describes  massless vector bosons. A mass term would violate the gauge 
symmetry. The total Lagrangian then reads 
\begin{eqnarray}
{\cal L}(\Phi,\,\partial_\mu\Phi,\,A_\mu,\, F_{\mu\nu})  = & 
-\frac{1}{4} F_{\mu\nu} F^{\mu\nu}
+D^\mu\Phi^\ast D_\mu\Phi
-V(\Phi^\ast \Phi)\,.
\end{eqnarray}
It can be interpreted as describing the quantum electrodynamics (QED) of 
charged scalar particles. This simple example illustrates nicely the 
essential idea that the postulate 
of a local gauge symmetry leads to the introduction of a massless vector 
field, the gauge field, as a fundamental interaction. It is well known that 
gauge invariance is essential in order to have a renormalizable theory. 

I will now consider the generalization to gauging a non-Abelian internal 
symmetry group as is the case, for example, for the weak isospin  $SU(2)_L$ of 
the electroweak interaction or the color  $SU(3)$ of  QCD. Let us take 
a Yang-Mills field which consists, for example, of $n$ complex 
components  $\psi_c$ 
\begin{eqnarray}
\psi = \left( \begin{array}{c} \psi_1 \\ \vdots \\ \psi_n \end{array}
\right)\,,
\end{eqnarray}
which form the fundamental representation of an internal $SU(n)$ 
symmetry group, possessing $l=n^2-1$  generators $\hat T_a$ 
($n\times n$ matrices), where $a=1,\dots ,l$ labels the generators. In the 
following, the Einstein convention for summation over equal indices is also 
used for the index $a$. 

A local gauge transformation has then the form
\begin{eqnarray}
\psi \,\,  \Longrightarrow \,\, \hat U(x) \psi\,,
\end{eqnarray}
where now 
\begin{eqnarray}
\hat U(x) = e^{-i\theta_a(x) \hat T_a}
\end{eqnarray}
is a  unitary  $n\times n$ matrix. 
The non-Abelian character of the symmetry group is reflected by the 
commutation relations of the generators 
\begin{eqnarray}
\Big[\hat T_a,\,\hat T_b\Big]= i f_{ab}^c \hat T_c\,,
\end{eqnarray}
where  $f_{ab}^c$ denote the  structure constants of the Lie algebra 
of the underlying symmetry group. 

Invariance of the Lagrangian under such a transformation is achieved by 
introducing in analogy to the Abelian case a vector 
field  $\hat A_\mu(x)$ which now is a  $n\times n$ 
 matrix field
\begin{eqnarray}
\Big(\hat A_\mu(x)\Big)_{c' c}=  A_\mu^a(x)\Big(\hat T_a\Big)_{c' c}\,.
\end{eqnarray}
Thus the number of independent gauge fields  $ A_\mu^a(x)$ 
equals the number of generators. The gauge field transforms according to
\begin{eqnarray}
\hat A_\mu(x)\,\, \Longrightarrow \,\,\hat U(x)\Big( \hat A_\mu(x)
-\frac{i}{g} \partial_\mu\Big) \hat U^{-1}(x)\,,
\end{eqnarray}
and correspondingly, the covariant derivative becomes a 
$n\times n$ matrix operator 
\begin{eqnarray}
\Big(\hat D_\mu\Big)_{c' c}= \delta_{c' c}\partial_\mu +ig 
A_\mu^a(x)\Big(\hat T_a\Big)_{c' c}\,,
\end{eqnarray}
where  $g$ denotes a coupling constant to be determined from experiment.
Also the field strength tensor is represented by a $n\times n$ matrix 
\begin{eqnarray}
\hat F_{\mu\nu}(x)\,& = &\,
\frac{1}{ig} \Big[\hat D_\mu,\,\hat D_\nu\Big]\nonumber\\
\, & =&\, F_{\mu\nu}^a(x)\hat T_a\,,
\end{eqnarray}
where
\begin{eqnarray}
F_{\mu\nu}^a(x)\,& = &\,
 \partial_\mu A_\nu^a(x) - \partial_\nu A_\mu^a(x)
 -g f^a_{bc} A_\mu^b(x) A_\nu^c(x)\,.
\end{eqnarray}
The essential difference to an Abelian gauge group lies in the fact, that the 
field strength tensor contains terms  quadratic in the gauge field. 
Consequently, already the free gauge Lagrangian, 
\begin{eqnarray}
{\cal L}_{g.f.} = -\frac{1}{4} F_{\mu\nu}^a F^{a,\,\mu\nu}\,,
\end{eqnarray}
describes  self-interactions of the gauge field which are absent in 
the Abelian case. They are driven by the three- and four-gluon vertices in 
Fig.\ \ref{figgluonself}.
\begin{figure}
\centerline {\psfig{file=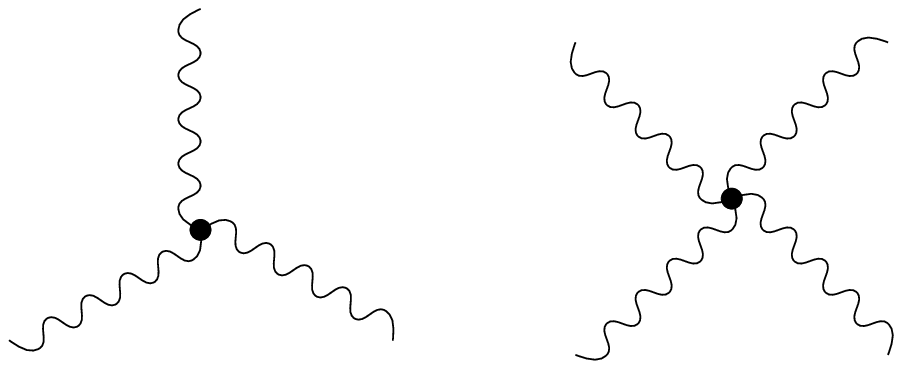,width=8cm,angle=0}}
\caption{figgluonself}
{Basic three- and four-gluon diagrams describing gluon self-interactions.}
\end{figure}

\subsection{Spontaneous Symmetry Breaking and Mass Generation}
%(Higgs-Kibble Mechanism)}
A very important step towards the unification of the e.m.\ and weak 
interactions was the observation that the massless gauge bosons can acquire 
effectively masses by the Anderson-Higgs-Kibble mechanism of spontaneous 
symmetry breaking via the coupling to massless scalar Higgs-fields.
I will illustrate this for the case of a three-component gauge field  
$\vec{A}_\mu$ (combined as a vector) which arises in gauging 
a  $SU(2)$  internal symmetry. In this case, the Higgs-field consists of 
two complex scalar fields constituting a  $SU(2)$  doublet 
\begin{eqnarray}
\phi = \left( \begin{array}{c} \phi_1 \\ \phi_2 \end{array}
\right)\,,
\end{eqnarray}
with a Lagrangian of the form 
\begin{eqnarray}
{\cal L} = 
-\frac{1}{4} \vec{F}_{\mu\nu} \cdot 
\vec{F}^{\,\mu\nu}+D^\mu\phi^\dagger D_\mu\phi
-V(\phi)\,,
\end{eqnarray}
where the covariant derivative is given by 
\begin{eqnarray}
D_\mu=\partial_\mu + \frac{ig}{2} \vec{A}_\mu(x) \,\cdot 
\vec{\tau}\,. 
\end{eqnarray}
The potential, which has first been considered in the  Ginzburg-Landau
theory of superconductivity and which later has also been used in the  
Goldstone model for the spontaneous breaking of a global symmetry, is given by 
the ansatz 
\begin{eqnarray}
V(\phi) = \frac{\lambda}{4}\Big(\phi^\dagger \phi -
\frac{\mu^2}{\lambda}\Big)^2\,,\quad \frac{\mu^2}{\lambda}>0\,.
\end{eqnarray}
It is illustrated in Fig.\ \ref{figgoldpot}. 
\begin{figure}
\centerline {\psfig{file=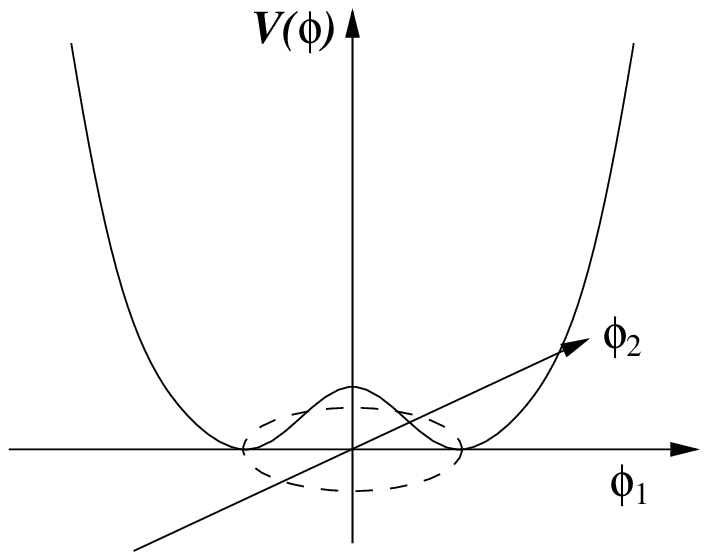,width=6cm,angle=0}}
\caption{figgoldpot}
{The Higgs potential.}
\end{figure}

This Lagrangian is invariant under the local gauge transformation
\begin{eqnarray}
\phi \,\,  \Longrightarrow \,\, e^{-\frac{i}{2}
\vec{\theta}(x) \cdot \vec{\tau}} \phi\,.
\end{eqnarray}
In view of the potential minimum at 
$\phi^\dagger \phi=v=\frac{\mu}{\sqrt{\lambda}}$, one 
makes for the Higgs field the ansatz
\begin{eqnarray}
\phi(x) = e^{-\frac{i}{2\sqrt{2}v}\vec{\tau}\cdot
\vec{\xi}(x)}
\left( \begin{array}{c} 0 \\ v+\frac{1}{\sqrt{2}}\eta(x) \end{array}
\right)\,.
\end{eqnarray}
In the case of a spontaneously broken global symmetry, the 
$\vec{\xi}(x)$-field would describe massless  Goldstone bosons. 
However, for a spontaneously broken local gauge symmetry, it 
can be transformed away by a proper  gauge fixing. Then one has 
\begin{eqnarray}
\phi(x) = 
\left( \begin{array}{c} 0 \\ v+\frac{1}{\sqrt{2}}\eta(x) \end{array}
\right)\,,
\end{eqnarray}
and the Lagrangian takes the form 
\begin{eqnarray}
{\cal L}\,&  = &
\,  -\frac{1}{4} \vec{F}_{\mu\nu} \cdot
\vec{F}^{\,\mu\nu} + \frac{g^2v^2}{8} 
\vec{A}_\mu \cdot \vec{A}^{\,\mu}
+ \frac{1}{2} \partial^\mu\eta \partial_\mu\eta -
\frac{\mu^2}{2}\eta^2 +\cdots\,.
\end{eqnarray}
This Lagrangian now describes  massive gauge bosons of mass 
$M_B=\frac{1}{2}gv$ as well as one  massive Higgs particle 
which appears as consequence of the spontaneously broken local gauge symmetry. 

\subsection{The Electroweak Lagrangian of the Standard Model}

To conclude this section, I will summarize the Lagrangian of the 
electroweak interaction of the standard model according to the Particle 
Data Group summary \cite{PDG96} (see also Chap.\ 6 of \cite{Hua92}). 
The standard model is based on the group  $SU(2)\times U(1)$ with 
three gauge bosons  $W^j_\mu$ ($j=1,2,3$) for  $SU(2)$ 
and one gauge boson  $B_\mu$ for  $U(1)$. The 
corresponding gauge coupling constants are denoted by  $g$ and $g'$, 
respectively. The  left-handed fermion fields 
(leptons and quarks) 
\begin{eqnarray}
\psi_{L\alpha} = \left( \begin{array}{c} \nu_{L\alpha} \\ 
l_{L\alpha}^- \end{array}
\right)\quad \mbox{and} \quad \left( \begin{array}{c} u_{L\alpha} \\ 
d_{L\alpha}' \end{array}
\right)
\end{eqnarray}
of the  $\alpha^{th}$ fermion family transform as  doublets 
under  $SU(2)$, where  $ d_{\alpha}'= \sum_\beta V_{\alpha\beta}d_\beta$ 
($V$ is the Cabibbo-Kobayashi-Maskawa mixing matrix). The right-handed fermion 
fields $\psi_{R\alpha}$ transform as singlets under $SU(2)$. In the minimal 
model, one considers three lepton and quark fermion families and one complex 
Higgs doublet. 
Spontaneous symmetry breaking results in three massive gauge bosons, two 
charged ones  $W^\pm=(W^1\mp W^2)/\sqrt{2}$ and one neutral 
 $Z=-B\sin\theta_W+ W^3\cos\theta_W$, and one massless boson, the 
photon  $A=B\cos\theta_W +W^3\sin\theta_W$, where 
 $\theta_W= \tan^{-1}(g'/g)$ denotes the weak angle. The weak boson masses 
are given by $M_{W^\pm}=80.22$ GeV and $M_{Z}=91.19$ GeV. 

After spontaneous symmetry breaking one arrives at the following 
fermion Lagrangian 
\begin{eqnarray}
{\cal L}_{F} &=& \sum_\alpha \bar \psi_\alpha^\dagger 
\Big( i\partial\!\!\!/ - m_\alpha -\frac{gm_\alpha H}{2M_W}\Big)\psi_\alpha
\nonumber\\
&&  -\frac{g}{2\sqrt{2}}\sum_\alpha \bar \psi_\alpha^\dagger 
\gamma^\mu(1-\gamma^5)
(T^+W_\mu^++T^-W_\mu^-)\psi_\alpha\nonumber\\
&& -e\sum_\alpha Q_\alpha \bar \psi_\alpha^\dagger \gamma^\mu\psi_\alpha A_\mu
 -\frac{g}{2\cos\theta_W}\sum_\alpha \bar \psi_\alpha^\dagger\gamma^\mu(
g_v^\alpha-g_a^\alpha\gamma^5)\psi_\alpha Z_\mu\,.\label{standmod}
\end{eqnarray}
$T^\pm$ denote weak isospin raising and lowering operators, respectively. 
The elementary electric charge is given by  $e=g\sin\theta_W$ and 
the vector and axial couplings are 
\begin{eqnarray} 
g_v^\alpha=t_{3}(\alpha) -2Q_\alpha\sin^2\theta_W\,,\qquad 
g_a^\alpha=t_{3}(\alpha)\,,
\end{eqnarray}
where  $t_{3}(\alpha)$ is the third component of the weak isospin of 
fermion $\alpha$ and  $Q_\alpha$ its charge. 
The second term describes the  charged-current weak interaction 
as, e.g., appears in beta decay ($G_F/\sqrt{2}=g^2/8M_W^2$), 
the third term the  e.m.\ interaction, and the last one the  
neutral-current weak interaction.

    % 

%\newpage
\setcounter{equation}{0}
\setcounter{figure}{0}
\setcounter{table}{0}

\section{Properties of the Electromagnetic Operators} 
\label{sec3}

In this Section I will give a brief review of the salient features of 
the e.m.\ current and two-photon operators which follow from the gauge 
invariance of the e.m.\ interaction. I will merely summarize the main 
results and refer the interested reader to my Schladming Lectures 
\cite{Are94}, where I have discussed the gauge conditions in greater detail. 

\subsection{Gauge Conditions for the Electromagnetic Operators}

In view of the fact, that the e.m.\ interaction is sufficiently weak, 
it is reasonable to assume that for the Hamiltonian of a hadronic system 
a Taylor expansion in the electromagnetic potential $A_\mu(x)$ exists, 
which reads up to second order 
\begin{eqnarray}
  H_{e.m.}(A,\,t) &=&
\int d^3x\,  j_\mu(x)A^\mu(x)|_{x_0=t}\nonumber\\
& &+{1\over 2}\int d^3x\,d^3y\, A^\mu(x)B_{\mu \nu}(x,\,y)
A^\nu(y)|_{y_0=x_0=t}+\cdots\,,
\end{eqnarray}
and which defines the e.m.\ current $j_\mu(x)$ and the two-photon operator 
$B_{\mu \nu}(x,\,y)$. Then gauge invariance leads to the following 
conditions for these operators: 

(i) The continuity equation of the electromagnetic current
\begin{eqnarray}
{\vec \nabla} \cdot {\vec j}\,(x)+i[ H_0,  \rho(x)]+
\partial_t \rho (x)=0\,,
\end{eqnarray}
or covariantly written 
\begin{eqnarray}
 \partial_\mu j^\mu (x)=0\,,
\end{eqnarray}
which implies conservation of the total charge of the system. 

(ii) The gauge condition for the  two-photon operator
\begin{eqnarray}
\partial^{\mu}_x B_{\mu \nu}(x,y)=i[ \rho(x),j_\nu(y)]\,,
\end{eqnarray}
which is essential for the derivation of the low energy theorem of the 
Compton amplitude. 
 
\subsection{Low Energy Theorems}

An important consequence of these gauge conditions are the  low-energy
theorems, which allow to derive very simple relations for the current and
two-photon operators  in the limit that the photon momenta go to zero. 
In the case of the current, it leads to the famous Siegert theorem while for 
the photon scattering amplitude such a theorem has been derived by 
Sachs and Austern \cite{SaA51}, and later on more general grounds by 
Low \cite{Low54} and Gell-Mann and Goldberger \cite{GeG54} 
for a spin-(1/2) particle. 

To this end one first introduces the  Fourier transforms of charge and
current densities and the two-photon operator as well, which are here 
assumed to be time independent for convenience, 
\begin{eqnarray}
 \tilde \rho ({\vec q}\,) &= &\, 
\int d^3x\, \rho({\vec x})e^{i{\vec q} \cdot {\vec x}}\,,\\
 {\vec J}({\vec q}\,) &=&\, \int d^3x\, {\vec j} ({\vec x})
e^{i{\vec q} \cdot {\vec x}}\,,\\
 \tilde B_{kl}({\vec q}^{\;\prime}, {\vec q}\,) &=&\,
  \int d^3x\, d^3y\, e^{i{\vec q^{\;\prime}} \cdot {\vec x}}
B_{kl}({\vec x},\,{\vec y})e^{i{\vec q} \cdot {\vec y}}\,.
\end{eqnarray}
Then the gauge condition for the current becomes 
\begin{eqnarray}
{\vec q} \cdot {\vec J} \,({\vec q}\,)= [H_0,\tilde \rho({\vec q}\,)]\,.
\end{eqnarray}
Assuming that the operators possess a Taylor expansion with respect to  
${\vec q}$, one finds up to first order
\begin{eqnarray}
J_l({\vec q}\,)= i [H_0,{\vec D}]-{1\over 2}[H_0,q_{l'}Q_{l'l}] -i({\vec q}
 \times {\vec M})_l\,,
\end{eqnarray}
where  ${\vec D},\,{\vec M}$ and  $Q_{l'l}$ denote 
respectively, electric and magnetic dipole, and electric quadrupole operators. 
In particular, one has for $\vec q =0$
\begin{eqnarray}
J_l(0)= i [H_0,{\vec D}]\,,
\end{eqnarray}
which is called the Siegert theorem. It is very useful in cases, when the 
current density is less well known than the charge density, because it 
allows to evaluate the current from the knowledge of the charge density 
alone in the low energy regime via the charge dipole operator. 
A similar low energy expansion is found for the two-photon operator
\begin{eqnarray}
\tilde B_{lk}({\vec q}^{\,\prime},\,{\vec q}\,)\,
&=&\, [D_k,[H_0,D_l]]
 +i[D_k, {1\over 2} q_{l^{\prime}}[H_0,Q_{l'l}]-i
({\vec q}\times {\vec M})_l]\nonumber\\
&& +i[D_l, {1\over 2} q_{l^{\prime}}^{\prime}[H_0,Q_{l'k}]
-i({\vec q}^{\,\prime}\times {\vec M})_k]\,,
\end{eqnarray}
yielding a corresponding Siegert theorem at $\vec q =\vec q^{\,\prime}=0$
\begin{eqnarray}
\tilde B_{lk}(0,0)= [D_k,[H_0,D_l]]\,.
\end{eqnarray}

\subsection{Multipole Decomposition}

As already mentioned, the observables are given in terms of the current 
matrix elements $J_{fi}^\mu$, which contain the information on the internal 
dynamics of the hadronic system. Since the intrinsic states of hadrons 
and nuclei can be classified according to the total angular momentum, 
it is very useful to decompose the charge and current operators in terms 
of multipole operators, which transfer a definite angular momentum. 
In this way one can take advantage of the angular momentum
selection rules and thus can separate the geometrical aspects 
from the  dynamical properties of the system. The other advantage of 
introducing these 
multipoles lies in the fact, that in principle, they can be determined 
from a complete set of observables and, thus, provide a convenient 
basis for the comparison between theory and experiment.

In detail, the multipole decomposition of charge and current 
density operators is given by 
\begin{eqnarray}
J_\lambda (\vec q\,) &=&\, (-)^\lambda \sqrt{2\pi(1+\delta_{\lambda 0})}
\sum_{LM}i^L\hat L {\cal O}^\lambda_{LM}D^L_{M\lambda}
(\phi_q,\theta_q,0)\,,
\end{eqnarray}
with
\begin{eqnarray}
 {\cal O}^\lambda_{LM} &=&\, 
\delta_{\lambda 0}C_{LM}+ \delta_{|\lambda|1}
(T_{LM}^{(e)}+\lambda T_{LM}^{(m)})\,,
\end{eqnarray}
with the charge multipole operators 
\begin{eqnarray}
C_{L M}=\int d^3x\,\rho({\vec x})j_L(qx) Y_{L M}(\Omega_x)\,,
\end{eqnarray}
and the transverse electric and magnetic multipole operators, respectively,
\begin{eqnarray}
T_{LM}^{(e)} &=&\, {1\over q}\int d^3x\, {\vec j}({\vec x})\cdot
({\vec \nabla}\times (j_L(qx){\vec Y}_{LLM}))\,,\\
 T_{LM}^{(m)} &=&\, \int d^3x\, {\vec j}({\vec x})\cdot
(j_L(qx){\vec Y}_{LLM})\,.
\end{eqnarray}
Here, the vector spherical harmonics are defined by
\begin{eqnarray}
{\vec Y}_{L\ell M}=\sum_{m \lambda}(\ell m 1\lambda|LM)Y_{\ell m}
{\vec e}_\lambda\,.
\end{eqnarray}

Since the multipole operators are irreducible spherical tensors of rank $L$, 
one can make use of the Wigner-Eckart theorem in evaluating them 
between states of definite angular momentum $|J M\rangle$: 
\begin{eqnarray}
\langle J_fM_f|J_\lambda({\vec q}\,)|J_i M_i\rangle\, &=&\,
  (-)^{\lambda +J_f-M_f}\sqrt{2\pi(1+\delta_{\lambda 0})}
\sum_{LM} i^L \hat L
\left(\matrix{J_f & L & J_i \cr-M_f & M & M_i \cr}\right)\nonumber\\
&& \langle J_f||{\cal O}_L^\lambda||J_i\rangle
D^L_{M\lambda}(\phi_q,\theta_q,0)\,.
\end{eqnarray}
The $3j$-symbol contains the  geometrical aspects, i.e., angular momentum 
selection rules, while the multipole strength is given by the  reduced
matrix element $\langle J_f||{\cal O}_L^\lambda||J_i\rangle$, which contain 
the information on the internal dynamics of the system.

\subsection{Generalized Siegert Theorem and Siegert Operators}\label{sec3sieg}

The above mentioned Siegert theorem can be generalized to the statement 
that in the low energy limit the transverse electric multipoles can be 
related to the charge multipoles of the same order.
This very important theorem for photo- and electronuclear theory 
allows then to calculate reliably within the limits of the theorem the
electric transition matrix elements from the charge density  
without explicit knowledge of the currents, e.g., exchange currents. 

The essential idea (for details see \cite{Are94}) rests on the observation 
that the  transverse electric multipole field ${\vec A}_{LM}(e)$ can be 
decomposed into a gradient field and a term which is two orders higher in $qx$
\begin{eqnarray}
 {\vec A}_{LM}(e)={\vec \nabla} \Phi_{LM}({\vec x},\,q) +
{\vec A}_{LM}^{\prime}(e;\Phi)\,,
\end{eqnarray}
where the specific form of ${\vec A}_{LM}^{\prime}(e;\Phi)$
depends on the choice of $\Phi_{LM}$ \cite{ScW90}. 

With the help of current conservation, one arrives at 
\begin{eqnarray}
\langle f|T_{LM}^{(e)}|i\rangle\, &=&\, i\int d^3x\,
\langle f|[H^{int},\rho ({\vec x})]|i\rangle
\Phi_{LM}({\vec x},q)\nonumber\\
&& +\int d^3x\,\langle f|{\vec j}_{int}({\vec x})|i\rangle \cdot
{\vec A}_{LM}^{\prime}(e;\Phi)\,,
\end{eqnarray}
where the first term defines a  Siegert operator for the electric
multipole whose form depends on the choice of  $\Phi_{LM}$. The sub- and 
superscript ``$int$'' refers to the internal motion \cite{ScW90}. 
A particularly important consequence is that a dominant part of 
meson exchange currents -- essentially in the electric dipole ($E1$) -- 
is already included implicitly in the Siegert operators \cite{Are81,GaH81}.

    % 

%\newpage
\setcounter{equation}{0}
\setcounter{figure}{0}
\setcounter{table}{0}

\section{Models for the Electromagnetic Operators} 
\label{sec4}
From the fermion part of the Lagrangian of the standard model 
in (\ref{standmod}) one obtains 
immediately the electromagnetic and neutral currents from the 
coupling of the fermions to photon and $Z$-boson, respectively, by writing 
these couplings in the form
\begin{eqnarray}
{\cal L}_F^\gamma + {\cal L}_F^Z= -e j^{(\gamma)}_\mu A^\mu -\frac{g}{2\cos 
\theta_W} \Big( j^{(Z_v)}_\mu - j^{(Z_a)}_\mu \Big)Z^\mu\,,
\end{eqnarray}
where one has for the e.m.\ current
\begin{eqnarray}
j^{(\gamma)}_\mu = \sum_\alpha Q_\alpha \bar \psi_\alpha^\dagger 
\gamma_\mu \psi_\alpha\,.
\end{eqnarray}
The neutral current splits into a vector 
\begin{eqnarray}
j^{(Z_v)}_\mu &=& \sum_\alpha g^\alpha_v \bar \psi_\alpha^\dagger 
\gamma_\mu \psi_\alpha\,,
\end{eqnarray} 
and an axial vector current
\begin{eqnarray}
j^{(Z_a)}_\mu &=& \sum_\alpha g^\alpha_a \bar \psi_\alpha^\dagger 
\gamma_\mu \gamma_5 \psi_\alpha\,.
\end{eqnarray}
Here, I will not consider the additional charged currents describing the 
coupling to the charged bosons $W^{\pm}$, which are also given in 
(\ref{standmod}). 

In principle, in order to get the hadronic currents, one has to evaluate 
these operators for the quark fields between the internal hadron wave 
functions given in terms of quark and gluon d.o.f. 
Unfortunately, these are not known yet due to the difficulties one 
encounters in solving QCD in the non-perturbative regime. Therefore one 
resorts either to effective hadron models, e.g., constituent quark models, 
or to a purely phenomenological description. In the latter case, one makes 
the most general ansatz for the one-body hadron currents 
which is allowed by Lorentz covariance, parity and time reversal invariance. 
The remaining form factors, undetermined functions of the only independent 
Lorentz scalar $q^2$ for on-shell particles, have to be taken from 
experiment. I will now briefly review the one- and two-body electroweak 
currents for nucleons. 

\subsection{One-Body Currents}

The on-shell nucleon vector currents (e.m.\ and neutral), fulfilling 
the requirements of Lorentz covariance, parity and time reversal invariance, 
have the well known form 
\begin{eqnarray}
\langle p'|J_\mu^V(0)|p\rangle = \bar u(p'\,)\Big[F_1(q^2)\gamma_\mu
+\frac{i}{2M}F_2(q^2)\sigma_{\mu\nu}q^\nu
%+\frac{1}{2M}F_3(q^2)q_\mu
\Big]u(p)\,,
\label{onebodycurr}
\end{eqnarray}
with  $q=p'-p$ and the Dirac and Pauli form factors, $F_1(q^2)$ and 
$F_2(q^2)$, respectively. The nucleon mass is denoted by $M$. 
For off-shell nucleons additional form factors 
arise which, however, are model dependent and not directly observable. 
It is convenient to introduce the Sachs form factors by 
\begin{eqnarray}
G_E(q^2) & =& 
F_1(q^2)+\tau F_2(q^2)\,,\quad\mbox{where } \tau=\frac{q^2}{4M^2}\,,\\
 G_M(q^2) & =&  F_1(q^2)+ F_2(q^2)\,, 
\end{eqnarray}
because these are directly observable, for example, in elastic electron 
nucleon scattering for the e.m.\ form factors. The simplest models for 
them are the purely phenomenological dipole form factors \cite{GaK71} or 
the ones of the vector meson dominance model (VMD) \cite{HoP76}.
Applying a $p/M$-expansion, one finds the corresponding nonrelativistic 
expressions 
\begin{eqnarray}
\langle p'|\rho(0)|p\rangle & =& 
\chi^{\prime\,\dagger} G_E(q^2)\chi\,,\\
\langle p'|{\vec j}(0)|p\rangle & =& 
\chi^{\prime\,\dagger} \frac{1}{2M}\Big[
G_E(q^2)({\vec p}^{\,\prime}+{\vec p}\,)+i G_M(q^2)
\vec \sigma\times {\vec q}\,\Big]\chi\,,
\end{eqnarray}
where $\chi'$ and $\chi$ denote the nucleon Pauli spinors. Note, that the 
current density is one order in $p/M$ higher than the charge density. 

Correspondingly, one finds for the axial currents for on-shell nucleons 
the general form 
\begin{eqnarray}
\langle p'|J_\mu^A(0)|p\rangle = \bar u(p'\,)\Big[G_A(q^2)\gamma_\mu
+\frac{G_P(q^2)}{M}q_\mu\Big]\gamma_5u(p)\,,
\end{eqnarray}
with the axial form factor $G_A(q^2)$ and the induced pseudoscalar form 
factor $G_P(q^2)$.
Again, the nonrelativistic reduction reads 
\begin{eqnarray}
\langle p'|\rho^A(0)|p\rangle & =& \chi^{\prime\,\dagger} 
\Big[\frac{G_A(q^2)}{2M} \vec \sigma
\cdot ({\vec p}+{\vec p}^{\,\prime}) \Big]\,
\chi\,,\\
\langle p'|{\vec j}^A(0)|p\rangle & =& -\chi^{\prime\,\dagger} 
\Big[G_A(q^2)\vec \sigma \Big]\chi\,. 
\end{eqnarray}
Here, it is the axial charge density which is one order in $p/M$ higher 
than the current density. A discussion of the neutral current form factors 
(vector and axial) and their phenomenological parametrizations is given in 
\cite{MuD94}.

\subsection{Two-Body Meson Exchange Currents} \label{secmec}
I can be very brief in this case, because this topic will be covered by the 
lectures of J.F.\ Mathiot at this workshop \cite{Mat97}. 
I only will give as most important example the explicit form of the pion 
exchange current in lowest, i.e., static order
\begin{eqnarray}
 \rho_{[2]}^{\,\,\pi}({\vec x}) & =& 0\,,\\
{\vec j}_{[2]}^{\,\,\pi}({\vec x})
 & =& -{f_\pi^2 \over m^2_\pi}
({\vec \tau}_1 \times {\vec \tau}_2)_3
\Big[\delta({\vec x} -{\vec r}_1){\vec \sigma}_1({\vec \sigma}_2\cdot {\vec 
\nabla}_2) J_{m_\pi}(r_{12}) - (1 \leftrightarrow 2)\nonumber\\
& & +({\vec \sigma}_1\cdot {\vec \nabla}_1)({\vec \sigma}_2\cdot {\vec 
\nabla}_2)
J_{m_\pi}({\vec r}_1 - {\vec x})\stackrel{\leftrightarrow}{\nabla}_x
J_{m_\pi}({\vec x} - {\vec r}_2)\Big]\,,\label{picurrent}
\end{eqnarray}
with the Yukawa function
\begin{eqnarray}
 J_{m_\pi}(r)={e^{-m_\pi r}\over 4\pi r}\,.
\end{eqnarray}
Pion mass and $\pi N$ coupling constant are denoted by $m_\pi$ and $f_\pi$, 
respectively. 
The vanishing of the charge density supports the early hypothesis of 
Siegert \cite{Sie37}, expressing the fact that the charge densities of the 
oppositely charged mesons, exchanged in opposite directions between proton 
and neutron and contributing with equal weight, cancel each 
other, whereas the currents associated with them do add. The first line of 
(\ref{picurrent}) describe the pair or contact current, and the second the 
pion current, where the photon interacts with the pion in flight (see Fig.\
\ref{figpionexch} for a diagrammatic representation). 
\begin{figure}
\centerline {\psfig{file=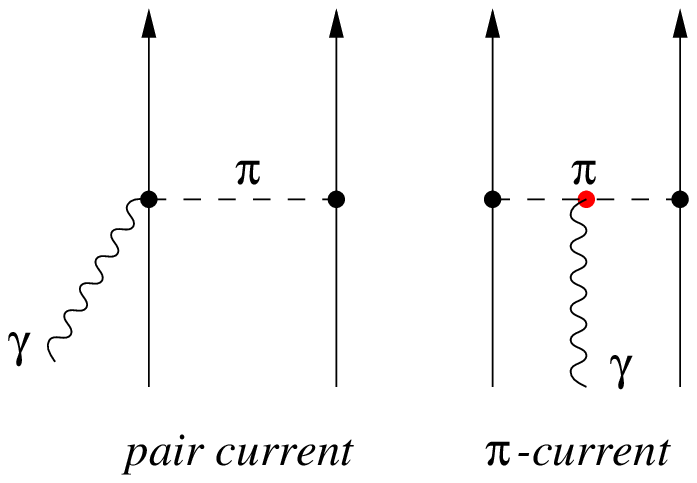,width=7cm,angle=0}}
\caption{figpionexch}
{Pair or contact and pion current contributions to the pion exchange current.}
\end{figure}
Analogous contributions arise from heavier meson exchanges 
($\sigma,\,\rho,\,\omega\,\dots$) and also  $\rho/\omega\pi\gamma$ 
\cite{AdT89,AdA97}. 
Furthermore, similar two-body contributions exist for the neutral currents 
\cite{HaH89}. 

Also for the two-photon operator, one will have corresponding exchange 
contributions. I show in Fig.\ \ref{figpiontwophoton} as an example 
the lowest order $\pi$ exchange contribution to the two-photon operator 
\cite{Fri76,Are80}. 
\begin{figure}
\centerline {\psfig{file=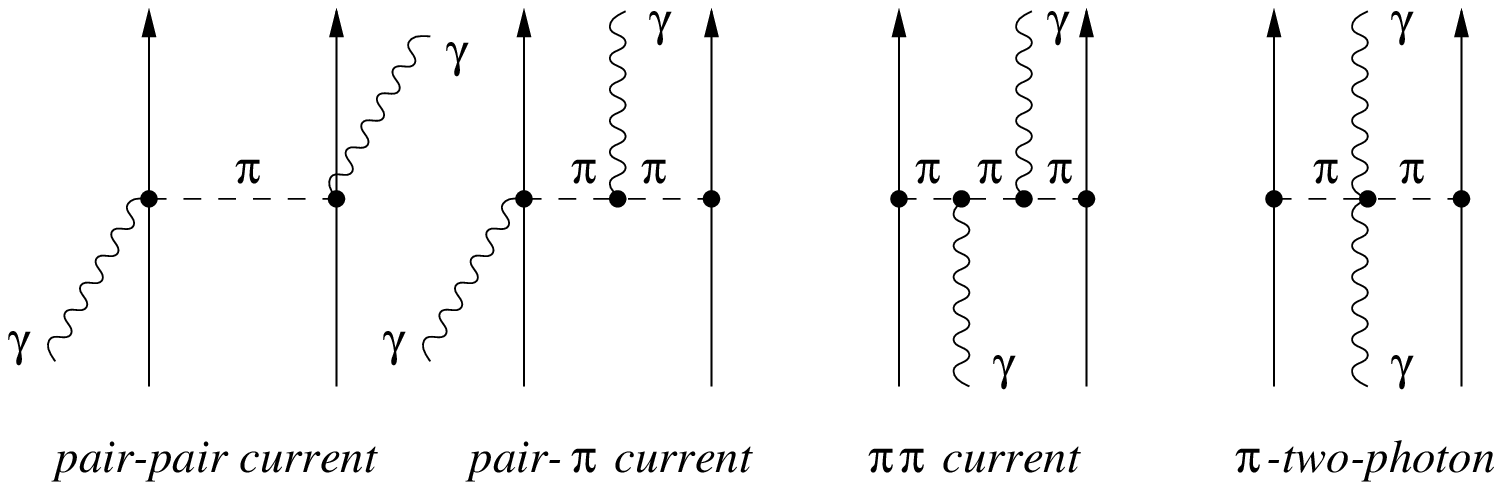,width=14cm,angle=0}}
\caption{figpiontwophoton}
{Various pair and pion current contributions to the pion exchange 
two-photon operator.}
\end{figure}

\subsection{Isobar Currents}
The existence of internal nucleon structure as manifest in the rich 
spectrum of internally excited nucleon resonances, or nucleon isobars, 
leads naturally to additional e.m.\ interaction operators in terms of 
nucleon-isobar transition currents as well as diagonal isobar currents. As 
most prominent example let me give the nonrelativistic expressions for the 
$\Delta(1232)$ resonance (for details see \cite{WeA78}).

(i)  $N\Delta$ transition current (nonrelativistic): 
\begin{eqnarray}
\langle \Delta\,p'|\rho (0)|N\,p\rangle & =& 0\,,\\
\langle \Delta\,p'|{\vec j}(0)|N\,p\rangle & =& 
\frac{1}{ 2M}\chi_\Delta^{\prime\,\dagger}\Big[
G_M(q^2){\vec \sigma_{\Delta N}}\times {\vec q}\,\Big]\chi\,,
\end{eqnarray}
where $\chi_\Delta^{\prime}$ denotes the final $\Delta$-(3/2)-spinor and 
$\sigma_{\Delta N}$ the $N\rightarrow\Delta$ transition spin operator. 
Here I have restricted myself to the dominant $M1$ part neglecting 
the small $C2$ and $E2$ contributions. 

(ii) diagonal $\Delta$ current (nonrelativistic):
\begin{eqnarray}
\langle \Delta\,p'|\rho(0)|N\,p\rangle & =&
\chi^{\prime\,\dagger}_\Delta G_E^\Delta(q^2)\chi_{_\Delta}\,,\\
\langle \Delta\,p'|{\vec j}(0)|N\,p\rangle & =&
\frac{1}{2M_\Delta}\chi_\Delta^{\prime\,\dagger} \Big[
G_E^\Delta(q^2)({\vec p}^{\,\prime}+{\vec p}\,)+i G_M^\Delta(q^2)
{\vec \sigma}_\Delta\times {\vec q}\,\Big]\chi_{_\Delta}\,,
\end{eqnarray}
where $M_\Delta$ denotes the $\Delta$ mass. The electric and magnetic form 
factors of the $\Delta$ are denoted by $G_E^\Delta(q^2)$ and 
$G_M^\Delta(q^2)$, respectively. Possible $E2$ and $M3$ contributions have 
been neglected for convenience. 
Analogous currents appear for higher resonances  ($N(1440),\,D_{13}(1520),\,
S_{11}(1535),\dots$) \cite{WeA78,ScW96}. 

These isobar currents can be incorporated either in terms of effective, 
non-local 
two-body operators where the intermediate appearance of an isobar is 
implicitly included (see Fig.\ \ref{figeffectivop})
\begin{figure}
\centerline {\psfig{file=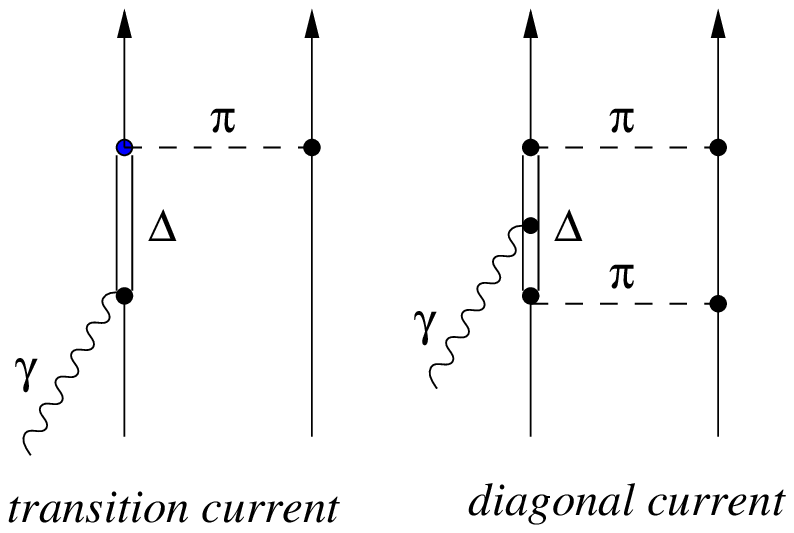,width=7cm,angle=0}}
\caption{figeffectivop}
{Isobar contributions to the e.m.\ current as effective two-body operators.}
\end{figure}
or in the framework of  nuclear isobar configurations (IC) 
as explicit constituents of the nuclear wave function \cite{ArW72,Gre76,WeA78}. 

\subsection{Meson Production Currents}
The internal nucleon structure becomes also manifest in the e.m.\ meson 
production on a nucleon. As an example, I will consider briefly pion 
photoproduction. More detailed reviews may be found in \cite{Dav94}. 
The lowest order tree diagrams for the elementary production 
process on the nucleon are shown in Figs.\ \ref{figresonanceterm} and 
\ref{figbornterm} for the resonance contribution and the Born terms. 
Restriction to these lowest order tree diagrams would violate unitarity. 
Therefore, the tree diagrams have to be either unitarized \cite{Dav94}
or used as driving terms in a dynamical model as represented diagrammatical 
in Fig.\ \ref{figdynmod}. A recent critical discussion of the various methods 
is given in \cite{WiW98}, where it is emphasized, that the resonance 
excitation multipoles cannot be extracted in a model independent way, and 
thus do not constitute observables in the strict sense \cite{WiW96} (see 
also the seminar by Th.\ Wilbois at this workshop). 
\begin{figure}[h]
\centerline {\psfig{file=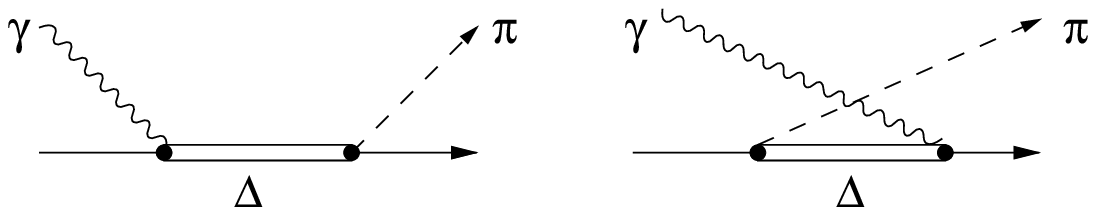,width=10cm,angle=0}}
\caption{figresonanceterm}
{Direct and crossed $\Delta$ resonance contributions to pion photoproduction.}
\end{figure}
\begin{figure}[h]
\centerline {\psfig{file=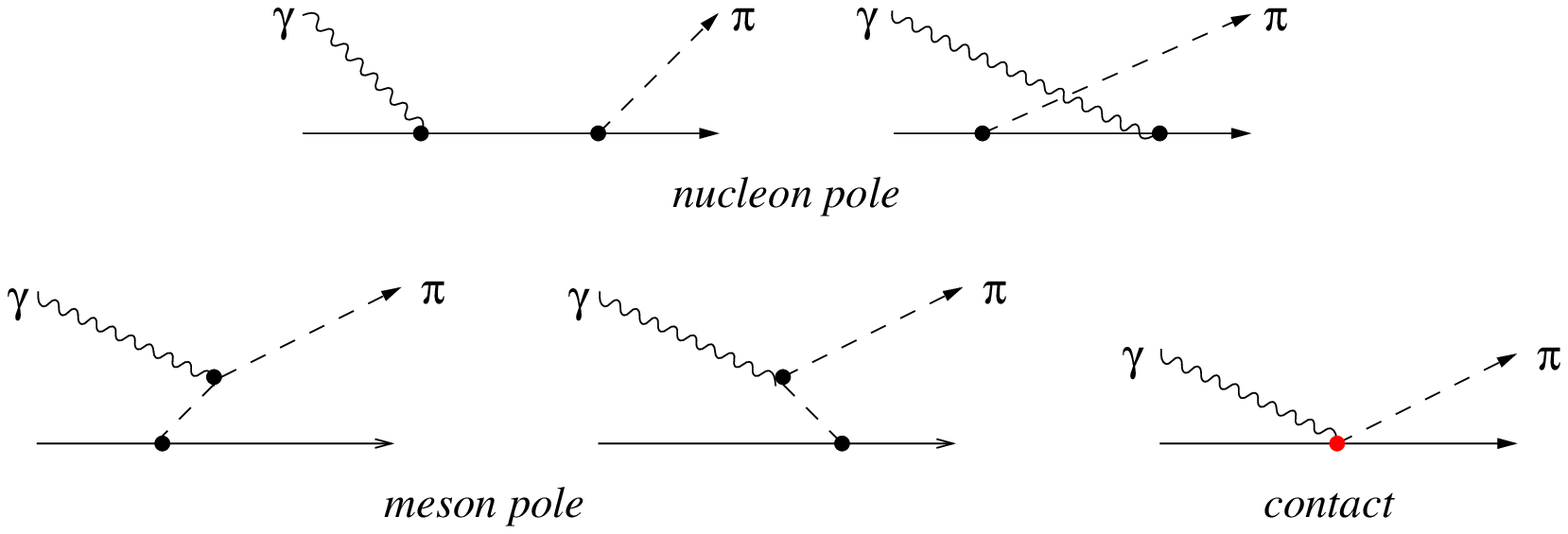,width=12cm,angle=0}}
\caption{figbornterm}
{Born contributions to pion photoproduction.}
\end{figure}
\begin{figure}
\centerline {\psfig{file=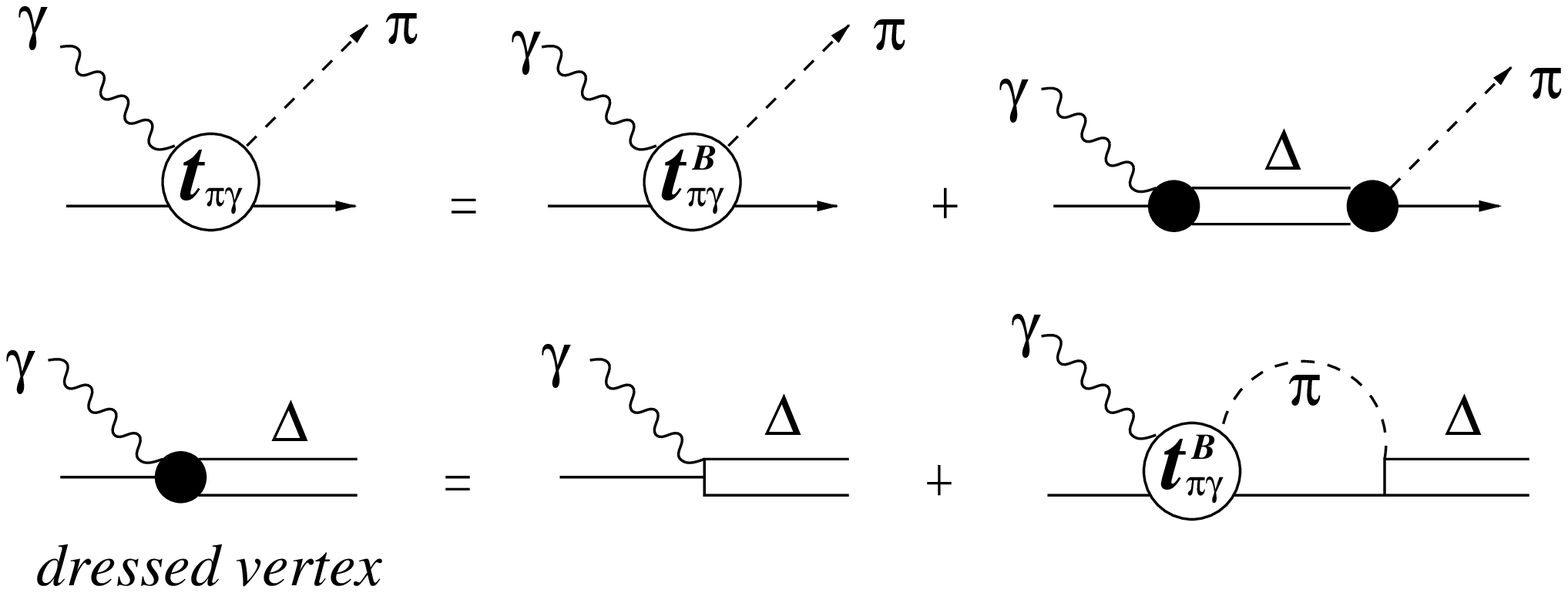,width=12cm,angle=0}}
\caption{figdynmod}
{Dynamical model of pion photoproduction.}
\end{figure}

\subsection{Relativistic Contributions}

As last topic in this section, I will briefly touch upon the role of 
relativity. It is clear that with increasing energy and momentum transfers, 
relativistic effects will become significant and, thus, have to be included. 
One may distinguish different types of relativistic contributions. First of all,
there are relativistic contributions to the electroweak operators. For example, 
from the relativistic expression in (\ref{onebodycurr}) on obtains in the 
$p/M$-expansion as leading order relativistic contribution the spin-orbit 
and Darwin-Foldy currents plus additional terms. Analogous contributions arise 
for the two-body currents, in particular from retardation in propagators 
of the exchanged mesons. Secondly, relativistic effects appear in the 
internal rest-frame wave function from  
the internal relativistic dynamics of the hadrons. Additional effects arise 
from the Lorentz boost of the internal wave function from the rest system 
to a moving frame taking into account the effect of Lorentz contraction and 
Thomas-Wigner rotation. Since the Lorentz boost is represented by a unitary 
operator acting on the rest frame wave function, its effect can be incorporated 
into the current density operators so that these are evaluated between rest 
frame wave functions. A large part of these operators is determined from the 
fact, that they have to fulfil the requirements of Lorentz covariance 
which is formulated in the form of commutator relations with the generators 
of the Poincar\a'e group. A systematic exploitation of these relations can 
be found in \cite{AdA96}, and a complete listing of all leading order one- 
and two-body currents for scalar, vector and pseudoscalar meson exchange is 
given in \cite{AdA97}, which incorporate also the boost contributions.

    % 

%\newpage
\setcounter{equation}{0}
\setcounter{figure}{0}
\setcounter{table}{0}

\section{Basic Electroweak Processes} 
\label{sec5}
Because of the small electroweak coupling constants, the most important 
processes are 
the  one- and two-boson processes, whereby free one- and two-boson 
processes are only realizable with photons. 

\subsection{One-Boson Processes}

It is useful to distinguish between real and virtual one-boson 
processes:

(i) Real photon reactions comprise absorption and emission 
by a hadronic system with internal excitation or deexcitation, 
respectively (see Fig.\ \ref{figonephotonprocess}a). 
Since real photons have transverse polarization, only the 
transverse current contributes. Energy and momentum transfers are 
not independent from each other but are related by $\omega= q$. 
\begin{figure}[h]
\centerline {\psfig{file=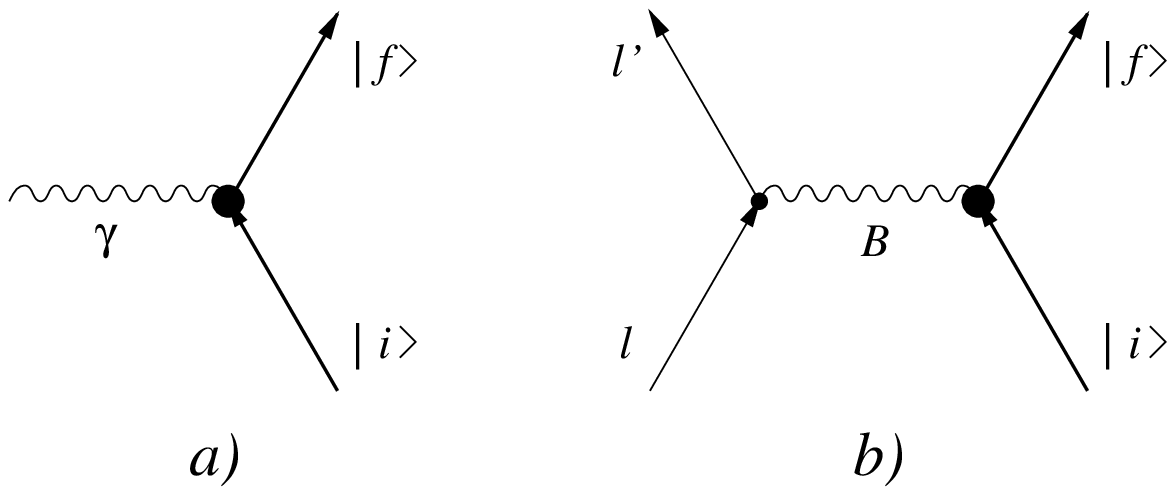,width=11cm,angle=0}}
\caption{figonephotonprocess}
{a) Photon absorption or emission by a hadronic system; b)
One-boson exchange diagram for lepton scattering off a hadron.}
\end{figure}

(ii) Virtual bosons appear in lepton scattering off a hadron, being exchanged 
between the lepton and the target as, for example, in elastic and inelastic 
electron and neutrino scattering off hadronic systems (see Fig.\ 
\ref{figonephotonprocess}b). Energy and momentum 
transfers can be varied independently within the range $\omega^2<q^2$, and in 
addition also longitudinal polarization of the virtual boson will contribute. 
For  $(e,\,e')$ and  $(\nu,\,\nu')$ one has a neutral current 
interaction  ($B=\gamma,\,Z$), while a charged current interaction ($B=W^\pm$) 
appears in $(e,\,\nu)$, for example. 
Since the weak gauge bosons couple to both, vector and axial vector 
currents, one has the interesting phenomenon of parity violation, 
e.g.\ in  $(e,\,e')$, which arises via the interference of $\gamma$- 
and  $Z$-exchange. Indeed, at present several experiments are underway to 
determine the  strange sea quark contribution to the nucleon form factors 
from the parity violating asymmetry of longitudinally polarized 
electrons (see Sect.\ \ref{sec6parity}). 

\subsection{Two-Boson Processes}

In the case of two-boson processes, a variety reactions are possible 
depending on the participation of either two real photons, one real 
photon and one virtual boson, or two virtual 
bosons. I have selected for illustration a few examples: 

(i) Fig.\ \ref{figcompton} shows the lowest order diagrams contributing to 
elastic and inelastic photon scattering (Compton and Raman scattering). 
\begin{figure}
\centerline {\psfig{file=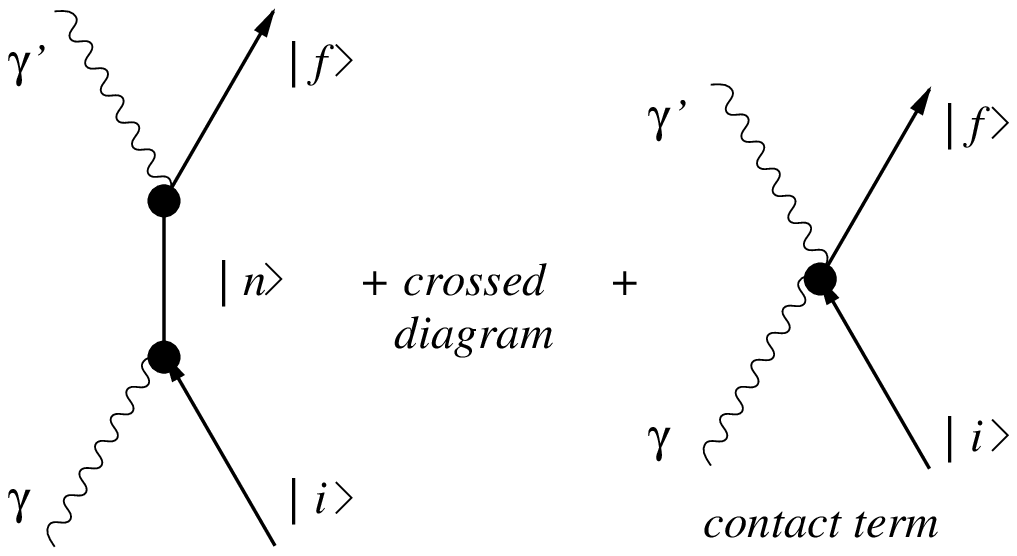,width=10cm,angle=0}}
\caption{figcompton}
{Two-step or dispersion (left) and one-step or contact (right) 
contributions to photon scattering.}
\end{figure}
I would like to emphasize, that the separation into two-step (dispersion 
diagram)  and one-step 
(contact diagram) contributions is gauge dependent, which means that the 
splitting into these two contributions is not observable. 

(ii) A bremsstrahlung process with one virtual boson exchange and a real 
photon emission on the hadronic side is depicted on the left side of 
Fig.\ \ref{figbremsdisp}. Here, the exchanged virtual boson can be both, 
a $\gamma$ or a $Z$. A corresponding process can occur on the leptonic side. 

(iii) Finally, a dispersion contribution to electron scattering as a 
second order correction to the leading one-photon exchange contribution 
is shown on the right side of Fig.\ \ref{figbremsdisp}. 
It corresponds to the Compton process with two virtual photons. Also here, 
$Z$-exchange is possible 
in both places. 
\begin{figure}[h]
\centerline {\psfig{file=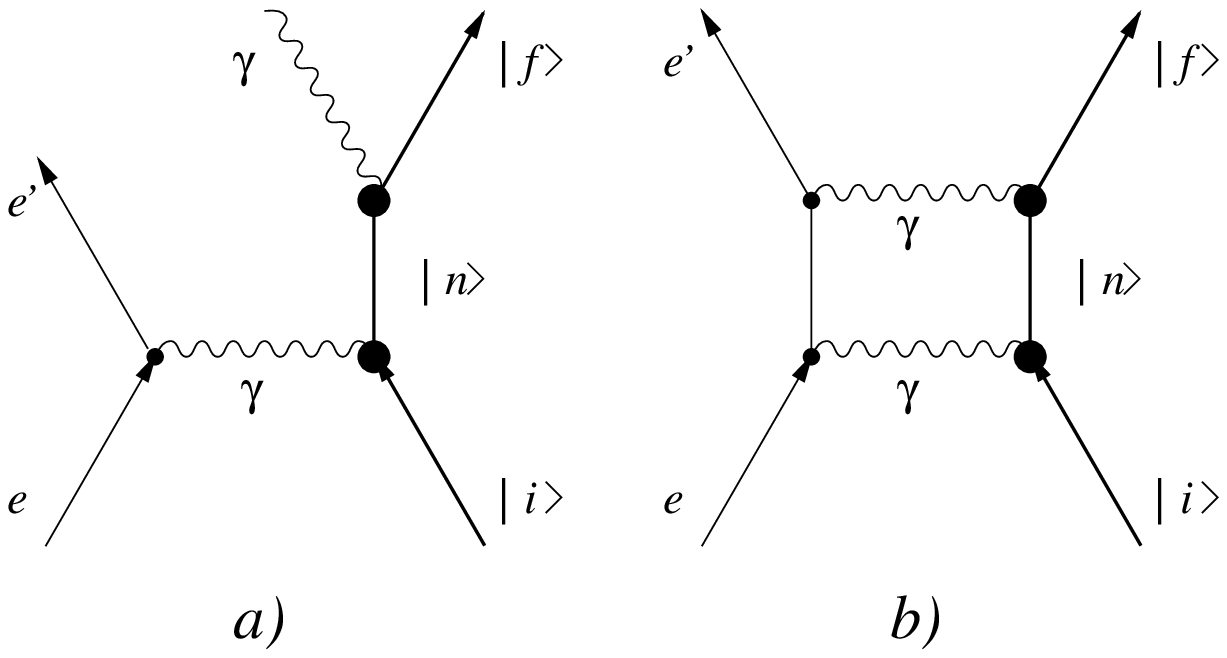,width=12cm,angle=0}}
\caption{figbremsdisp}
{Left: Bremsstrahlungs process at the hadronic leg; Right: 
Dispersion contribution to electron scattering.}
\end{figure}

\subsection{Cross Section for Electron Scattering}

Electron scattering is a particularly important example of a one-photon 
process. Evaluation of the diagram in Fig.\ \ref{figphotonexch} leads to 
the well-known expression for the cross section 
\begin{eqnarray}
d\sigma_{fi} & = & (2\pi)^{-5}
\delta^{(4)}(P_f-q-P_i)
\,\frac{m_e^2\,d^{3}k_f}{4k_{i\, 0}k_{f\,0}M_{i}}
\,\sum_f tr({\cal M}^\dagger_{fi}\hat \rho^f{\cal M}_{fi}\hat \rho^e 
\hat \rho^i)
\,,
\end{eqnarray}
where  $q=k_i-k_f$ denotes the momentum transfer. The notation for 
the momenta of electron and hadron is explained in Fig.\ 
\ref{figphotonexch}. 
\begin{figure}
\centerline {\psfig{file=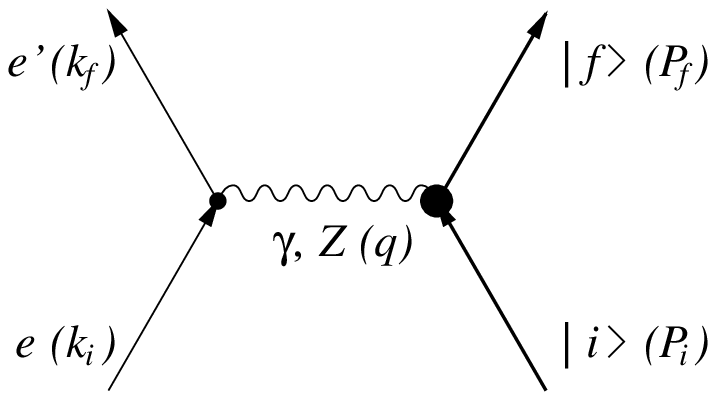,width=8cm,angle=0}}
\caption{figphotonexch}
{One-photon exchange diagram for electron scattering.}
\end{figure}
The initial electron and hadron polarization density matrices are denoted 
by  $\hat \rho^e$ and  $\hat \rho^i$, respectively. The 
phase space density of the final states including possible polarization 
analysis is described by  $\hat \rho^f$. Its specific form depends 
on the experimental conditions, whether one considers an inclusive 
process with no specific analysis of the final states, or an  
exclusive reaction with a more detailed analysis of the final hadronic 
system. 

In the one-boson-exchange approximation, the invariant matrix element 
${\cal M}_{fi}$ contains two contributions from virtual  $\gamma$ and 
 $Z$  exchange with the latter naturally being
strongly suppressed at not too high momentum transfers 
($-q_\mu^2 \ll M^2_Z$), 
\begin{eqnarray}
 {\cal M}_{fi}= \frac{e^2}{q_\mu^2}\ j^{(\gamma)\,\mu}J_{fi,\,\mu}^{(
\gamma)}
 + \sqrt{2}G_F \,j^{(Z)\,\mu}J_{fi,\,\mu}^{(Z)} \,.
\end{eqnarray}
The lepton and hadron currents are denoted by  $j^{(\gamma/Z)}_{\mu}$ and
 $J_{fi,\,\mu}^{(\gamma/Z)}$, respectively. The superscripts ``$\gamma$'' 
and ``$Z$'' indicate the electromagnetic and weak neutral current 
contributions. Allowing for  longitudinal electron polarization of degree 
$h$, one finds
\begin{eqnarray}
 m_e^2 \delta^{(4)}(P_f-q-P_i)
&\sum_f &tr({\cal M}^\dagger_{fi}\hat \rho^f
{\cal M}_{fi}\hat \rho^e \hat \rho^i)\nonumber\\
&=& \Big(\frac{e^2}{q_\mu^2}\Big)^2 
\eta_{\mu\nu}^{(\gamma \gamma)}(h)\,W^{(\gamma \gamma)\,\mu\nu}_{fi}
+ 2 G_F^2\,\eta_{\mu\nu}^{(Z Z)}(h)\,W^{(Z Z)\,\mu\nu}_{fi}
\nonumber\\
&& +\sqrt{2} G_F\, \frac{e^2}{q_\mu^2} \,\eta_{\mu\nu}^{(\gamma Z)}(h) \,
\Big(W^{(\gamma Z)\,\mu\nu}_{fi}
+ W^{(Z \gamma)\,\mu\nu}_{fi}\Big)\,.
\end{eqnarray}
Here, the various  lepton tensors are given by
\begin{eqnarray}
\eta_{\mu\nu}^{(\gamma \gamma)}(h) & =&
\eta_{\mu\nu}^{v v}(h)\,,\\
 \eta_{\mu\nu}^{(\gamma Z)}(h) & =&
g_V \eta_{\mu\nu}^{v v}(h) + g_A \eta_{\mu\nu}^{v a}(h)\,,\\
 \eta_{\mu\nu}^{(Z Z)}(h) & =& (g_V^2 + g_A^2) 
\eta_{\mu\nu}^{v v}(h) + 2 g_V g_A \eta_{\mu\nu}^{v a}(h) \,.
\end{eqnarray}
Since in contrast to the $\gamma$, the $Z$ couples to both vector and 
axial vector current, the additional tensor $\eta_{\mu\nu}^{v a}$ arises in 
$\eta_{\mu\nu}^{(\gamma Z)}$ and $\eta_{\mu\nu}^{(Z Z)}$, whereas 
the pure electromagnetic case is described by the tensor 
$\eta_{\mu\nu}^{v v}$ alone. Both tensors can be parametrized as 
\begin{eqnarray}
 \eta_{\mu\nu}^{v v}(h) & =& 
  \eta_{\mu\nu}^0 + h \eta_{\mu\nu}^{\prime}\,,\\
 \eta_{\mu\nu}^{v a}(h) & =& 
\eta_{\mu\nu}^{\prime} + h \eta_{\mu\nu}^{0}\,,
\end{eqnarray}
where in the high energy limit, i.e., electron mass  $m_e\rightarrow 0$, 
the tensors $\eta_{\mu\nu}^0$ and $\eta_{\mu\nu}^{\prime}$ have the 
explicit form
\begin{eqnarray}
 \eta_{\mu\nu}^0 & =& 
(k_{i\,\mu} k_{f\,\nu} + k_{f\,\mu} k_{i\,\nu})
- g_{\mu \nu} k_i\cdot k_f\nonumber\\
& =& \frac{1}{2}(k_\mu k_\nu - q_\mu q_\nu + 
g_{\mu\nu} q_\rho^2)\,,\\
 \eta_{\mu\nu}^{\prime} & =&i 
\varepsilon_{\mu \nu \alpha \beta}
k_i^{\alpha} k_f^{\beta}\nonumber\\
& =& \frac{i}{2} \varepsilon_{\mu \nu \alpha \beta}k^\alpha q^\beta\,,
\end{eqnarray}
where  $k=k_i+k_f$.
The  hadronic tensors are given by the
electromagnetic and weak current matrix elements 
\begin{eqnarray}
 W^{(B B')\,\mu\nu}_{fi}
= \sum_f tr(J_{fi}^{(B)\, \mu\,\dagger}\hat \rho^f 
J_{fi}^{(B')\, \nu}\hat \rho^i)\delta^{(4)}(P_f-P_i-q)
\,,
\end{eqnarray}
where  $B,\,B' \in \{\gamma,\,Z\}$, and 
the trace refers to the spin quantum numbers of the hadronic initial 
state. 

Proceeding now as in the pure electromagnetic case, one obtains for the
differential cross section including both beam and target polarization
\begin{eqnarray}
\frac{d^{3}\sigma^{\gamma +Z}}{dk_f^{lab} d \Omega_{k_f}^{lab}}\, 
& =&  \frac{\alpha}{2 \pi^2}
\frac{ k_f^{lab}}{k_i^{lab}q_\mu^4}
\sum_{\lambda \lambda ^{\prime}=1}^3\sum_{ m_i m_i^{\prime}}
\rho_{m_i m_i^{\prime}}^i 
\Big( \rho_{\lambda \lambda ^{\prime}}^{vv}
\sum_f J_{f \lambda^{\prime}m_i^{\prime}}^{(\gamma)\, \ast}
\hat \rho^f J_{f \lambda m_i}^{(\gamma)} \nonumber\\
 & & + \frac{G_F}{\sqrt{2}}\frac{q_\mu^2}{\pi \alpha}
  (g_V \rho_{\lambda \lambda ^{\prime}}^{vv}
  +g_A \rho_{\lambda \lambda ^{\prime}}^{va})
\Re e\Big[  
\sum_f J_{f \lambda^{\prime}m_i^{\prime}}^{(Z)\, \ast}
\hat \rho^f J_{f \lambda m_i}^{(\gamma)} \Big] \Big)\, .
\end{eqnarray}
The terms quadratic in the weak amplitude
($J^{(Z)\,\ast}_\mu J^{(Z)}_\nu$) have been omitted since
they are of order  ${{\cal O}} (q_\mu^4/M_Z^4)$ compared to the
electromagnetic process. 
The index  $\lambda$ of the  hadronic current matrix 
elements $J_{f \lambda m_i}^{(\gamma/Z)}$
refers for $\lambda = \pm 1$ to the transverse current components (with
respect to  ${\vec q}\,$), while for 
 $\lambda = 0$ the component is given by a linear combination of 
 charge and  longitudinal current 
\begin{eqnarray} 
J_0 = -\frac{|{\vec q}\,|^2}{q_\mu^2}
(\rho - \frac{\omega}{|{\vec q}\,|^2}
{\vec q} \cdot {\vec J}\,)
= \rho - \frac{\omega }{q_\mu^2}(\omega\rho - {\vec q} 
\cdot {\vec J}\,)\,,
\end{eqnarray}
which reduces to the charge density  $\rho$ for a conserved 
current. 

The spherical components of the two types of virtual boson density matrices 
are 
\begin{eqnarray}
  \rho_{\lambda \lambda^{\prime}}^{vv} & =&
 \rho_{\lambda \lambda^{\prime}}^0 + h
 \rho_{\lambda \lambda^{\prime}}^{\prime} \ ,\\
  \rho_{\lambda \lambda^{\prime}}^{va} & =&
 \rho_{\lambda \lambda^{\prime}}^{\prime} + h
 \rho_{\lambda \lambda^{\prime}}^0 \ ,
\end{eqnarray}
and they obey the symmetry relations
\begin{eqnarray}
 \rho_{\lambda \lambda^{\prime}}^{vv/va}\, &  = &
 \rho_{\lambda^{\prime} \lambda}^{vv/va} \ ,\\
  \rho_{-\lambda -\lambda^{\prime}}^0\,
 &  = & (-)^{\lambda +\lambda^{\prime}}
    \rho_{\lambda \lambda^{\prime}}^0 \ , \\
  \rho_{-\lambda -\lambda^{\prime}}^{\prime}\, &  = &
 (-)^{\lambda +\lambda^{\prime}+1}\rho_{\lambda 
\lambda^{\prime}}^{\prime} \ .
\end{eqnarray}
The nonvanishing components for unpolarized electrons are (note $q^2<0$)
\begin{eqnarray}
  \rho_L=\rho_{00}^0=
-\beta^2 q_{\nu}^2\frac{\xi^2}{2\eta}\,,&&\quad 
 \rho_T =\rho_{11}^0=
-\frac{1}{2}q_{\nu}^2\,\Big(1+\frac{\xi}{2 \eta} \Big) \ ,\\
 \rho_{LT}=\rho_{01}^0=
-\beta q_{\nu}^2 \frac{\xi}{\eta}\,
 \sqrt{\frac{\eta+ \xi}{8}}\,,&&
\quad \rho_{TT}=\rho_{-11}^0=
q_{\nu}^2\frac{\xi}{4 \eta}\,, 
\end{eqnarray}
and for  longitudinally polarized electrons 
\begin{eqnarray}
  \rho_{LT}^{\prime}=\rho_{01}^{\prime}=
 -\frac{1}{2}\,\beta\frac{q_{\nu}^2}{\sqrt{2\eta}}\,\xi\,,\quad 
\rho_T^{\prime}=\rho_{11}^{\prime}=
  -\frac{1}{2}q_{\nu}^2\, \sqrt{\frac{\eta+\xi}{\eta}} \ ,
\end{eqnarray}
with
\begin{eqnarray}
 \beta = {|{\vec q}^{\,lab}| \over |
{\vec q}^{\,c}|}\,,\quad
\xi = -\frac{q_{\nu}^2}{|{\vec q}^{\,lab}|^2} \,, \quad
\eta = \tan^2\frac{\theta_e}{2}\ ,
\end{eqnarray}
where  $\beta$ expresses the boost 
from the lab system to the frame in which
the hadronic tensor is evaluated and 
 ${\vec q}^{\,c}$ denotes the
momentum transfer in this frame.  
The relation of the kinematic functions  $\rho_{\alpha}^{(\prime)}$
to the kinematic functions  $v_{\alpha^{(\prime)}}$ of \cite{MuD94} are 
simply given by 
\begin{eqnarray}
 \rho_\alpha^{(\prime)} = -\frac{q_\mu^2}{2\eta}\, v_{\alpha^{(\prime)}}
\,,
\end{eqnarray}
where  $\alpha \in \{ L,\, T,\, LT,\, TT\}$. 

The final result for the cross section for an unpolarized target is then
\begin{eqnarray}
\frac{d^{3}\sigma^{\gamma +Z}}{dk_f^{lab} d \Omega_{k_f}^{lab}}
& =&
  c \Big\{ \rho _L (W_L^{\gamma_v}+ a_v W_L^{Z_v})
 + \rho_T ( W_T^{\gamma_v} + a_v W_T^{Z_v})
\nonumber\\
 &&+ \rho_{LT} \Big( (W_{LT}^{\gamma_v} + 
a_v W_{LT}^{Z_v})  
+ (W_{LT}^{\gamma_a} + a_v W_{LT}^{Z_a} )\Big)
\nonumber\\
 &&+ \rho_{TT} \Big( (W_{TT}^{\gamma_v} 
+ a_v W_{TT}^{Z_v})  
+ (W_{TT}^{\gamma_a} + a_v W_{TT}^{Z_a} )\Big)
\nonumber\\
 &&-\rho'_T a_a W_T^{\prime Z_a}
 - \rho'_{LT} {a_a} \Big(W_{LT}^{\prime Z_a} 
                  -W_{LT}^{\prime Z_v} \Big)
\nonumber\\
 &&+h \Big[\rho _L 
{a_a} W_L^{Z_v} + \rho_T {a_a} W_T^{Z_v}
 + \rho_{LT} {a_a} (W_{LT}^{Z_v}  
                 + W_{LT}^{Z_a})
\nonumber\\
 &&+ \rho_{TT} {a_a} (W_{TT}^{Z_v}  
                 + {W_{TT}^{Z_a}} )
- \rho'_T (W_T^{\prime \gamma_a}  + a_v W_T^{\prime Z_a}
 )\nonumber\\
&&-\rho'_{LT} \Big((W_{LT}^{\prime \gamma_a}
 +a_v W_{LT}^{\prime Z_a} )
       -(W_{LT}^{\prime \gamma_v} + a_v W_{LT}^{\prime Z_v}) 
\Big)\Big]
\Big\}\,,
\end{eqnarray}
where $c$ is related to the Mott cross section $\sigma_M$ by 
\begin{equation}
c = {\alpha \over 6 \pi^2} {k^{lab}_2 \over k_1^{lab} q_{\nu}^4}
= -\frac{\eta}{2q_{\nu}^2}\frac{\sigma_M}{3\pi^2 \alpha}\,.
\end{equation}
The  structure functions  $W_{\alpha}^{(\prime)\,B}$ 
contain the information on the internal hadron structure with the notation
 $\alpha \in \{L,T,LT,TT\}$ and $ B \in \{\gamma_v,Z_v,\gamma_a,Z_a\}$, 
where the subscripts ``$v/a$'' denote a parity conserving/violating 
contribution. Also  parity violation in the  hadronic sector has been 
allowed for, and the corresponding structure functions are denoted by 
a superscript ``$\gamma_a$''. It leads to the same types of structure 
function as via the weak axial hadron current {($Z_a$)} \cite{KuA97}. 

In case, no analysis of the final state is performed, one sums over all 
final states. Then the interference terms  $LT$ and  $TT$ vanish, and one 
obtains for the inclusive cross section for an unpolarized target  
\begin{eqnarray}
 \Sigma(h)\,& =&
d^{2}\sigma^{\gamma +Z}/dk_2^{lab} d \Omega_{k_2}^{lab}\nonumber\\
 &=& c \Big\{ \rho _L (F_L^{\gamma_v}+ a_v F_L^{Z_v})
 + \rho_T ( F_T^{\gamma_v} + a_v F_T^{Z_v})
-\rho'_T {a_a F_T^{\prime Z_a}}\nonumber\\
&&  +h \Big[\rho _L {a_a} F_L^{Z_v} + \rho_T 
{a_a} F_T^{Z_v}
- \rho'_T ({F_T^{\prime \gamma_a}} + a_v 
{F_T^{\prime Z_a}}) \Big]
\Big\}\,.
\end{eqnarray}

One should note that the  helicity dependent part of this expression 
is a direct measure of the total parity violating contribution which arises 
from both, hadronic parity violation and from electroweak interference. 
It can be determined from the  longitudinal asymmetry 
\begin{eqnarray}
  {\cal A}  =  \frac{1}{2h\Sigma_0}(\Sigma_+-\Sigma_-) \,,
\end{eqnarray}
where
\begin{eqnarray}
 \Sigma_0 = \frac{1}{2}\Big(\Sigma_+ +\Sigma_-\Big),\quad\mbox{ and }\quad 
\Sigma_{\pm} = \Sigma(\pm h)\,.
\end{eqnarray}
One may split the asymmetry into contributions from hadronic and electroweak 
parity violation, resulting in 
\begin{eqnarray}
 {\cal A}= {\cal A}^{\gamma_a}+ {\cal A}^{Z}\,,
\end{eqnarray}
where 
\begin{eqnarray}\label{asymmg}
 {\cal A}^{\gamma_a}&=& -\frac{ \rho_T^{\prime}W_T^{\prime\,\gamma_a}}
{ \rho_L W_L^{\gamma_v}+\rho_T W_T^{\gamma_v}}\,, \\
  {\cal A}^{Z}&=&
\frac{ \rho_L {a_a} W_L^{Z_v} +
\rho_T {a_a} W_T^{Z_v}-\rho_T^{\prime}a_v W_T^{\prime\,Z_a}}
{ \rho_L W_L^{\gamma_v}+\rho_T W_T^{\gamma_v}}\nonumber\\
 &=& \frac{G_F}{2\sqrt{2}}\frac{q_\mu^2}{\pi \alpha}\,\,
\frac{ \rho_L W_L^{Z_v} +
\rho_T W_T^{Z_v}+\rho_T^{\prime}(1-4\sin^2\theta_W)
W_T^{\prime\,Z_a}}
{ \rho_L W_L^{\gamma_v}+\rho_T W_T^{\gamma_v}}\,.\label{asymmz}
\end{eqnarray}
The term proportional to $\rho_T^{\prime}$ is largely suppressed, since 
$\sin^2 \theta_W=0.232$. 
    %

%\newpage
\setcounter{equation}{0}
\setcounter{figure}{0}
\setcounter{table}{0}

\section{Selected Reactions on the Deuteron}
\label{sec6}
In order to illustrate the various facets of electroweak processes, I have 
chosen the deuteron, which one might consider a personal bias. However the 
two-body system has an undisputable preference, namely, its simple structure 
allows the most exact treatment, at least in the nonrelativistic 
domain. In the following, 
I will discuss a variety of specific reactions on the 
deuteron with special emphasis on the study of subnuclear degrees of freedom. 
I will begin with the electromagnetic deuteron break-up where 
these subnuclear d.o.f.\ appear either implicitly in the case of meson 
d.o.f.\ in the form of effective operators, the meson exchange currents
(MEC), or explicitly in the case of isobar d.o.f.\ as nuclear 
isobar configurations (IC). Later on, also explicit meson d.o.f.\ will appear, 
when I will discuss specific meson production reactions. 

\subsection{Electromagnetic Break-up}
The special role of the electromagnetic deuteron break-up is
a consequence of (i) the already mentioned simple structure, and
(ii) the specific features of the electromagnetic
probe as discussed in the preceding section. The continued interest in
this process has persisted over more than sixty years because
the two-body system is in fact a unique laboratory \cite{ArS91}. 
With respect to the study of subnuclear d.o.f., 
one may summarize the evidence for MEC and IC
in $\gamma^{(\ast)}+d\rightarrow p+n$ as follows: 

(i) The strongest MEC contributions appear in $E1$ in $d(\gamma,N)N$ where 
they dominate the cross section above 100 MeV, but
they are mostly covered by the Siegert operator (see Sect.\ \ref{sec3sieg}). 

(ii) A clear signature of MEC is furthermore observed in $M1$ in $d(e,e')np$
near break-up threshold, particularly strong if not dominant at higher 
momentum transfers. 

(iii) In the $\Delta$ region, one finds the strongest manifestation of IC.\\
Now I will illustrate these findings by a few examples.

\subsubsection{Backward Electrodisintegration near Threshold}
I will start with an early case of clear evidence of $\pi$ meson 
exchange current in electrodisintegration near threshold, 
namely the inclusive process $d(e,e')pn$, where no analysis
is made of the hadronic final state \cite{SiB79}. The threshold 
region is dominated by the excitation
of the antibound $^1S_0$ resonance in $NN$ scattering at
very low energies. It is a process inverse to thermal $n$-$p$ radiative
capture, which proceeds via $M1$ transition and where MEC  and IC
give almost a  10  percent  enhancement \cite{RiB72,Are81b}.  The
advantage of electron scattering, allowing the momentum transfer to be 
varied independently, becomes apparent here, 
since  the  relatively  small  effect
of subnuclear d.o.f.\ in the real photon process can be amplified
considerably. The reason for this lies in the
fact that with increasing momentum transfer the one-body contribution drops
rapidly due to a destructive interference of S- and D-wave contributions
and thus the distribution of the momentum transfer onto both nucleons via 
two-body operators becomes more favourable. This can be seen in Fig.\ 
\ref{figfigsimon}, 
where the longitudinal and transverse form factors are shown as obtained 
from a Rosenbluth separation \cite{SiB79}. While the longitudinal form factor
is well described by the classical theory, supporting the Siegert hypothesis 
that the charge density is little affected by exchange effects,
one finds a large discrepancy in $F_T$ which is only resolved if MEC and IC
are added.

\begin{figure}
\centerline {\psfig{file=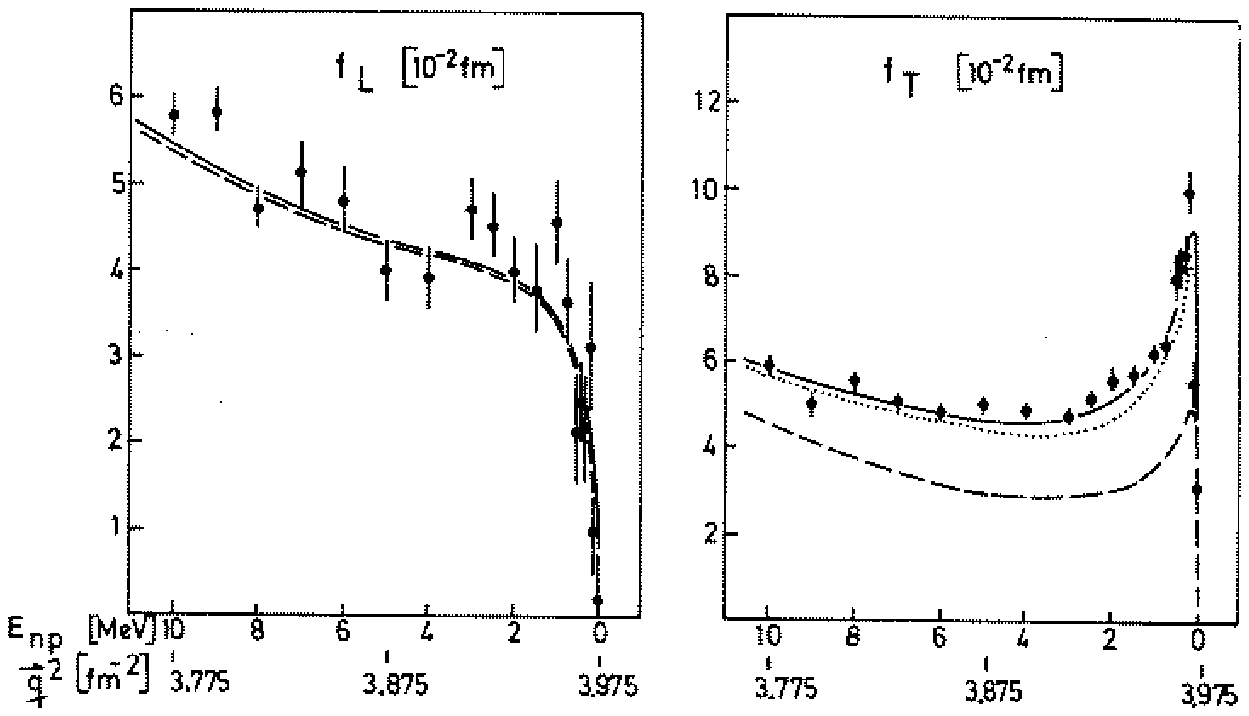,width=7.5cm,angle=-90}}
\caption{figfigsimon}
{The longitudinal (left panel) and transverse (right panel) 
deuteron form factors at 
$\vec q^{\,2}\approx 4\mbox{ fm}^{-2}$ compared to calculations
with the Hamada-Johnston potential for the normal theory (dashed) 
and additional $\pi$ MEC (dotted) and further added IC (full) 
(from \cite{SiB79}).}
\end{figure}
\begin{figure}
\centerline {\psfig{file=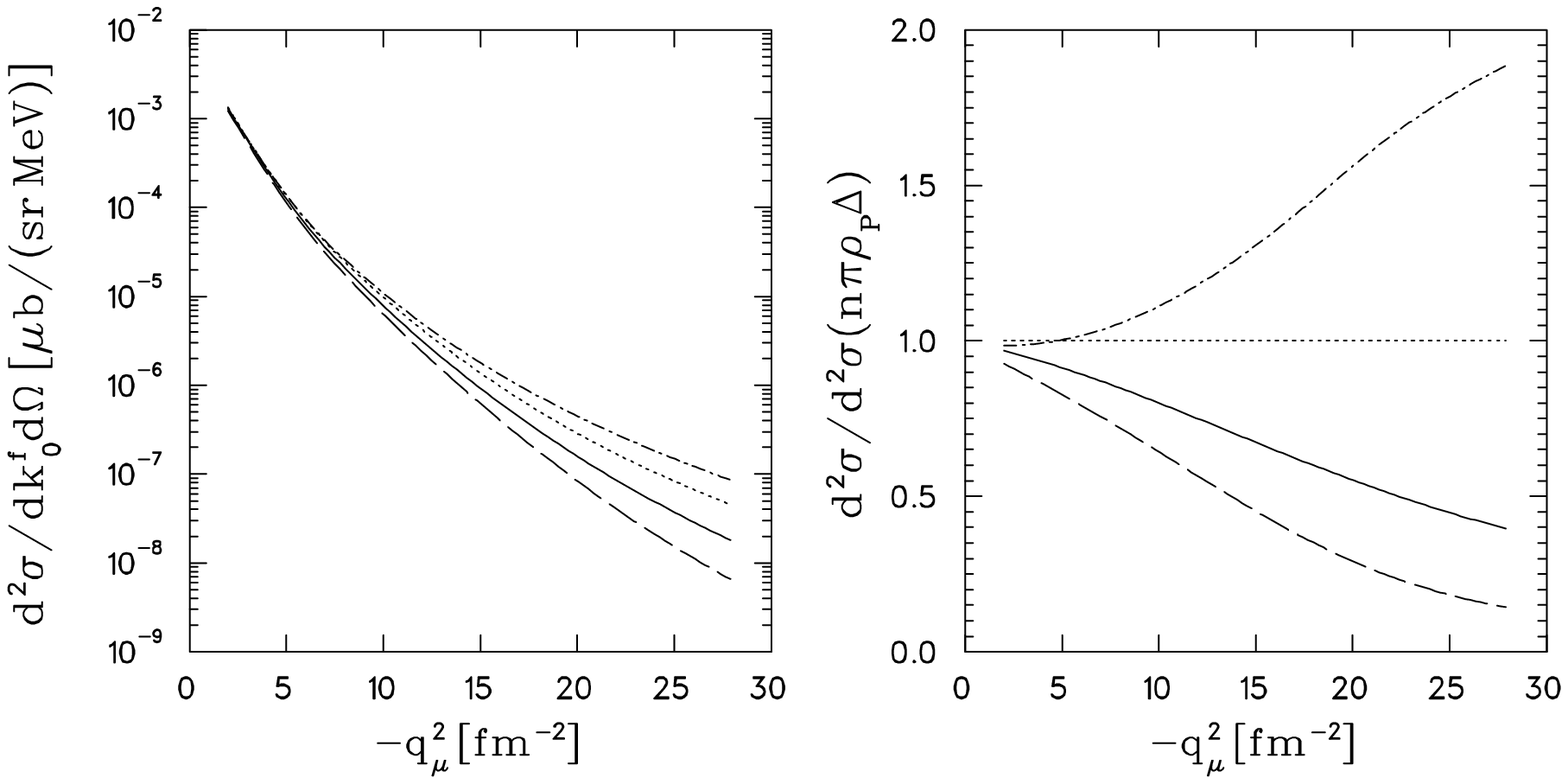,width=13cm,angle=0}}
\caption{figfigritz4}
{Deuteron electrodisintegration near threshold for
$E_{np}=1.5\,\mbox{MeV}$ at backward angles ($\theta_e=155^\circ$).
Left: absolute values; Right: relative with respect to the nonrelativistic 
theory including the conventional $\pi$ and $\rho$ MEC and $\Delta$ 
contribution. 
Notation of the curves: dotted: nonrelativistic theory including the 
conventional $\pi$ and $\rho$ MEC and $\Delta$ contribution; 
dash-dotted: in addition relativistic one-body contributions 
including kinematic boost; 
dashed: further added all relativistic contributions to  $\pi$ 
MEC; full: complete calculation (from \cite{RiG97}).
}
\end{figure}

The situation for higher momentum transfers is shown in Fig.\ 
\ref{figfigritz4} where the theoretical inclusive cross section is plotted 
at backward angles for moderate momentum transfers as recently calculated 
by Ritz et al.\ \cite{RiG97}. In contrast to the theoretical results shown 
in Fig.\ \ref{figfigsimon}, here the theoretical treatment of MEC is 
completely consistent, including heavy meson exchange, with the potential 
model, namely the Bonn OBEPQ-B model \cite{MaH87,Mac89}, and, in addition, 
all leading order relativistic contributions are taken into account. 
The latter can be obtained in a $(p/M)$ expansion starting from a 
covariant approach \cite{AdT89}. As briefly discussed at the end of 
Sect.\ \ref{sec4}, these comprise relativistic terms both in the current 
operators and in the wave functions including boost effects. 
It is obvious that for a conclusive 
interpretation, one has to include all corrections of the same order
consistently. One readily sees in Fig.\ \ref{figfigritz4} again the 
strong enhancement due to MEC, but in addition one notes sizable relativistic
contributions.
It is clear from this result that already at low excitation energies
relativistic effects may become important and have to be considered in a
quantitative comparison of theory with experiment. 

\begin{figure}
\centerline {\psfig{file=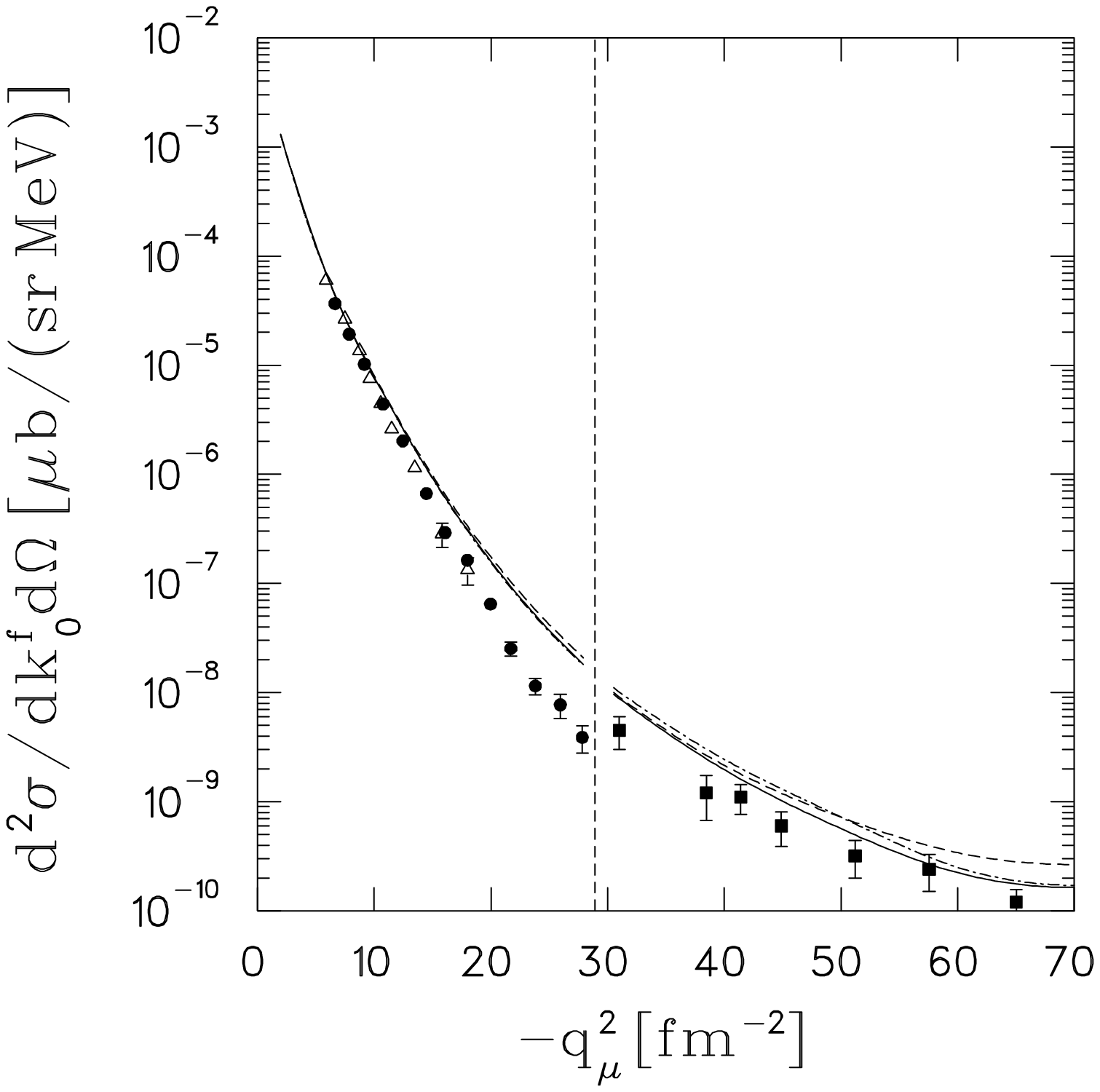,width=8cm,angle=0}}
\caption{figfigritz6}
{
Potential model dependence of deuteron electrodisintegration near 
threshold and comparison with experiment.  Experimental data points:
open triangles: \protect{\cite{BeJ81}}, filled circles:
\protect{\cite{AuC85}} ($\theta_e=155^\circ$, averaged over energies
$0 \leq E_{np}\leq 3\mbox{\,MeV}$; theory for
$E_{np}=1.5\mbox{\,MeV}$); filled squares: \protect{\cite{ArB90}}
($\theta_e=180^\circ$, averaged over energies $0\mbox{\,MeV}\leq
E_{np}\leq 10\mbox{\,MeV}$; theory for $E_{np}=5\mbox{\,MeV}$).
Notation of the curves: full: Bonn OBEPQ-B potential, dashed:
OBEPQ-A potential, dash-dotted: OBEPQ-C potential 
(from \cite{RiG97}).}
\end{figure}

Such a comparison, based on the results for three OBEPQ versions of the 
Bonn potential with experimental data, is shown in Fig.~\ref{figfigritz6}. 
Here the calculation has been extended to the high momentum data of 
\cite{ArB90}, although this kinematic region is beyond the limits of
validity of the $(p/M)$ expansion. Between
$-q_\mu^2=10$ and $30\,\mbox{fm}^{-2}$ one finds a
systematic and increasing overestimation of the data by the theory, 
whereas above $30\,\mbox{fm}^{-2}$ the overestimation is much less
pronounced and more constant.  The variation of the different
potential versions is comparably small, except for the very highest
momentum transfers considered. 

In summary, a good agreement with experimental data is achieved at low 
and intermediate momentum transfers. Relativistic contributions are sizable 
and have to be included. In particular, the noted difference in 
nonrelativistic calculations between taking the e.m.\ nucleon form factors
in the Sachs or in the Dirac-Pauli parametrization disappears essentially 
\cite{WiB93,BeW94}. 
But at higher momentum transfers, a completely covariant approach without
resort to a $(p/M)$ expansion is necessary.

\subsubsection{Signature of a $\Delta\Delta$ Component in Electrodisintegration}

One important consequence of the internal nucleon structure is the presence
of small nuclear wave function components where one or more nucleons are 
internally excited, i.e., exist in an isobar state like a $\Delta(1232)$ or 
a Roper resonance $N(1440)$ \cite{ArW72,Gre76,WeA78}. This is a consequence 
of the fact, that during a collision 
of two nucleons there is a nonvanishing probability for the internal excitation 
of one or both nucleons. Because of the relatively large excitation energy 
($\ge 300$ MeV) compared to typical nuclear energies, such virtual excitations 
have a correspondingly short lifetime. Thus they will modify the nuclear wave 
function mainly in the shorter range domain\cite{WeA78}. 
Since these isobar configurations 
(IC) have a rather small probability, their influence is difficult to detect. 
In any case, evidence for them will only be indirect since they are not 
directly observable. However, under certain favourable conditions they may 
lead to sizable effects in processes which are sensitive to the shorter range 
region, i.e., in higher momentum transfer processes. I would like to show one 
example for the double $\Delta$ component in the deuteron, namely the 
charge structure function $f_L$ of $d(e,e'p)n$. In lowest order, one has no 
contribution to the charge density from meson exchange and the relativistic 
contributions are very small. Thus the largest effect from subnuclear 
d.o.f.\ arises from IC. 
\begin{figure}[h]
\centerline{\psfig{file=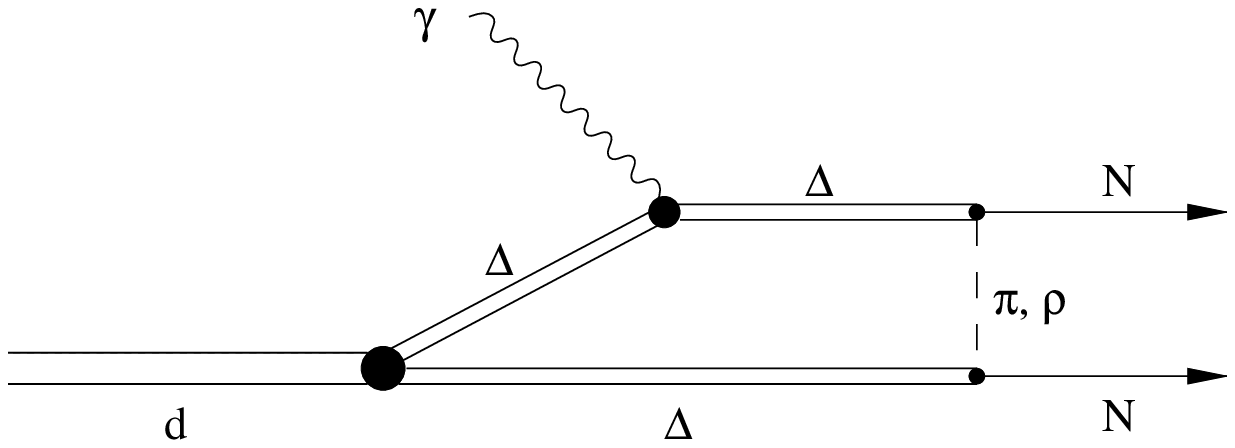,width=9cm,angle=0}}
\caption{figdd_component}
{Diagrammatic representation of the $\Delta\Delta$ contribution to the
longitudinal structure function $f_{L}$. 
}
\end{figure}
\begin{figure}[h]
\centerline{\psfig{file=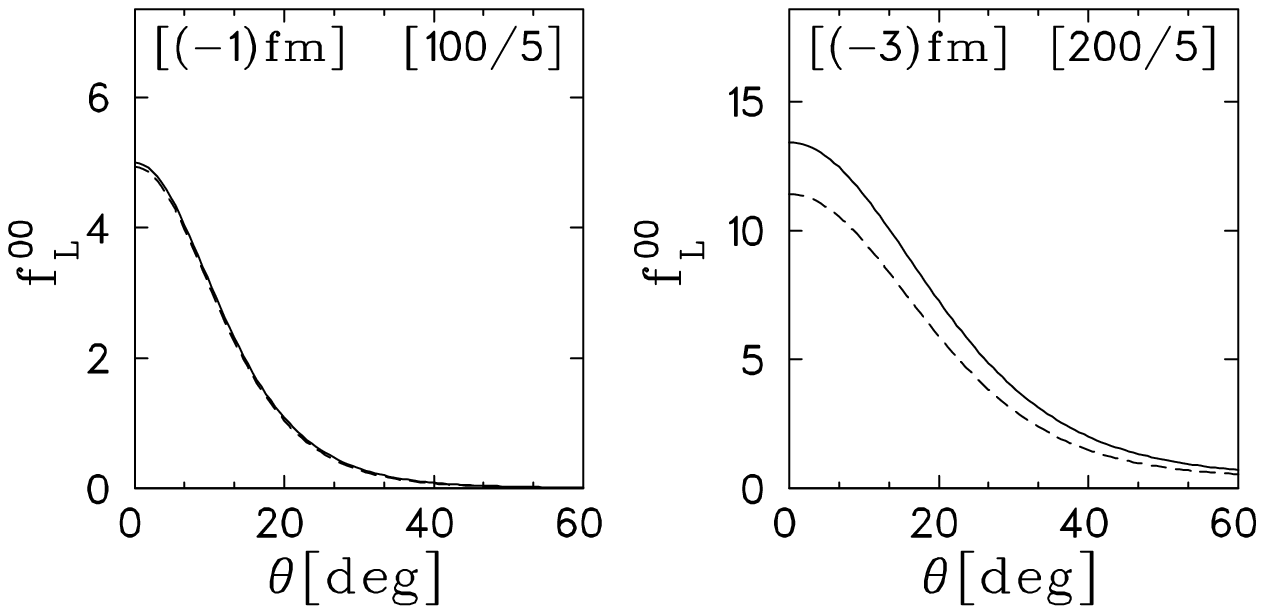,width=12cm}}
\caption{figfig_dd}
{
Longitudinal structure function in the $\Delta$ region for
$E_{np}=100$ and 200 MeV at $q^{\,2}=5$ fm$^{-2}$ without (dashed) and with
(full) $\Delta\Delta$ component. The top left inset ``[(-n) fm]'' indicates 
the unit [$10^{-n}$ fm] for $f^{00}_L$ and the top right inset means 
``[$E_{np}$/$q_{c.m.}^2$]'', where $E_{np}$ in [MeV] and $q^2$ in 
[fm$^{-2}$].
}
\end{figure}

Since the lowest order $\pi$-MEC charge density vanishes (see Sect.\ 
\ref{secmec}), effects from subnuclear d.o.f.\ can arise only from IC. 
In view of the fact, that the charge excitation of a $\Delta$, which has to
proceed via $C2$ is largely suppressed, the only contribution of IC to the 
longitudinal structure function $f_{L}$ comes from the diagonal $\Delta$ 
charge density,
states as sketched in the diagram of Fig.\ \ref{figdd_component}. 
At low energy transfer, the contribution is expected to be relatively 
small. 
However, with increasing energy transfer its relative importance 
is enhanced considerably because of the increase of the double $\Delta$ 
component in the final state. This behaviour is shown in Fig.\ 
\ref{figfig_dd}. Although the $\Delta\Delta$ components in the deuteron 
have a total admixture probability of less than 1 percent, 
they produce about a 15 percent enhancement of $f_L$ at 200 MeV, while at 
100 MeV their effect is still negligible. 

\subsubsection{Heavy Meson Exchange and Consistent Relativistic Effects 
in Photodisintegration below Pion Threshold}

Now I will turn to a discussion of the influence of heavy 
meson exchange on deuteron photodisintegration with inclusion of 
competing relativistic effects in the one-body and pion exchange sector 
as presented recently by Ritz et al.\ \cite{RiA97}. The details of the 
calculational framework can be found in \cite{RiG97}. As an example, I show in 
Fig.\ \ref{figritzdsdo} the unpolarized differential cross section at four 
representative photon energies
$E_\gamma=4.5$, $40$, $100$, and $140$ MeV covering the region between the
maximum of the total cross section and pion production threshold. 
\begin{figure}
\centerline {\psfig{file=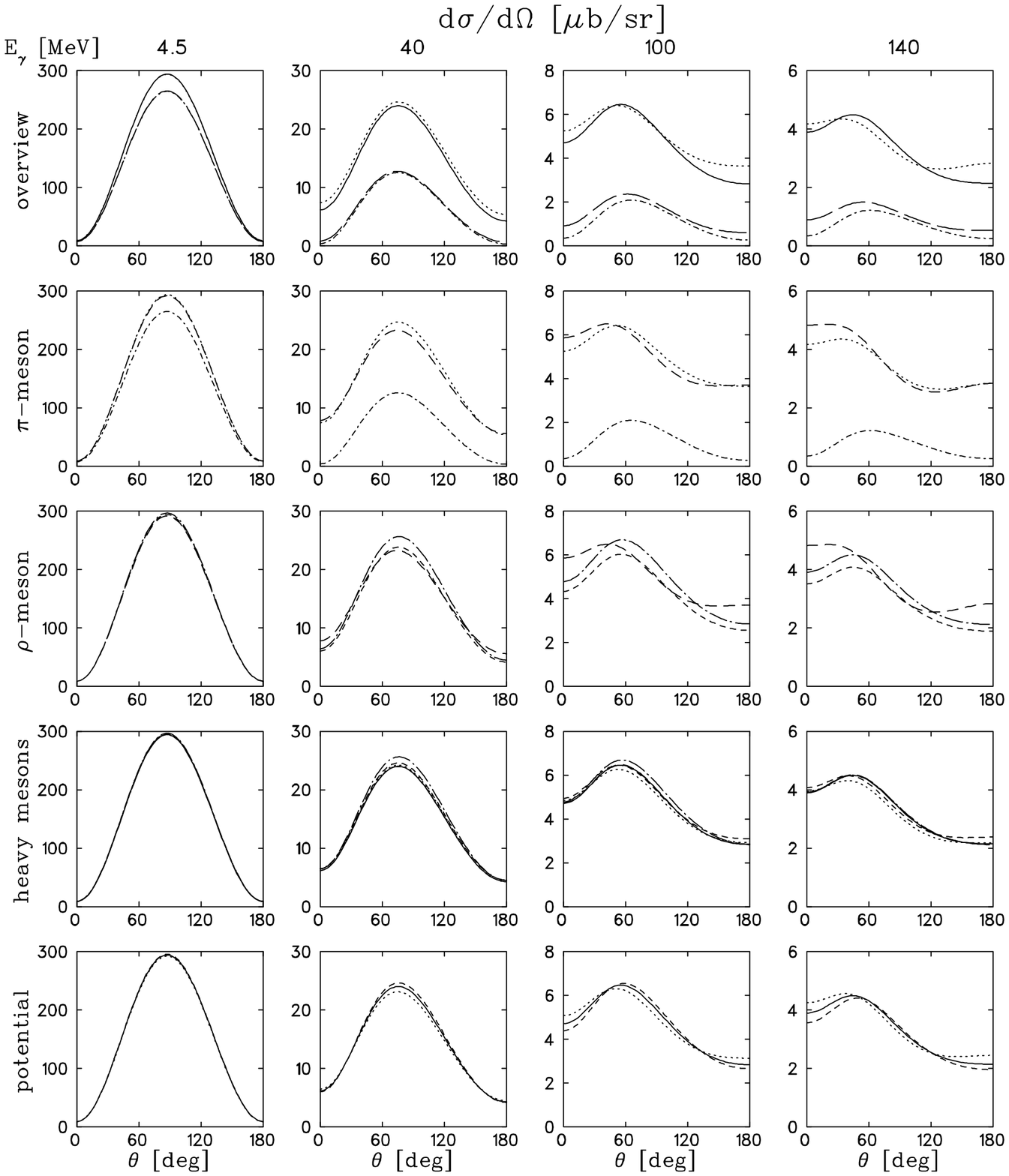,width=14cm,angle=0}}
\caption{figritzdsdo}
{
The differential cross section of $d(\gamma,p)n$ 
for various laboratory photon energies.
Notation of the curves:
(1) overview:
nonrelativistic one-body current (long-dashed);
relativistic one-body current (dash-dotted);
nonrelativistic $\pi$ MEC added (dotted);
total result (full);
(2) $\pi$ meson:
relativistic one-body current (dash-dotted);
nonrelativistic $\pi$ MEC added (dotted);
relativistic $\pi$ MEC including retardation (dashed);
(3) $\rho$ meson:
relativistic one-body plus complete $\pi$ MEC (dashed);
Pauli MEC (short-dashed);
Dirac MEC (long-dashed-dotted);
(4) heavy meson:
relativistic one-body current plus complete $\pi$ and full
$\rho$ MEC (long-dash-dotted);
$\delta$ MEC (dotted);
$\omega$ MEC (short-dashed);
$\sigma$ MEC (dashed);
$\eta$ MEC (full);
(5) potential:
Bonn OBEPQ version B (full);
version A (short-dashed);
version C (dotted).
}
\end{figure}
In order to distinguish the different influences from pion, rho, and other
heavy meson exchanges, their effects are shown in separate panels for each
observable and each energy. In addition, an overview is shown as well as the 
potential model dependence with respect to the versions A, B, and C of the Bonn 
OBEPQ potential model \cite{MaH87,Mac89}.

The overview shows that in the maximum of the total cross section,
at $4.5$ MeV, the differential cross section is dominated by the nonrelativistic
one-body current, while only the nonrelativistic $\pi$ MEC gives a 10 
percent enhancement, predominantly in $E1$. Obviously, relativistic effects 
and heavy mesons are negligible as well as the potential model dependence.

\begin{figure}
\centerline {\psfig{file=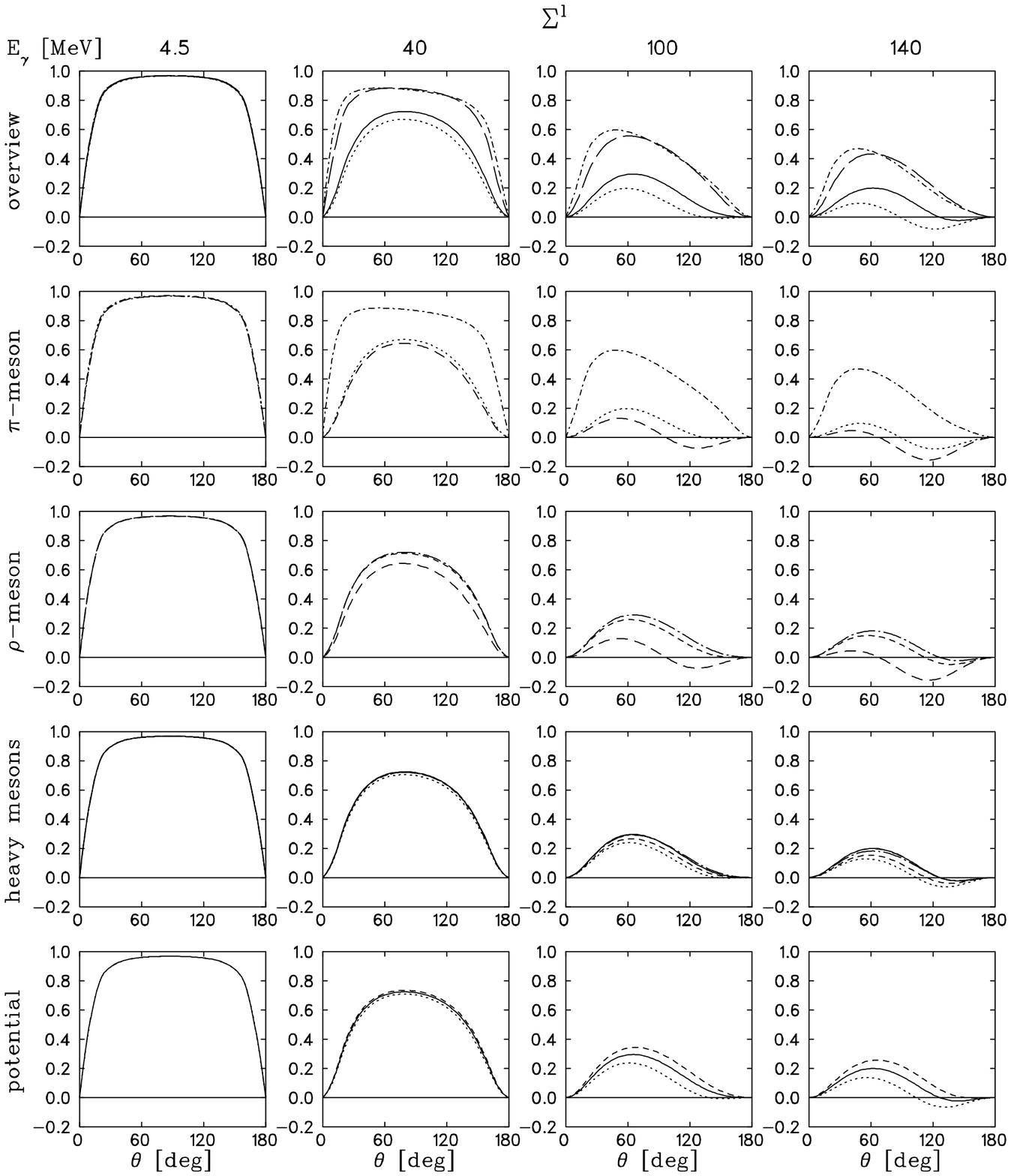,width=14cm,angle=0}}
\caption{figritzsigma}
{
The photon asymmetry $\Sigma^l$ for $d(\vec \gamma,p)n$. 
Notation of the curves as in Fig.~\protect{\ref{figritzdsdo}}.
}
\end{figure}
At the next higher energy ($40$ MeV), the nonrelativistic $\pi$ MEC becomes
comparable to the one-body current. All other contributions give a small
overall reduction, somewhat more pronounced in forward and backward
direction. However, if one looks at the separate contributions, one notices
a subtle destructive interference of different larger effects. First,
relativistic $\pi$ MEC gives a slight reduction in the maximum but leaves
the forward and backward directions almost unchanged. Next, with respect 
to the $\rho$, one sees a strong forward and backward reduction from the 
Pauli current (see \cite{RiG97} for the distinction between Pauli and Dirac 
current for the $\rho$ MEC), whereas the Dirac contribution, which is often 
neglected, mainly leads to a sizable enhancement in
the maximum which, however, is largely cancelled by the additional heavy
mesons. Finally, one finds a small potential model dependence of a few percent.

Considering now the two higher energies ($100$ and $140$ MeV), one readily
notices a dramatic increase of relativistic effects. First a sizable
reduction appears from the relativistic one-body current showing the well 
known effect of diminishing the differential cross section at forward and
backward angles, which comes mainly from the dominant spin-orbit 
current \cite{CaM82}. The further reduction from the remaining contributions 
is again the result of a strong destructive interference of larger 
contributions. In fact, first the
relativistic $\pi$ MEC surprisingly enhances the cross section in forward
direction, while then the $\rho$ Pauli current results in a drastic 
reduction at both extreme angles. The Dirac contribution gives again an 
overall enhancement but of smaller size. This effect of the Dirac $\rho$ 
current is somehow surprising, because it is roughly of the same size as 
that of the Pauli current, whereas from the size of the coupling constants 
one would have expected a suppression by a factor of about $50$. 
The additional heavy mesons beyond the $\rho$ meson 
show a much smaller influence.
Their individual contributions are remarkably big (up to $\sim{}5$\%),
in particular compared to their role in the parametrization of the $NN$ 
force and their importance in the electrodisintegration of the deuteron 
\cite{RiG97}. Most prominent is the effect of the $\delta$ meson leading 
to a reduction of the differential cross section. However, looking at the 
net result, the heavy meson exchanges tend to cancel each other.
With respect to the potential model
dependence, one sees now a larger variation, in particular also at forward
and backward angles which increases with energy.

The photon asymmetry for linearly polarized photons $\Sigma^l$, shown in 
Fig.\ \ref{figritzsigma}, is very
sensitive to two-body effects, as is long known \cite{ArS91}.
At $4.5$ MeV only the nonrelativistic one-body current contributes to
$\Sigma^l$ and no potential model dependence appears. Then at $40$ MeV, 
the nonrelativistic $\pi$ MEC becomes sizable as well as the Pauli $\rho$
current. All other effects, relativistic one-body and $\pi$ MEC, Dirac
$\rho$ and additional heavy meson effects are very small, as is the
potential model dependence. At higher energies the relativistic one-body 
current as well as the relativistic $\pi$ MEC become important, too.
The first leads to a sizable reduction of the photon asymmetry,
the latter to a smaller increase. The $\rho$ MEC increases the photon
asymmetry, of which the Pauli current is the most dominant part while 
the Dirac current is comparably small, although its size increases with 
the photon energy. The influence of the various heavy meson
exchanges are much more pronounced than in the differential cross section,
mainly coming from the $\delta$ MEC\@. But again the various heavy mesons 
tend to interfere destructively. The potential dependence is quite 
large at $100$ and $140$ MeV, where the OBEPQ version A yields the 
biggest asymmetry, version B intermediate values, and version C the 
lowest photon asymmetry. Thus one might be tempted
to single out one potential against the others by comparison with experimental
data. However, one has to be careful in such a reasoning \cite{BlB91},
because, before drawing definite conclusions as to which model should be
preferred, one has to study in detail the remaining theoretical 
uncertainties 
due to the strength of the dissociation and isobar currents. Here, the
additional independent measurement of the unpolarized cross section and the
vector target asymmetry $T_{11}$ could help in fixing the respective 
strengths of these contributions.

\subsubsection{Pion Retardation in Photodisintegration above $\pi$ Threshold}
% (M.\ Schwamb et al., MKP-T-97-24)\\

Photodisintegration in the $\Delta$ resonance region is of particular interest 
for the study of the $N \Delta$ interaction. However, all models developed so 
far are unable to describe in a satisfactory manner the experimental data 
over the whole $\Delta$ resonance region (for a review see \cite{ArS91}).
Among the most sophisticated approaches  are the  unitary three-body model 
of Tanabe and Ohta \cite{TaO85} and the coupled channel approach (CC) of 
Wilhelm and myself \cite{WiA93}. In both models, all  free parameters
were fixed in advance  by fitting $NN$  and $\pi N$ scattering, and 
$\pi$ photoproduction on the nucleon. Consequently, no adjustable parameters 
remained for deuteron photodisintegration. However, it turned out that both 
approaches considerably underestimated the total cross section in the 
$\Delta$ region  by about 20-30\% \cite{TaO85,WiA93}. Another failure 
was the wrong prediction of the shape of differential cross section and photon 
asymmetry, especially at photon energies above 300 MeV 
\cite{TaO85,WiA93,BlB95}.

In these calculations, one of the principal problems is the question of 
how to fix the magnetic $\gamma N \Delta$ strength 
$G^{M1}_{\Delta N}$. In \cite{TaO85} 
and \cite{WiA93} it has been fitted to the $M_{1+}(3/2)$ multipole of pion
photoproduction on the nucleon including resonance and Born contributions. 
When embedded into the two-nucleon system, these Born terms become part
of the retarded two-body recoil and $\pi$ meson currents, respectively
(Fig.\ \ref{figvergleich}). 
\begin{figure}[h]
\centerline {\psfig{file=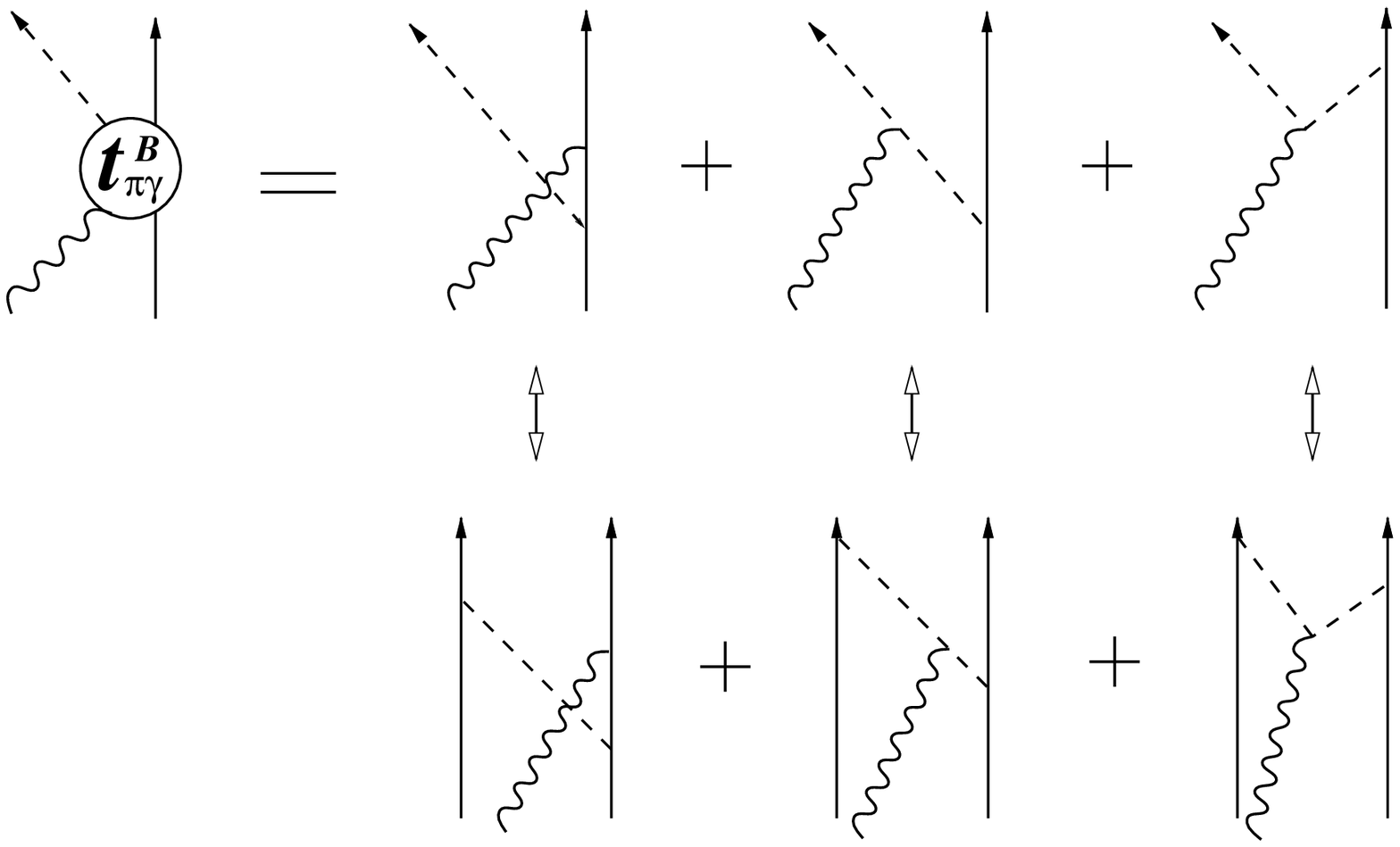,width=10cm,angle=0}}
\caption{figvergleich}
{Correspondence of Born terms of the $M_{1+}^{3/2}$ multipole of pion 
photoproduction and meson exchange current diagrams.
}
\end{figure}
However, in the usual static calculations of MEC, 
the retardation of the pion is completely neglected. Furthermore, the recoil 
current is not present due to its cancellation against the wave function 
renormalization \cite{GaH76}. Thus, there is an inconsistency in the 
treatment of pion d.o.f. 
Indeed, it had already been conjectured in \cite{WiA93} that this 
inconsistency of using static MEC may be the origin of the observed 
underestimation of the total cross section in the coupled channel approach, 
because by incorporating the Born terms effectively into an increased 
$M1$-$\Delta$ excitation
strength, a satisfactory agreement with the data could be achieved.

In order to avoid these shortcomings, recently 
M.\ Schwamb et al.\ \cite{ScA97} have included for the first time 
in a coupled channel approach complete retardation 
in the $\pi$ exchange contributions to potential and MEC. For the retarded 
$NN$ potential an improved version of the energy dependent Bonn OBEPT has 
been chosen, which had been developed by Elster et al.\ \cite{ElF88}. It 
has to be renormalized via 
subtraction of a $N \Delta$ box graph \cite{GrS79,PoS87}. The transitions 
between $NN$ and $N \Delta$ space are mediated by retarded $\pi$ and 
furthermore 
$\rho$ exchange, whose form factors are fixed by fitting the $^1D_2$
$NN$ partial wave. In order to ensure unitarity up to the $2 \pi$ threshold, 
the formation of an intermediate $NN$ state with the quantum numbers of the 
deuteron and  a pion as spectator (denoted for simplicity by $\pi d$ channel)
has been considered in addition. 

The results for the total cross section 
are shown in Fig.\ \ref{figschwambwqtot}. Similar to \cite{WiA93},
the static calculation considerably underestimates the data.
Inclusion of retardation in the hadronic interaction even lowers further 
the cross section, which is more than compensated by  retardation in
the $\pi$ MEC giving a strong  enhancement
which can be traced back essentially to the inclusion of
recoil current contributions. The inclusion of the $\pi d$ channel
and the $\rho \pi \gamma / \omega \pi \gamma$ MECs  enhances the
cross section further, so that our full calculation now  gives
quite a good agreement with  experimental data over the whole energy range.
In Figs.\ \ref{figschwambwqdiff} and \ref{figschwambphoasy}, differential 
cross section and photon asymmetry for various energies are shown. 
Whereas the differential cross  section is in satisfactory agreement with 
the data, the absolute size of the asymmetry is slightly underestimated. 
However, in contrast to \cite{{TaO85},{WiA93}}, the shape of these two 
observables at higher energies is reproduced quite well. In summary, 
complete 
inclusion of pion retardation is necessary, showing a strong influence on 
cross section and polarization observables and yielding a much improved 
description of experimental data. 
\begin{figure}
\centerline {\psfig{file=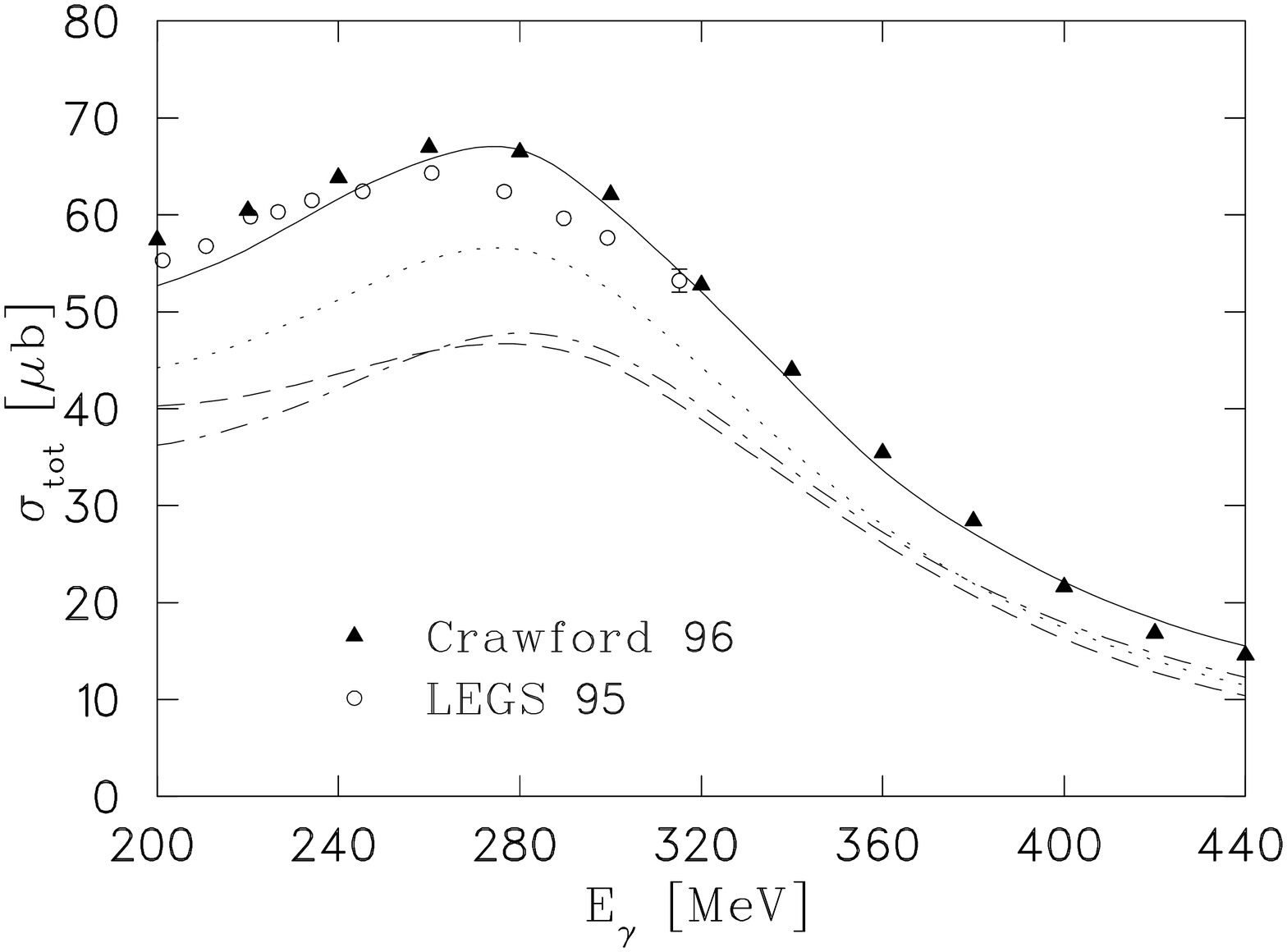,width=6.5cm,angle=90}}
\caption{figschwambwqtot}
{Total cross section of $d(\gamma,p)n$. 
Notation:  Dashed: static calculation of \cite{WiA93};  
Dotted: static OBEPR calculation; Dash-dot: 
retardation switched on in the hadronic part only, but static MECs; 
Full: complete calculation including $\pi d$ channel and 
$\rho \pi \gamma / \omega \pi \gamma$ MECs (from \cite{ScA97}).
}
\end{figure}
\begin{figure}
\centerline {\psfig{file=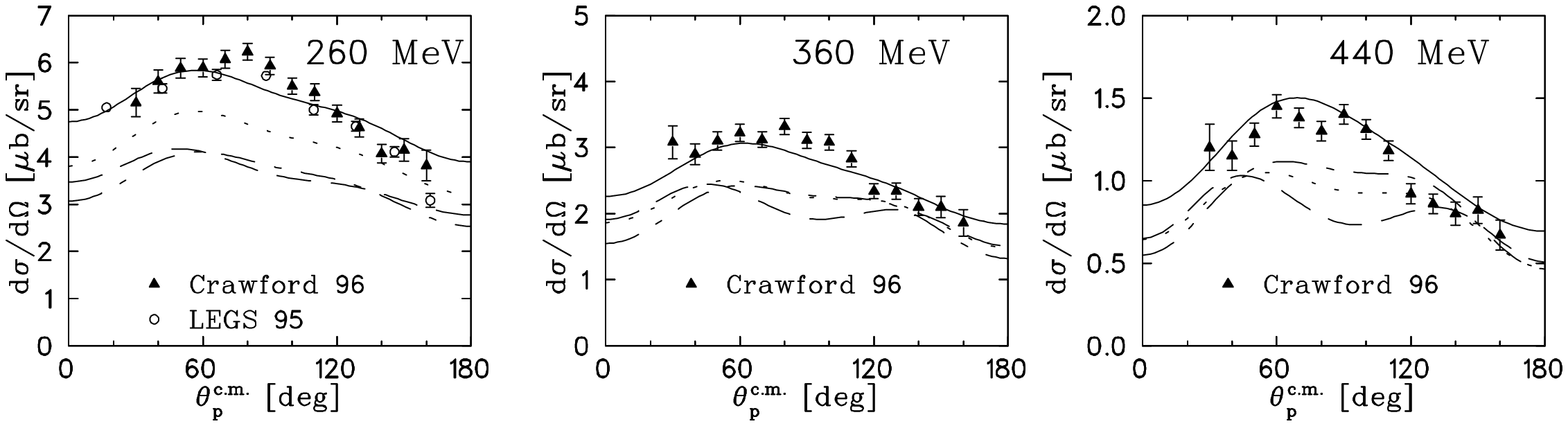,height=15cm,angle=90}}
\caption{figschwambwqdiff}
{
Differential cross section of $d(\gamma,p)n$. Notation as in Fig.\ 
\ref{figschwambwqtot} (from \cite{ScA97}). 
}
\end{figure}
\begin{figure}
\centerline {\psfig{file=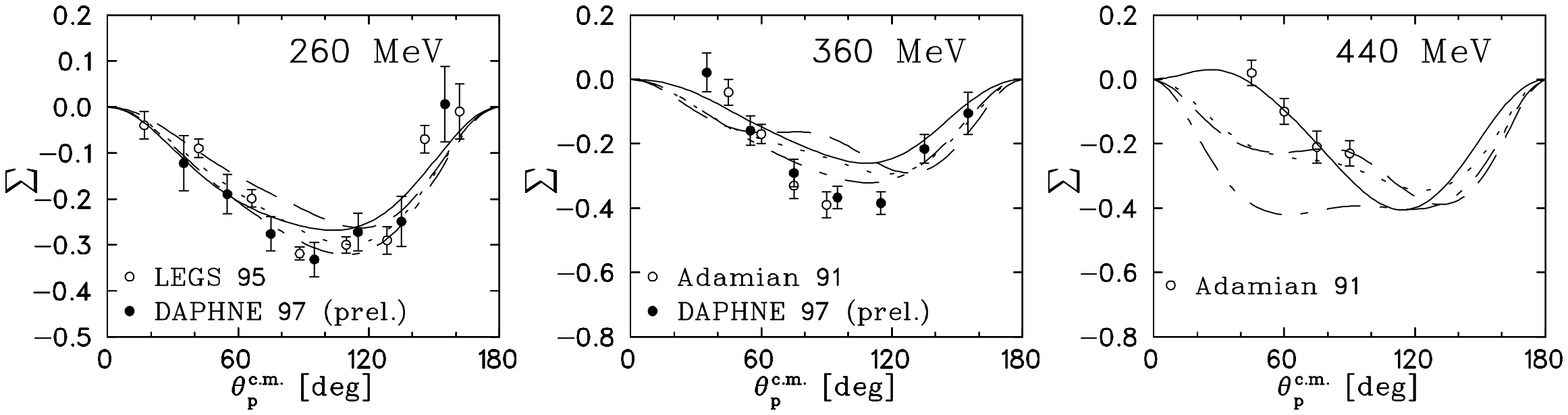,height=15cm,angle=90}}
\caption{figschwambphoasy}
{
Photon asymmetry $\Sigma^l$ of $d(\vec\gamma,p)n$. Notation of the curves as in 
Fig.~\protect{\ref{figschwambwqtot}} (from \cite{ScA97}).
}
\end{figure}

\subsection{Photoproduction of Mesons}\label{sec6meson}

Meson photoproduction is the primary absorptive process on the nucleon. It 
proceeds mainly through the intermediate excitation of a nucleon resonance 
and gives important information on the internal nucleon structure. 
Therefore, it provides stringent tests for any kind of hadron models. But 
one has to keep in mind, that the elementary process contains besides the 
resonance contributions also background or Born terms and that both 
contributions are coupled  dynamically. In view of the fact, that the 
background is largely phenomenological, the extraction of the 
interesting resonance properties is not unique \cite{WiW96,WiW98}.
% (see also the seminar of Th.\ Wilbois). 
The reason for this is, that one can arbitrarily 
shift resonance contributions to the background and vice versa via unitary 
transformations without changing the observable quantities. 

Meson photo- and electroproduction on light nuclei is primarily motivated 
by the following possibilities: (i) study of the elementary 
neutron amplitude in the absence of a neutron target, (ii) investigation 
of medium effects, i.e., possible changes of the production 
operator in the presence of other nucleons, and (iii) it provides an 
interesting means to study nuclear structure. As an illustration of these 
various aspects, I will present recent results on coherent and incoherent 
pion and eta photoproduction on the deuteron.

\subsubsection{Coherent Pion Photoproduction in the $\Delta$(1232) Region}
%(P.\ Wilhelm et al., Nucl.\ Phys.\ A 593 (1995) 435, 
%A 609 (1996) 469)\\}

Recently, Wilhelm and myself have studied coherent $\pi^0$ production on 
the deuteron in the $\Delta$ region. In the first work in \cite{WiA95}, we had 
neglected rescattering effects in order to study systematically the details 
of the elementary production amplitude on the one hand, and to investigate 
the influence of genuine two-nucleon mechanisms on the other hand. However, 
the comparison of theoretical predictions and experimental data gave
clear indication that rescattering effects may be important. 
These have then been investigated in \cite{WiA96} using a theoretical 
concept which with respect to the hadronic part
is based mainly on the developments of Sauer and collaborators
\cite{Sau86,PoS87} and is similar to the treatment of Lee et al.\ 
\cite{BeL81,Lee83,LeM85}. The model includes explicit pion, nucleon
and delta degrees of freedom in a system of coupled equations for the 
$N\Delta$, $NN\pi$ and $NN$ channels. For other approaches see the 
references cited in \cite{WiA95} and the recent work of Kamalov et al.\ 
\cite{KaT97}. 

I show in Fig.\ \ref{figgdpi0tot_amp} all diagrams of the process considered 
in the calculation of \cite{WiA96}. They include the impulse approximation
(IA), namely the direct $\Delta$ excitation (denoted $\Delta$[1] in 
Fig.\ \ref{figgdpi0tot_amp}), the direct and crossed nucleon pole terms 
(NP[1] and NC[1], respectively), and the disconnected direct and crossed 
two-nucleon processes (NP[2] and NC[2], respectively). These are the first 
five diagrams on the rhs of Fig.\ \ref{figgdpi0tot_amp}. The remaining 
diagrams of Fig.\ \ref{figgdpi0tot_amp} describe two-body currents and 
rescattering effects considered in \cite{WiA96}. 

\begin{figure}[h]
\centerline {\psfig{file=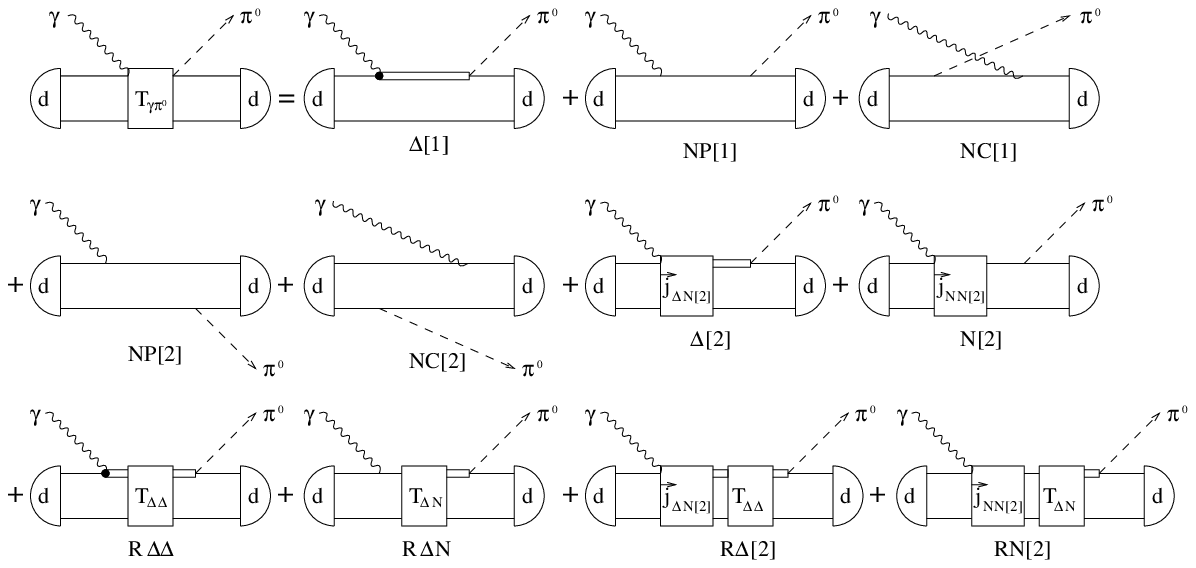,width=14cm,angle=0}}
\caption{figgdpi0tot_amp}
{Diagrammatic representation of the impulse approximation, two-body 
and rescattering contributions to \protect{$d(\gamma,\pi^0)d$} considered 
in \cite{WiA96}.}
\end{figure}

The common feature of the IA diagrams in Fig.\ \ref{figgdpi0tot_amp} is
that the hadronic intermediate state, either $\Delta N$, $NN$ or $\pi NN$,
propagates freely.  In principle each diagram will be accompanied
by a corresponding one, where the intermediate state is subject to the
hadronic interaction. Furthermore, the interaction will allow
couplings between different intermediate states. However, in \cite{WiA96} 
the interaction has been restricted to the most important intermediate 
$N\Delta$ state resulting in the rescattering amplitude R$\Delta\Delta$ and 
to the $NN$-$N\Delta$ coupling R$\Delta$N in Fig.\ \ref{figgdpi0tot_amp}.

As first result, I show in Fig.\ \ref{figgdpi0tot} the total cross 
section. The dotted curve corresponds to the impulse approximation.
Adding MECs ($\Delta$[2] and N[2]) gives a slight increase of a few
percent in the maximum (dash-dotted curve). But by far much more
important are the other rescattering mechanisms. They reduce the cross
section significantly and shift the maximum to a slightly lower
position.  Furthermore, a comparison of the full to the dashed curve
clearly demonstrates that the perturbative treatment (Born 
approximation) is certainly insufficient. It underestimates the full
dynamical effect by more than half and thus can provide a qualitative
description at best.
\begin{figure}
\centerline {\psfig{file=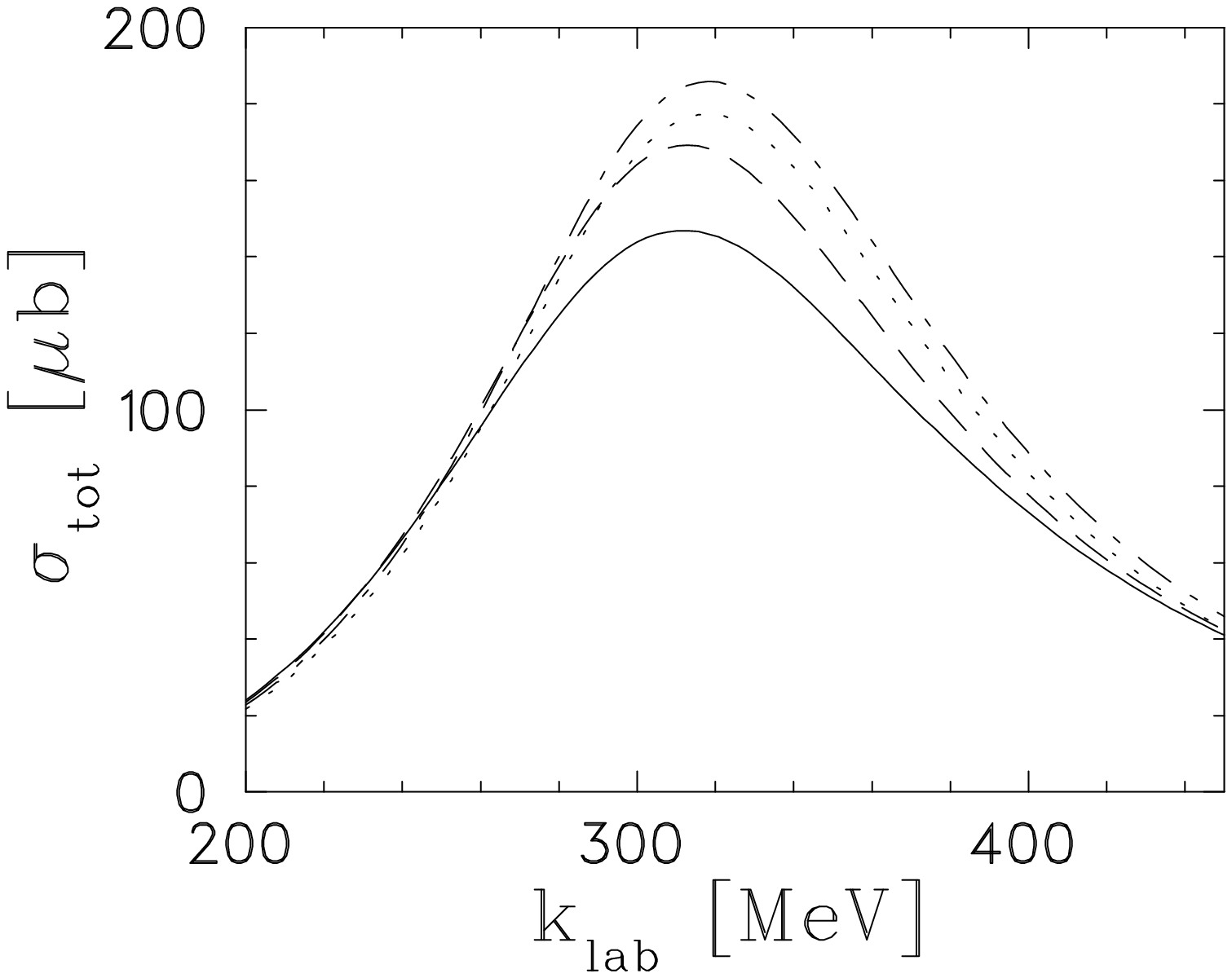,width=8cm,angle=0}}
\caption{figgdpi0tot}
{ Total cross section of \protect{$d(\gamma,\pi^0)d$}. Notation of the 
curves: 
 Dotted: impulse approximation;
 Dash-dot: IA plus MEC;
 Dashed: rescattering in Born approximation;
 Full: complete calculation (from \cite{WiA96}).}
\end{figure}

\begin{figure}
\centerline {\psfig{file=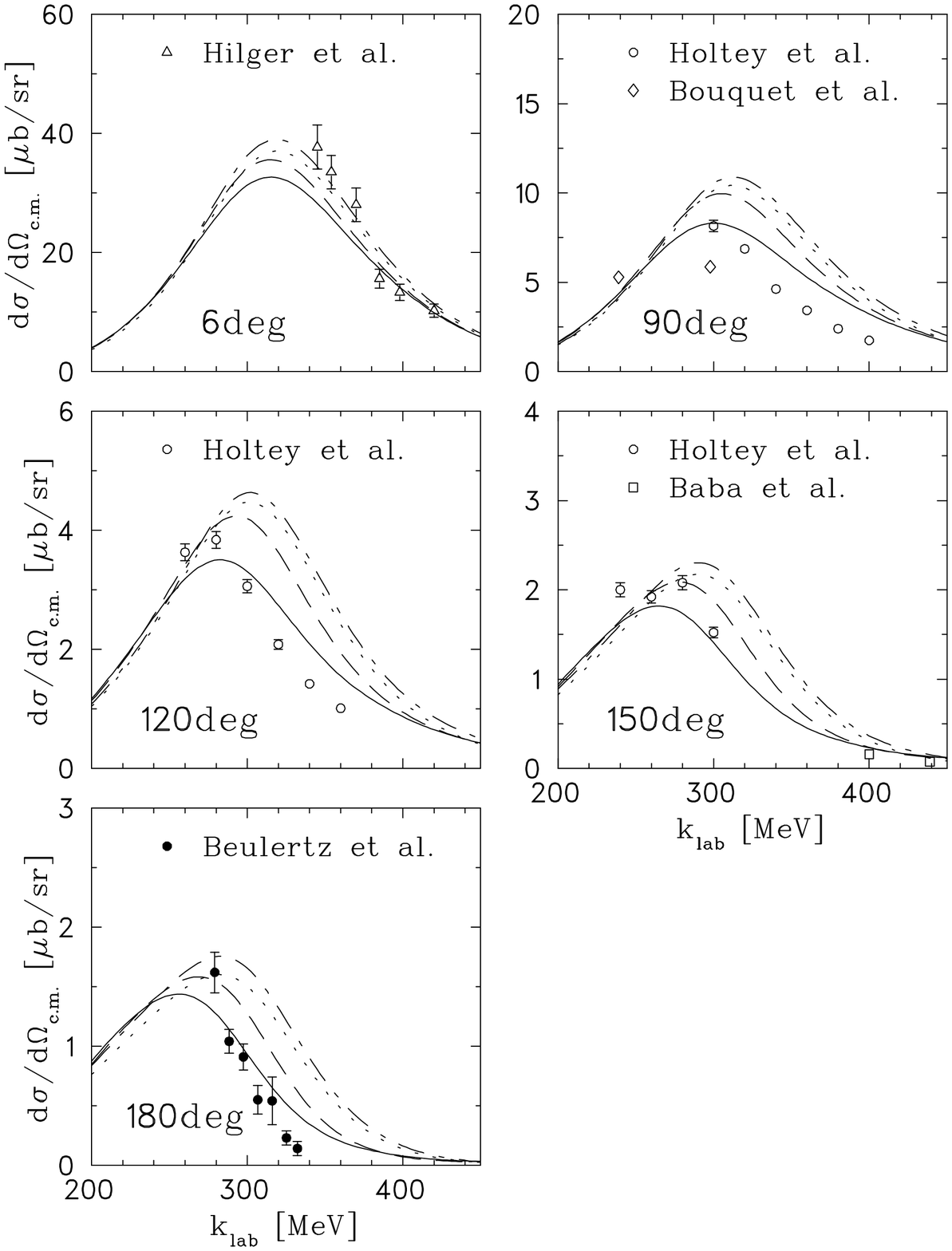,width=10cm,angle=0}}
\caption{figgdpi0diff}
{Differential cross section of $d(\gamma,\pi^0)d$. Notation of the curves:
 Dotted: impulse approximation;
 Dash-dot: IA plus MEC;
 Dashed: rescattering in Born approximation;
 Full: complete calculation (from \cite{WiA96}).}
\end{figure}

The differential cross sections for fixed pion angles, plotted in Fig.\
\ref{figgdpi0diff}, show the same features.  For energies in the $\Delta$
region, MECs lead to a slight increase but additional pion rescattering
reduces the cross section at all angles.  Its influence strongly
grows with the pion angle.  This qualitatively agrees with what one
finds in pion deuteron elastic scattering.  Intuitively, one would
always expect that rescattering mechanisms become more important at
higher momentum transfers, i.e., for larger scattering angles at fixed
energy, since rescattering provides a means to share the
momentum transfer between the two nucleons.  

These results clearly show that pion rescattering is significant and
reduces the cross section in the resonance region. 
This means, in particular with respect to a test of theoretical models
for pion production amplitudes on the neutron, that one needs a
reliable description for the rescattering process.  Compared to
experimental data, one readily finds that the sizable discrepancies 
without rescattering are largely reduced and that a reasonable agreement 
with the data is achieved. In addition polarization observables have 
been studied in \cite{WiA96}. Most of them are less sensitive to
rescattering than the cross section. An exception is the tensor target 
asymmetry $T_{20}$ which may serve as a special tool to
disentangle different reaction mechanisms (for details see \cite{WiA96}).

\subsubsection{Coherent Eta Photoproduction in the $S_{11}$(1535) 
Region}

The special interest in the
electromagnetic production of $\eta$ mesons on the nucleon is based on the
fact that, being an isoscalar meson, it constitutes a selective filter for
isospin $I=\frac{1}{2}$ nucleon resonances $N^{*}$.
Furthermore, the e.m.\ $\eta$ production is dominated by the intermediate
excitation of the $S_{11}$(1535) resonance. Thus this reaction is an ideal
tool for investigating the characteristic features of this resonance, which
usually is difficult to separate  from the other resonances because of 
their
large widths. For example, one can study its electromagnetic transition
multipoles and its decay channels, providing thus a good test for hadron
models.

In a recent exploratory study, Breitmoser and myself have 
investigated $d(\gamma,\pi^0)d$ from threshold through the $S_{11}$ 
resonance \cite{BrA97}. As a first step, we had restricted 
ourselves to the plane wave impulse approximation (PWIA) in order to study
details of the elementary reaction amplitude with respect to the yet
unknown neutron properties and to study different ways of implementing the
elementary amplitude in a bound system. With respect to the latter, one
has to face several problems which are analogous to the corresponding pion 
production process on the deuteron, just discussed before. First of all, 
the energy of the struck nucleon on which the reaction takes place is not 
well defined in a bound nonrelativistic system. This leads to a severe 
uncertainty of the invariant mass of the $\gamma N$ subsystem of the 
elementary reaction, on which the elementary $t$ matrix depends. 
For example, the invariant mass determines the decay width of the 
resonance, to which the resulting cross section is very sensitive. Secondly,
the elementary reaction amplitude, which usually is given
in the $\gamma N$ c.m.\ frame, has to be transformed to an arbitrary
reference frame. This may be done either by a Lorentz boost of all four
momenta on which the elementary amplitude depends or by calculating the
$t$ matrix with respect to an arbitrary frame right from the beginning. 
The last method is more general because one does not loose any terms 
which vanish in the $\gamma N$ c.m.\ frame. But in both cases, one faces 
again the problem of assigning an energy to the bound nucleon.

In the following, I will 
exclusively discuss the problem of assigning an invariant energy to the 
subsystem. For a bound system of two nucleons, the general expression 
for $W_{\gamma N}$ is, assuming the reaction to take place at nucleon ``1'',
\begin{eqnarray}
W_{\gamma N} = \sqrt{(p_{10}+k_{0})^{2}-(\vec{p}_{1}+\vec k)^{2}}
 = \sqrt{(p_{10}^{\prime}+\omega_{q})^{2}-(\vec{p}_{1}\,^{\prime}+\vec
q\,)^{2}}\,,
\end{eqnarray}
where $p_{1\mu}^{(\prime)}=(p_{10}^{(\prime)},\vec p_1^{\,(\prime)})$ 
denotes its initial (final) four momentum. The photon momentum is denoted 
by $k$. In general, one has $p_{10}^{(\prime)}\not=\sqrt{M^{2}+\vec{p}
\,^{\,(\prime)2}}$ because the bound nucleon is off-mass-shell. 
In fact, as already mentioned, the energy of an 
individual nucleon, bound in a nonrelativistic many-particle system, is not
defined at all, i.e., there is no operator which allows to determine the 
energy of a bound nucleon with momentum $\vec p$. Only the total sum of the 
energies of all nucleons is a well defined quantity, e.g., for the deuteron
$E_d^{(\prime)}=p_{10}^{(\prime)}+p_{20}^{(\prime)}$.

One of many possible choices is to distribute the total energy of the 
deuteron equally on each of the two nucleons (Blankenbecler-Sugar choice) 
in the deuteron rest system, i.e., there each nucleon has the energy 
$M_{d}/2$, independent of the momentum. This assignment for the initial 
or final deuteron state will be denoted by $W_{\gamma N}^{BS}$ and 
$W_{\gamma N}^{BS\prime}$, respectively, in the following discussion. 
Another possibility is to take the active nucleon on-shell, either before 
or after the interaction. The corresponding invariant masses will be 
denoted by $W_{\gamma N}^{N}$ and $W_{\gamma N}^{N\prime}$, respectively. 
As last choice, denoted by $W_{\gamma N}^{S}$, one may put the spectator 
nucleon on-shell, i.e., $p_{20}=E_{p}$. This choice has been used, for 
example, in coherent $\pi^0$ photoproduction in 
\cite{WiA95,WiA96}. It may be justified by the fact that the deuteron 
is only loosely bound, and hence the spectator acts nearly like a free 
nucleon. 

\begin{figure}
\centerline {\psfig{file=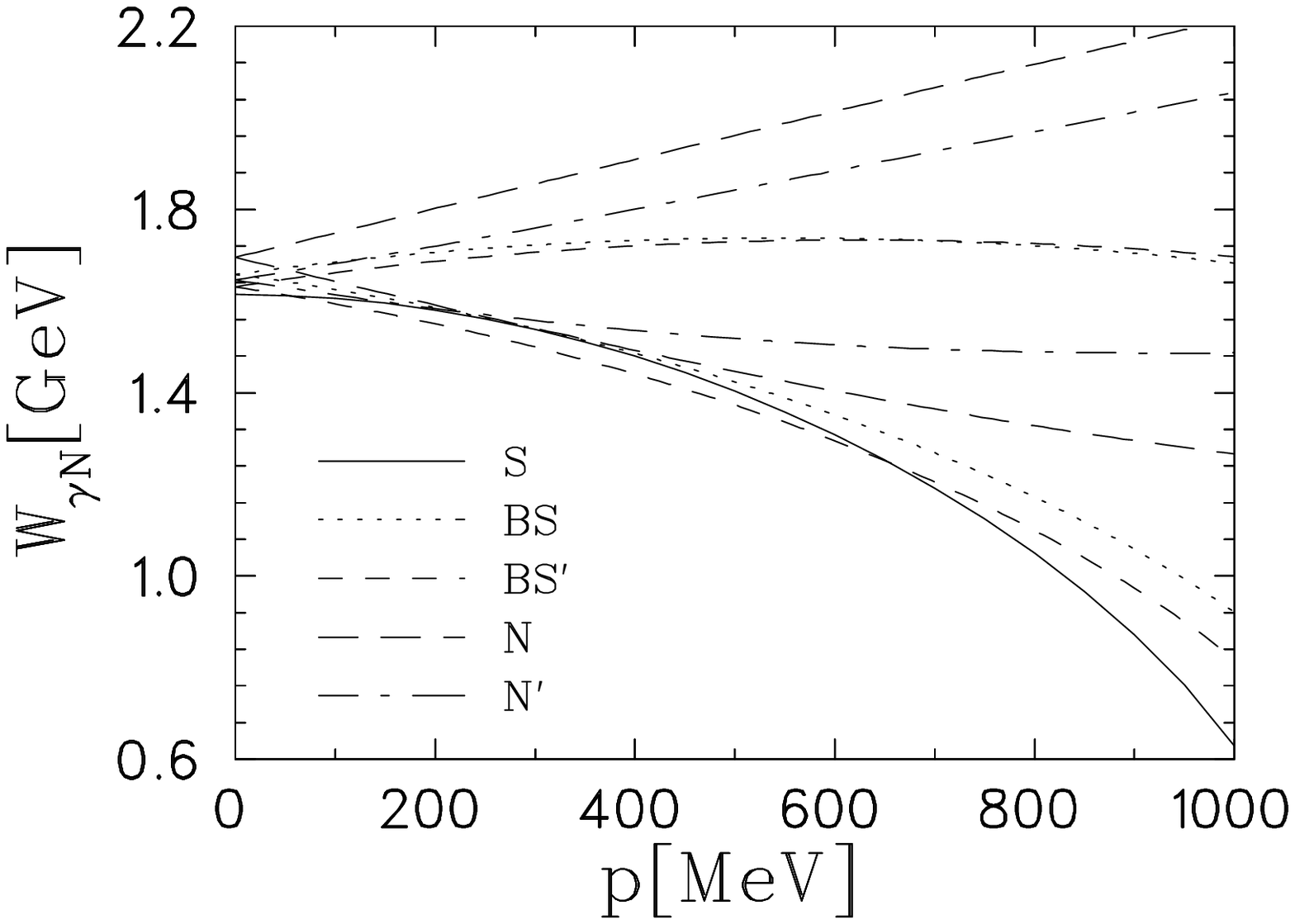,width=10cm,angle=0}}
\caption{figfig5}
{
Invariant mass $W_{\gamma N}$ of the active photon-nucleon
subsystem in $d(\gamma,\eta)d$ 
as a function of the spectator momentum $p$. Photon energy
$E_{\gamma}^{Lab}=800$ MeV. 
Full:  spectator on-shell $W_{\gamma N}^{S}$. 
Dotted:  both initial nucleons half off-shell $W_{\gamma N}^{BS}$. 
Short dashed:
both final nucleons half off-shell $W_{\gamma N}^{BS\prime}$. 
Long dashed:  active initial nucleon on-shell $W_{\gamma N}^{N}$. 
Dash-dotted:  active final nucleon on-shell $W_{\gamma N}^{N\prime}$. 
For the invariant mass assignments $W_{\gamma N}^{BS}$,
$W_{\gamma N}^{BS'}$, $W_{\gamma N}^{N}$, and $W_{\gamma N}^{N'}$, two
curves show the borderlines of  the invariant mass region 
(from \cite{BrA97}).
}
\end{figure}

\begin{figure}
\centerline {\psfig{file=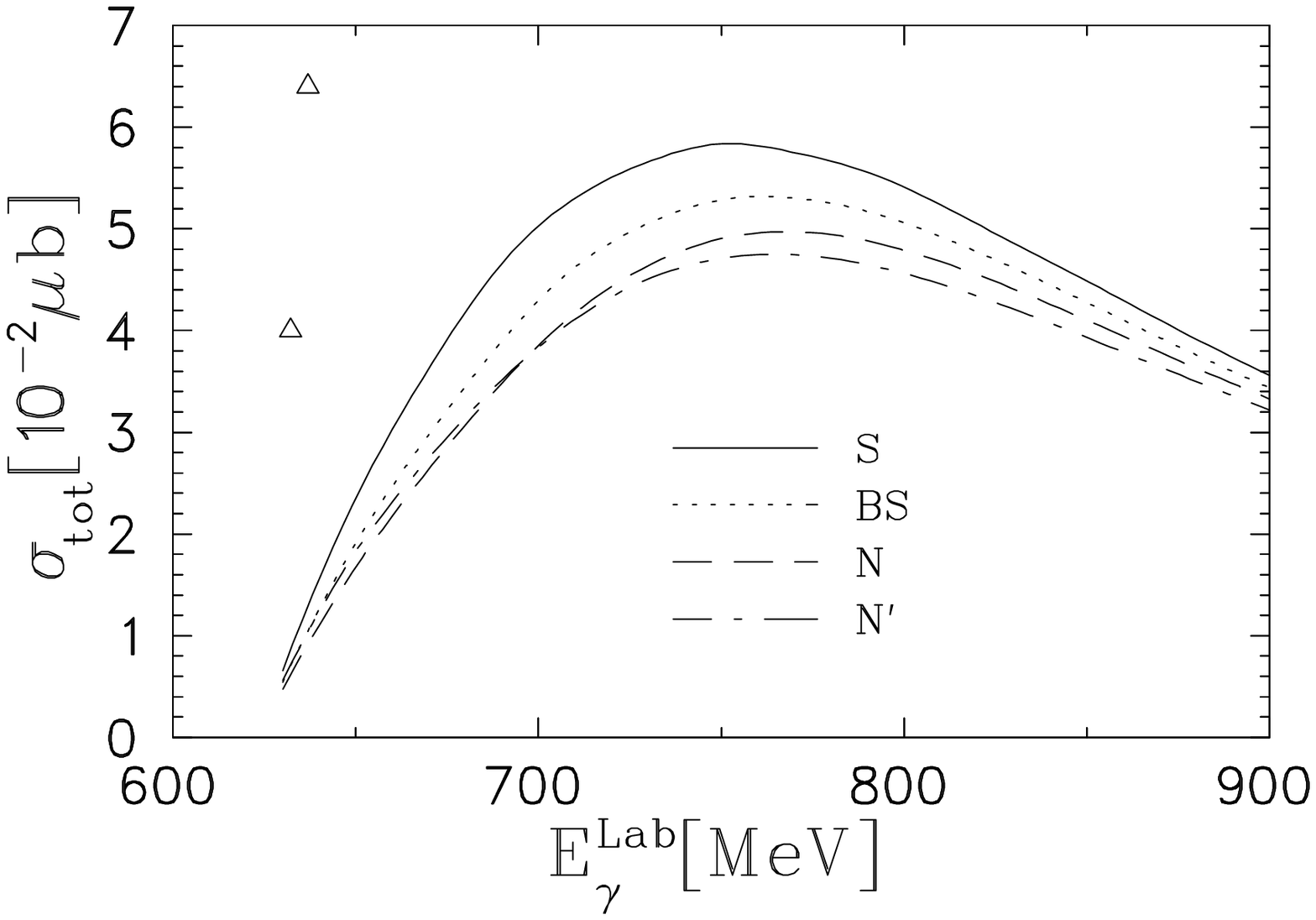,width=10cm,angle=0}}
\caption{figfig11}
{Total cross section for $d(\gamma,\eta)d$ including only the resonance 
$S_{11}$ for different choices of the invariant
mass of the active subsystem. 
Full:  spectator on-shell $W_{\gamma N}^{S}$.
Dotted:  both initial nucleons half off-shell $W_{\gamma N}^{BS}$.
Long dashed:  active initial nucleon on-shell $W_{\gamma N}^{N}$.
Dash-dotted:  active final nucleon on-shell $W_{\gamma N}^{N'}$.
Triangles represent the upper limit of \cite{Beu94} (from \cite{BrA97}). 
}
\end{figure}

Fig.\ \ref{figfig5} shows the invariant mass $W_{\gamma N}$ for these
different choices as function of the spectator momentum $\vec p$ at fixed
photon lab energy $E_{\gamma}^{Lab}=800$ MeV. For the first four choices
($W_{\gamma N}^{BS}$, $W_{\gamma N}^{BS'}$, $W_{\gamma N}^{N}$
and $W_{\gamma N}^{N'}$), the boundaries of the range spanned by the 
angle dependence are represented by two curves corresponding to $\vec p$ and  
$\vec k$ or $\vec q$ parallel (upper curve) and antiparallel (lower curve).
One readily notes that $W_{\gamma N}^{N}$ spans the largest range, while
the smaller ranges of $W_{\gamma N}^{BS}$, $W_{\gamma N}^{BS'}$
and $W_{\gamma N}^{N'}$ are compatible with each other. However, the 
average
invariant masses nearly coincide
for $W_{\gamma N}^{N}$ and $W_{\gamma N}^{N'}$, and they show a slight
increase with increasing spectator momentum, whereas they decrease for both
$W_{\gamma N}^{BS}$ and $W_{\gamma N}^{BS'}$ which, moreover, show a very
similar behaviour. Finally, $W_{\gamma N}^{S}$ gives the lowest invariant
mass with the strongest decrease with increasing $p$.
One has to keep in mind that the main contribution to the total cross
section originates from momenta
$p$ below 400 MeV. But even in this region one notes a sizable difference
between the various choices for the invariant mass of the active $\gamma N$
subsystem.

The influence of different choices for $W_{\gamma N}$ for the 
$\gamma N$ subsystem on the total cross section is shown in Fig.\ 
\ref{figfig11}, where only the $S_{11}$ resonance alone has been considered. 
The result using $W_{\gamma N}^{BS'}$ is not shown because it is very 
similar to the one with $W_{\gamma N}^{BS}$.
One readily notes considerable differences for the various prescriptions.
The largest total cross section is obtained with the spectator on-shell, 
$W_{\gamma N}^{S}$, having its maximum at 750 MeV. If one puts the active
nucleon on shell, i.e., takes $W_{\gamma N}^{N}$ or $W_{\gamma N}^{N'}$
instead, the maximum
is decreased by about 18\% and slightly shifted towards higher energies.
This decrease and shift can be understood as a result of the assignment
of a higher invariant mass to the $\gamma N$ subsystem and the additional
smearing due to the dependence on the angle between the spectator momentum
and the photon, respective $\eta$ momentum (see Fig.\ \ref{figfig5})
which leads to a larger effective width. The result is a slight upshift of
the resonance position and a broadening, thus lowering of the maximum.
One notes also only a little difference between putting the active nucleon 
before or after the interaction on-shell.
From the foregoing discussion it is apparent that the curve for the
Blankenbecler-Sugar assignment $W_{\gamma N}^{BS}$ is about halfway
between the spectator on shell and the active nucleon on shell, because,
according to Fig.\ \ref{figfig5},  $W_{\gamma N}^{BS}$ lies in between the
spectator and active nucleon assignments.

In view of these results, one has to conclude that the choice for the 
invariant mass of the $\gamma N$ subsystem has a significant influence
on the cross section. In order to obtain a cross section of the same 
magnitude for two different choices of $W_{\gamma N}$, one has to
change the elementary helicity amplitude, too. This introduces a 
systematic uncertainty with respect to the determination of the elementary 
neutron amplitude, which can only be removed by a proper relativistic 
treatment. 

\subsubsection{Final State Interaction in Incoherent Eta Photoproduction}
%(A.\ Fix et al., MKPH-T-97-15)

As already mentioned above, one major motivation for studying 
$\eta$ photoproduction on the deuteron is to obtain information on 
the elementary process on the neutron. This is of particular interest 
with respect to the isospin dependence of the production amplitude, i.e., 
the relative size of its isoscalar ($t_{\gamma\eta}^{(s)}$) to its 
isovector ($t_{\gamma\eta}^{(v)}$) part. The method of extracting the 
neutron amplitude is based on the so-called spectator-nucleon model, in
which the pure quasifree production is considered
as the only mechanism for the knock-out reaction $d(\gamma,\eta N)N$.
Therefore, a careful investigation of the validity of this approximation,
i.e., the estimation of possible other competing effects is necessary. 
One can expect that such effects become important close to the reaction 
threshold. In this region, the smallness of the excitation energy in 
the final $np$ system and the large momentum transfer (which is of the 
order of the $\eta$ mass in the $\gamma d$ c.m.\ frame) lead to a kinematical
situation, where the two final nucleons move primarily
together with a large total, but small relative momentum.
For this kinematics, the spectator model is expected to give
a very small cross section since the momenta of both nucleons are large
and, on the other hand, the corrections due to the strong
$NN$ interaction may be significant. 

Recently, such final state interaction effects in the 
$\gamma d\!\rightarrow\!\eta np$ process have been studied by Fix and 
myself \cite{FiA97}. As full amplitude we have included one-loop 
$NN$ and $\eta N$ rescattering terms as shown in Fig.\ \ref{figgraf} 
besides the pure impulse approximation (IA). 
\begin{figure}
\centerline {\psfig{file=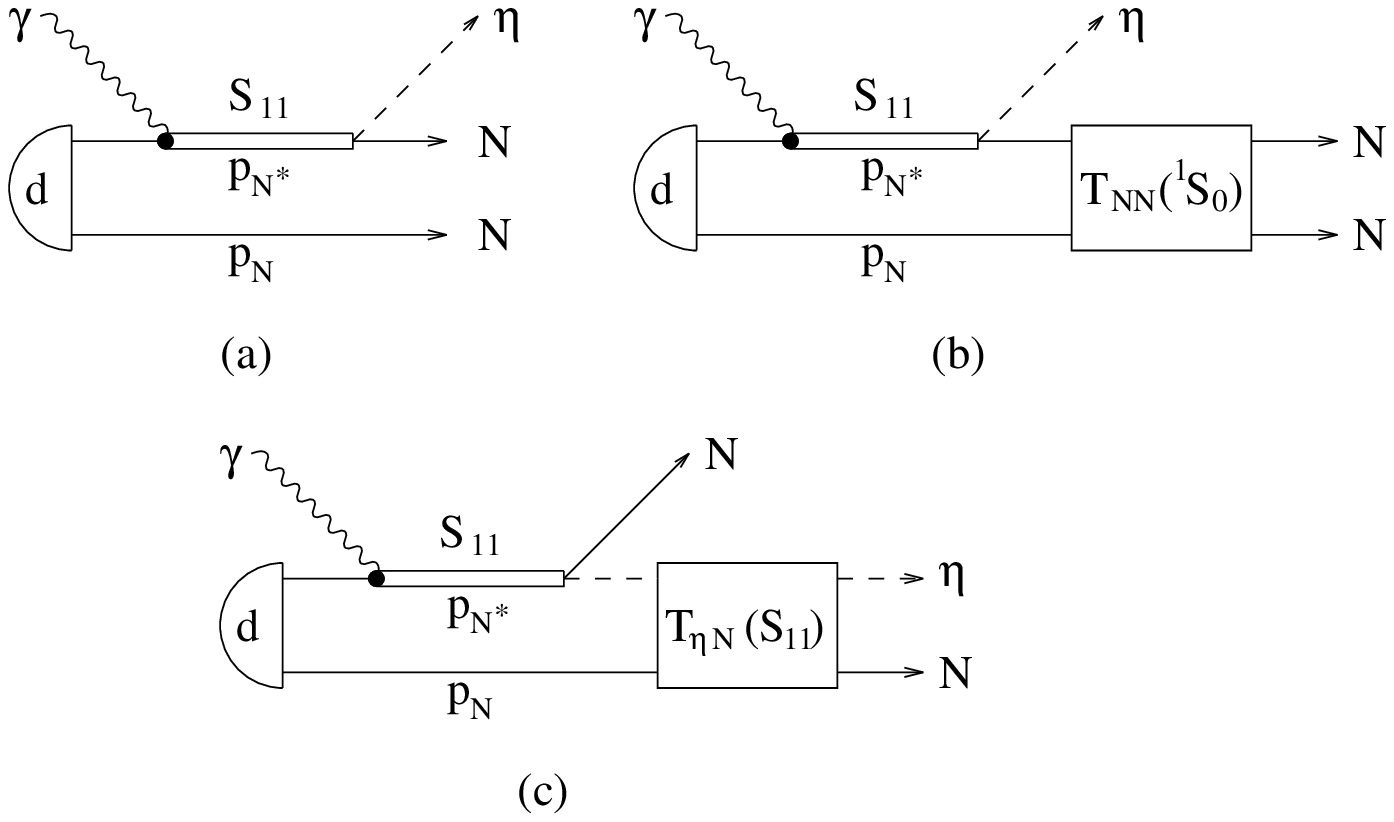,width=9cm,angle=0}}
\caption{figgraf}
{Diagrammatic representation of the contributions to the 
$\gamma d\!\rightarrow\!\eta np$ amplitude considered in \cite{FiA97}: 
(a) impulse approximation, (b) $NN$ rescattering and 
(c) $\eta N$ rescattering.
}
\end{figure}
The resulting total cross section is 
shown in Fig.\ \ref{figtot}. One readily notes, that the simple spectator 
approach cannot describe the experimental data close to the threshold  
(see also \cite{SaF95}). As has been mentioned above, at small photon 
energies, the $\gamma d\!\rightarrow\!\eta np$ process is governed in 
the spectator model by the high Fourier components of the 
deuteron wave function, which have only a small probability. On the other 
hand, the final state interaction provides a mechanism for bypassing 
this suppression. Indeed, quite a significant contribution from 
$NN$ rescattering is found. It turns out to be even dominant in the 
vicinity of the threshold, and at $\omega=720$ MeV it still increases 
the IA result by about 10$\%$. The $\eta N$ interaction is
relatively less important, but is also significant, 
mainly through the constructive interference between $NN$ and
$\eta N$ rescattering contributions.
With inclusion of both rescattering effects, we are able to reproduce
the experimental cross section with a ratio of isoscalar to proton amplitude 
$\alpha=0.11$.

\begin{figure}
\centerline {\psfig{file=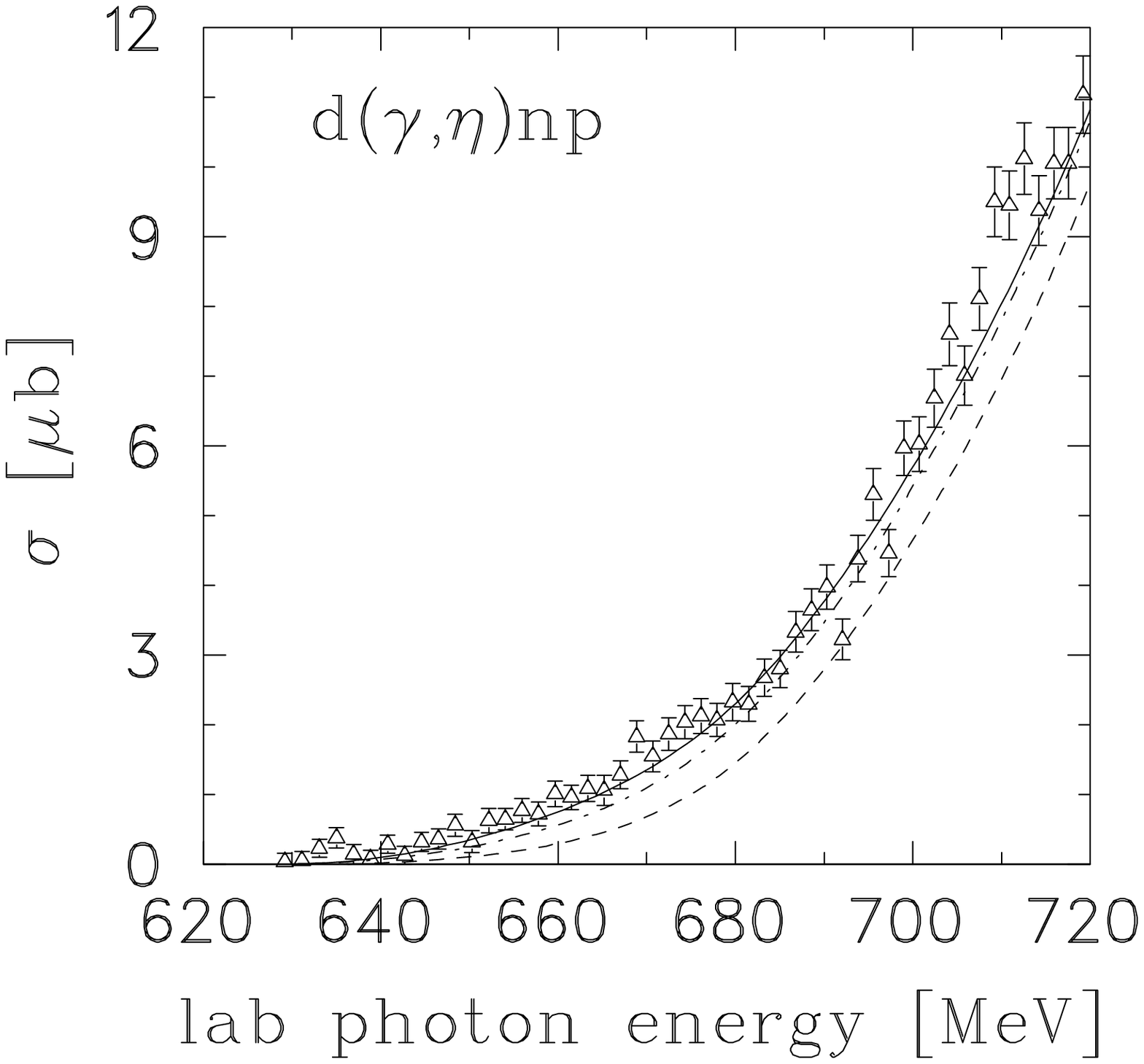,width=7.5cm,angle=0}}
\caption{figtot}
{Total 
$\gamma d\!\rightarrow\eta np$ cross section compared to inclusive
$\gamma d\!\rightarrow\eta X$ experimental data \protect{\cite{Kru95}}. 
The full and dashed lines represent the results obtained with and without
allowance for rescattering of the final particles, respectively.
The dash-dotted line includes only IA and $NN$ rescattering 
(from \protect{\cite{FiA97}}).
}
\end{figure}

\begin{figure}
\centerline {\psfig{file=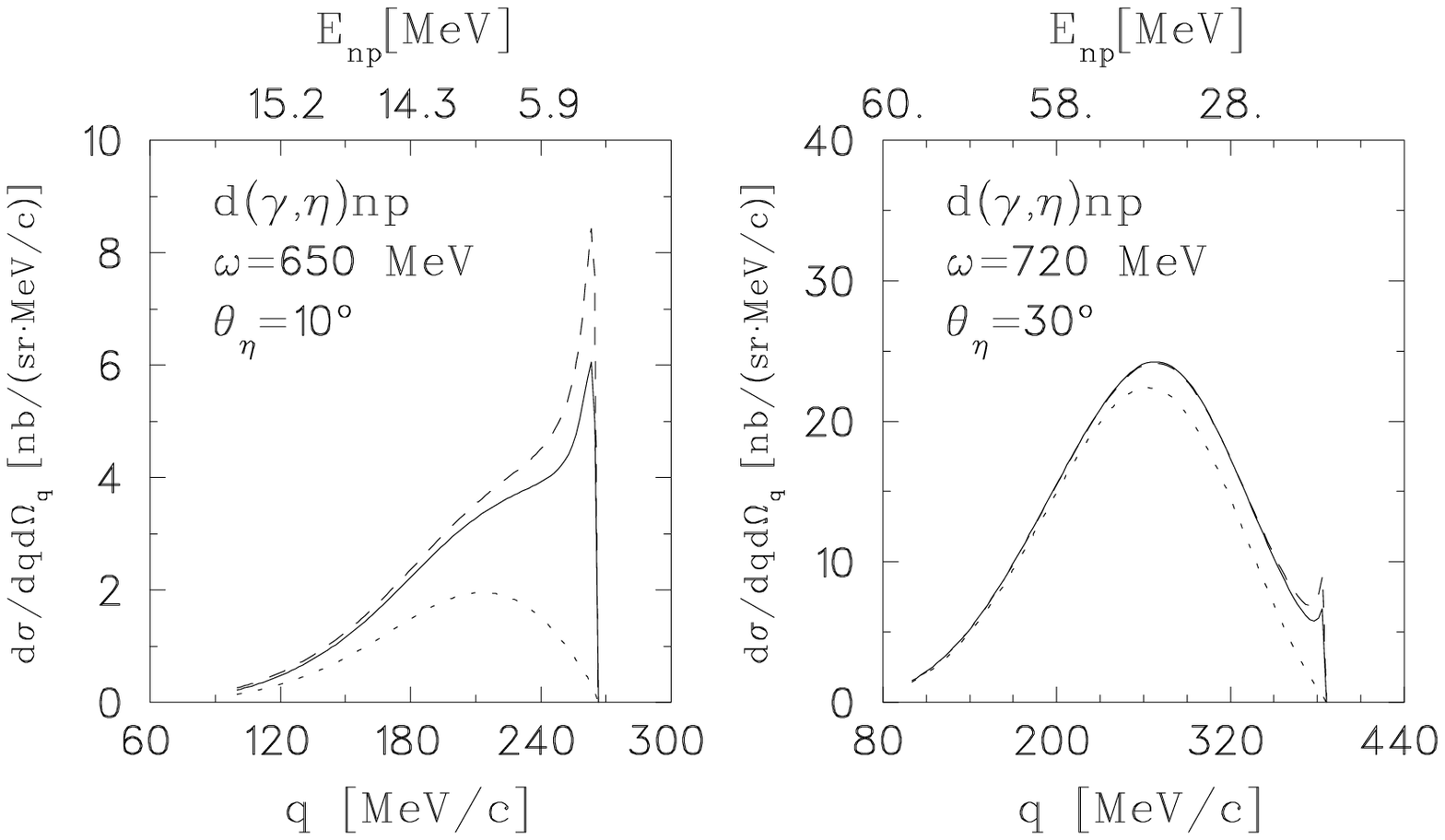,width=12cm,angle=0}}
\caption{figThQ}
{
The $\eta$ meson spectra at forward emission angles
for $d(\gamma,\eta)np$ at two different photon energies
and angles.
The dotted curves show the pure
impulse approximation, whereas the full curves include
the interaction between the outgoing nucleons.
The dashed curves represent the results obtained without
the $D$-wave contribution to the
$NN$ rescattering amplitude.
The excitation energy $E_{np}$ in the final $NN$ system is indicated at the
top abscissa (from \protect{\cite{FiA97}}).
}
\end{figure}

The influence of $NN$ rescattering on  the $\eta$ momentum distribution 
at a fixed angle $\theta_{\eta}$ is shown in Fig.\ \ref{figThQ}. 
As expected, the strong interaction between the final
nucleons in the $^1S_0$ state changes drastically the cross section for
large $\eta$ momentum values. When $q$ reaches its maximum, the excitation
energy $E_{np}$ in the $np$ pair becomes very small, and thus the resonant
$^1S_0$ state appears as a rather narrow peak. The same effect appears in 
charged $\pi$ photo- and electroproduction on the deuteron 
\cite{Lag78,KoA87} as well as in 
deuteron electrodisintegration \cite{FaA76}.
In principle, the experimental observation of this peak in the high
$\eta$ momentum spectrum may serve as another evidence for the
isovector nature of the $S_{11}$ photoexcitation. In the hypothetical
case that the isoscalar amplitude is much larger than the isovector one, 
%$t_{\gamma\eta}^{(s)}\gg t_{\gamma\eta}^{(v)}$, 
the low-energy $NN$ rescattering would be dominated by the $^3S_1$ state,
which does not exhibit any resonant behaviour at $E_{np}\!\approx$ 0.
In conclusion, one sees that the role of $NN$ rescattering is quite
important, especially in the threshold region. 
At higher energies, the main part of the cross section
is dominated by the IA, which gives a rather broad quasifree bump,
where the role of
$NN$ rescattering is expected to be of minor importance.

The effect of $\eta N$ rescattering is demonstrated in Fig.\ \ref{figTH}
for the $\eta$ angular distribution in the $\gamma d$ c.m.\ frame,
where the theory is compared with the experimental data for the
inclusive reaction $\gamma d\!\rightarrow\!\eta X$ \cite{Kru95}.
In view of the small isoscalar part $t_{\gamma\eta}^{(s)}$ of the
elementary amplitude and the large momentum mismatch, the contribution from
the coherent $\gamma d\!\rightarrow\!\eta d$ process is expected to be
negligible and, thus, the inclusive $\eta$ spectrum is dominated by the
deuteron break-up channel. In the backward direction,
the increase from
$\eta N$ rescattering is in part kinematically enhanced.
At these angles, the nucleons leave the interaction region
with large momenta. Therefore, the spectator model gives a very
small cross section underestimating the
data by roughly a factor of \,3\, for $\omega=720$ MeV.
In this situation, the
$\eta N$ rescattering mechanism, allowing the large transferred momentum
to be shared between the two participating nucleons, becomes much more 
effective.
The resonant character of the $\eta N$ interaction appears more pronounced
at forward angles. Although its strong inelasticity decreases the cross 
section at high photon energies, close to threshold this effect is more than
compensated by the attraction in the $\eta N$ system.

\begin{figure}
\centerline {\psfig{file=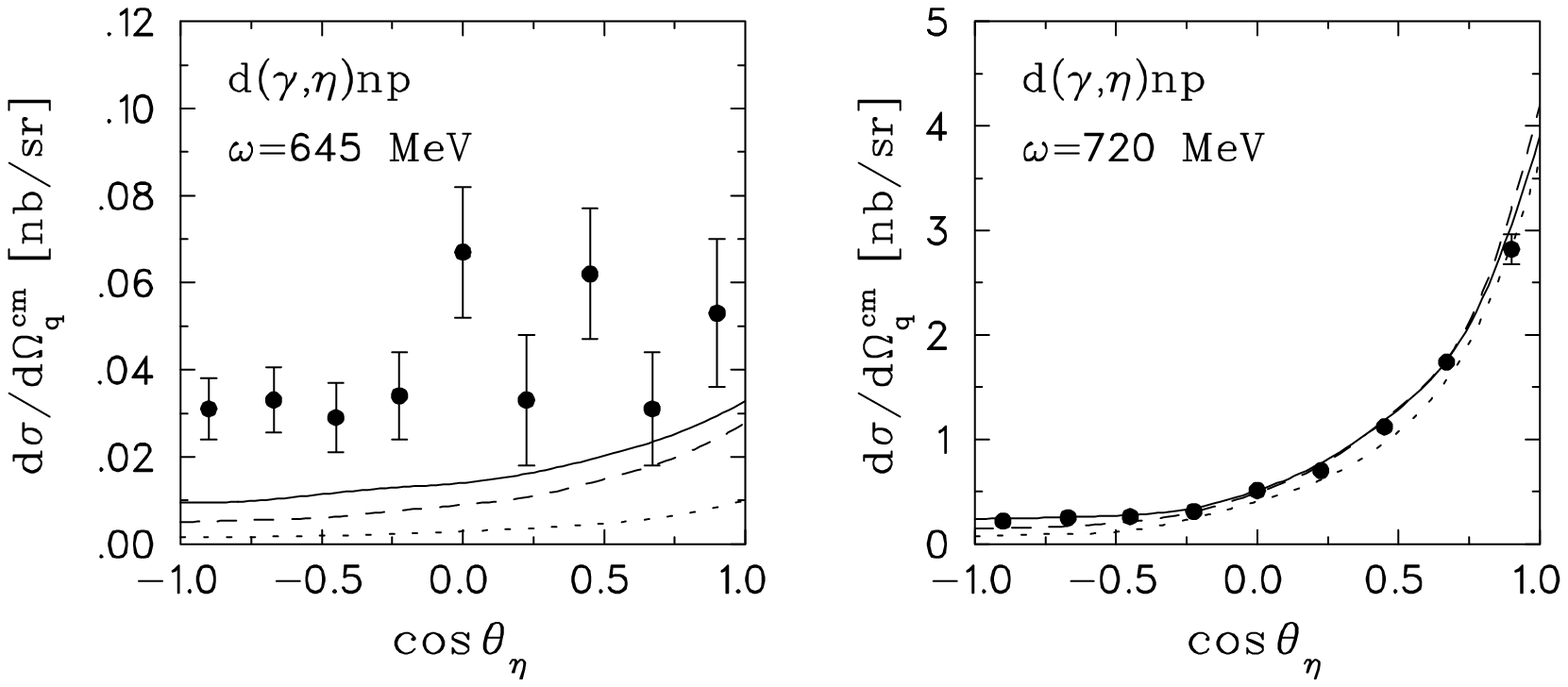,width=12cm,angle=0}}
\caption{figTH}
{The differential $d(\gamma,\eta)np$ cross section, calculated
in the $\gamma d$ c.m.\ frame (from \protect{\cite{FiA97}}). Shown are the IA
prediction (dotted lines), the successive addition of $NN$ (dashed
lines) and $\eta N$ (full lines) rescattering.
The experimental points represent the inclusive $\gamma d\!\rightarrow\eta 
X$ measurements from \protect{\cite{Kru95}}.
}
\end{figure}

\subsection{The Gerasimov-Drell-Hearn Sum Rule}

The Gerasimov-Drell-Hearn (GDH) sum rule connects the anomalous magnetic 
moment of a particle with the energy weighted integral - henceforth denoted by
$I^{GDH}(\infty)$ - from threshold up to infinity over the spin asymmetry of 
the total photoabsorption cross section, i.e., the difference of the total 
photoabsorption cross sections for circularly polarized photons on a target
with spin parallel and antiparallel to the spin of the photon.
In detail it reads for a particle of mass $M$, charge $eQ$, anomalous
magnetic moment $\kappa$ and spin $S$
\begin{eqnarray}
I^{GDH}(\infty)=\int_0^\infty \frac{dk}{k}\Big(\sigma^P(k)-\sigma^A(k)\Big)
= 4\pi^2  \kappa^2\frac{e^2}{m^2}S \,,
\label{igdh}
\end{eqnarray}
where $\sigma ^{P/A}(k)$ denote the total absorption cross sections for
circularly polarized photons on a target with spin parallel and antiparallel
to the photon spin, respectively, and the anomalous magnetic moment is
defined by the total magnetic moment operator of the particle
\begin{eqnarray}
{\vec M} &=& (Q+  \kappa)\frac{e}{m}{\vec S}\,.
\end{eqnarray}
This sum rule gives a very interesting relation between a
magnetic ground state property of a particle and an integral property of
its whole excitation spectrum. In other words, this sum rule shows that the
existence of a nonvanishing anomalous magnetic moment points directly to an
internal dynamic structure of the particle. Furthermore, because the rhs
of (\ref{igdh}) is positive, it tells us that the integrated, energy-weighted
total absorption cross section of a circularly polarized photon on a particle 
with its spin parallel to the photon spin is bigger than the one on a target with
its spin antiparallel, if the particle possesses a nonvanishing anomalous
magnetic moment. The recent interest in this sum rule stems from the
study of the spin dependent structure functions in deep inelastic
scattering \cite{Dre95}.

The GDH sum rule has first been derived by Gerasimov \cite{Ger65} and,
shortly afterwards, independently by Drell and Hearn \cite{DrH66}. It is
based on two ingredients which follow from the general principles of
Lorentz and gauge invariance, unitarity, crossing symmetry and causality
of the forward Compton scattering amplitude of a particle. The first one is the
low energy theorem of Low \cite{Low54} and Gell-Mann and Goldberger
\cite{GeG54} for a spin one-half particle which later has been generalized
to arbitrary spin \cite{LaC60,Sai69,Fri77}.
The low energy theorem for the forward amplitude for elastic scattering 
of a photon with momentum $\vec k$ on a target with spin $\vec S$ and an 
anomalous magnetic moment $\kappa$ reads
\begin{eqnarray}
T_{\lambda \lambda}(\vec k, \vec S\,) &=&  - e^2 \frac{Q^2}{m} + 
\lambda  \kappa^2 \frac{e^2}{m^2} \langle {\bf S}\cdot {\bf k}\rangle 
+{\cal O}({\bf k}^2)\,,
\end{eqnarray}
where the first term describes the classical Thomson amplitude. It is 
important to note that the spin term is already of relativistic order. 

The second ingredient is the assumption of an unsubtracted dispersion
relation for the difference of the elastic forward scattering
amplitudes for circularly polarized photons and a completely polarized
target with spin parallel ($S_P$) and antiparallel ($S_A$) to the photon spin 
\begin{eqnarray}
\Re e f(k) &=& \frac{{\cal P}}{\pi}
\int_{-\infty}^{\infty} dk'\frac{\Im m f(k)}{k'-k}\,,
\end{eqnarray}
where 
\begin{eqnarray}
f(k)&=& T_{\lambda \lambda}(k,\,S_P) - T_{\lambda \lambda}(k,\,S_A)\nonumber\\
&=& 2 \kappa^2\frac{e^2}{m^2}k\,S + {\cal O}(k^2) \,.
\end{eqnarray}
Certainly, this assumption appears to be the weakest point on which the GDH 
sum rule is based. A hand waving argument in support of it is, that with 
increasing energy, due to the increased phase space and increased 
number of produced 
particles, the dependence of the total cross section on the target spin 
orientation will decrease, so that the spin asymmetry converges more 
rapidly than the unpolarized cross section. 

Using crossing symmetry and the optical theorem, one obtains then 
\begin{eqnarray}
\Re e f(k) &=&
\frac{k}{2\pi^2} {\cal P}\int_0^{\infty} dk' k' 
\frac{ \sigma^P(k')-\sigma^A(k')}{k^{\prime\,2} -k^2}\,.
\end{eqnarray}
Finally, the derivative at $k=0$ yields the GDH sum rule. 

\subsubsection{The GDH Sum Rule for the Nucleon}
I will first consider the GDH sum rule for the nucleon. 
Since proton and neutron have large anomalous magnetic moments, one finds
correspondingly large GDH sume rule predictions for them, i.e., 
\begin{eqnarray}
I^{GDH}_p(\infty) =  204.8\, \mu\mbox{b}\,, \quad \mbox{ and } \quad
I^{GDH}_n(\infty) =  233.2\, \mu\mbox{b}\,.
\end{eqnarray}
Although this sum rule is known for more than 30 years, it has never been
evaluated by a direct integration of experimental data on $\sigma^P(k')-
\sigma^A(k')$.  
The absorptive processes to be included are for the proton 
(analogously for the neutron)
\begin{eqnarray*}
\gamma + p \,& \rightarrow & \,p + \pi^0\,,\\
\gamma + p \,& \rightarrow & \,n + \pi^+\,,\\
\gamma + p \,& \rightarrow & \,N + \pi + \pi\,,\,\, \mbox{etc.}
\end{eqnarray*}
Early evaluation by Karliner \cite{Kar73} of the finite  GDH integral
\begin{eqnarray}
I^{GDH}(k) \,& = &
\int_0^k \frac{dk'}{k'}\Big( \sigma^P(k')-\sigma^A(k')\Big)\,,
\end{eqnarray}
based on a multipole analysis of experimental data on single pion
photoproduction on the nucleon, did not give
conclusive results due to the lack of data at higher energies, and even
present day data do not allow a definite answer as to its validity (see 
e.g. \cite{SaW94}). The contribution to the GDH sum rule for the proton from 
single pion production is shown in Fig.\ \ref{figgdh_proton}. 
\begin{figure}
\centerline {\psfig{file=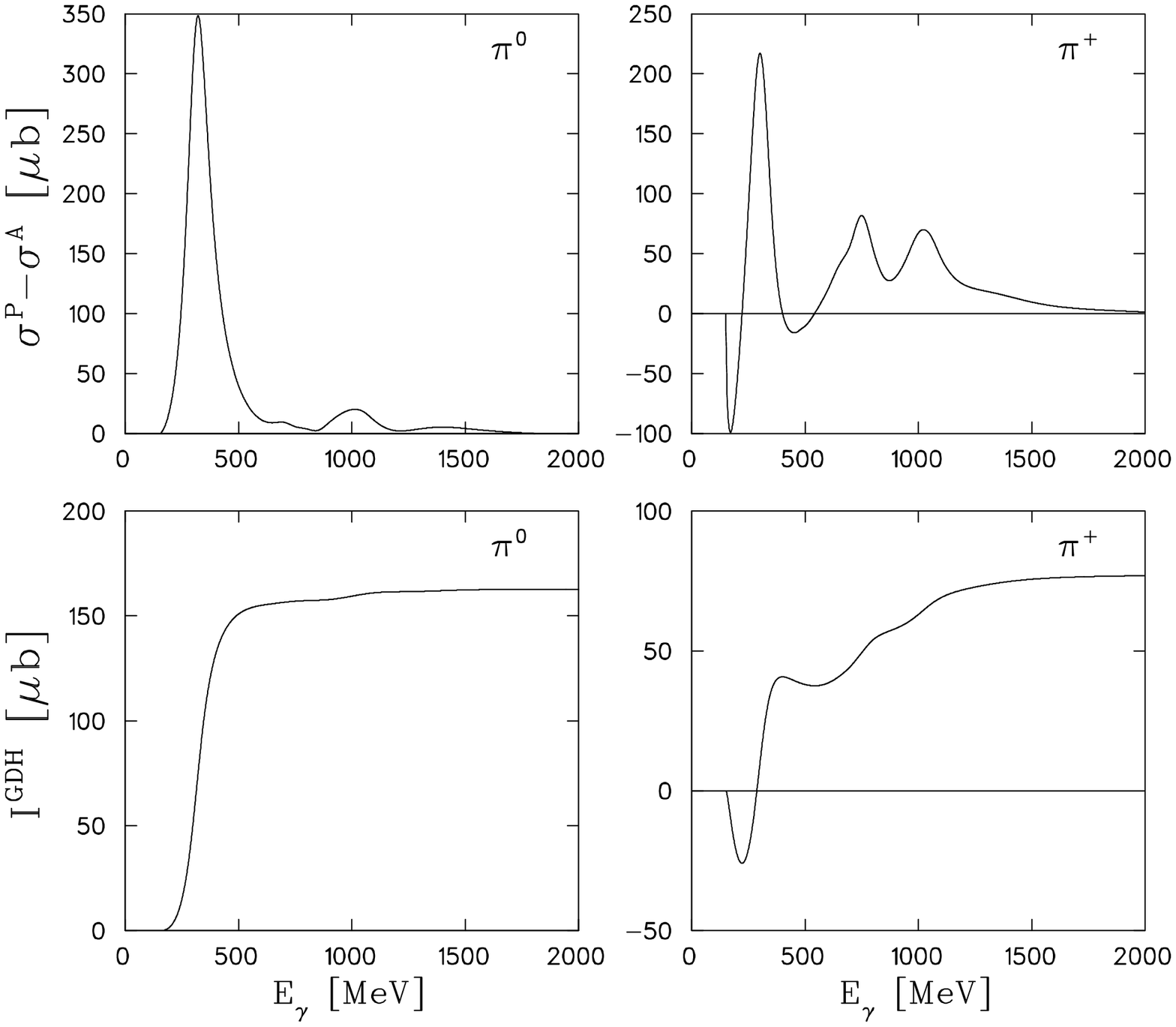,width=10cm,angle=0}}
\caption{figgdh_proton}
{The spin asymmetry $\sigma^P-\sigma^A$ and the finite GDH integral as function 
of the upper integration energy for the proton from the VPI multipole 
analysis (fit SM95) \cite{SAID} (from \cite{ArK97}).}
\end{figure}
Explicit integration up to 2 GeV gives the values \cite{ArK97} (see also 
\cite{SaW94}) 
\begin{eqnarray}
I^{GDH}_p(2\,\mbox{GeV})_{\pi}=  239\, \mu\mbox{b} \quad \mbox{ and } \quad 
I^{GDH}_n(2\,\mbox{GeV})_{\pi}=  168\, \mu\mbox{b} \,.
\end{eqnarray}
Furthermore, an estimation of the $2\pi$ contribution up to 1.7 GeV 
from Karliner \cite{Kar73}
\begin{eqnarray}
I^{GDH}_p(1.7\,\mbox{GeV})_{\pi\pi}= 65\, \mu\mbox{b} \quad \mbox{ and } \quad 
I^{GDH}_n(1.7\,\mbox{GeV})_{\pi\pi}= 35\, \mu\mbox{b} \,,
\end{eqnarray}
worsens the comparison with the sum rule values. 

\subsubsection{The GDH Sum Rule for the Deuteron} 

Applying the GDH sume rule to the deuteron, one finds a very interesting
feature. On the one hand, the deuteron has isospin zero, ruling out the
contribution of the large nucleon isovector anomalous magnetic moments to
its magnetic moment, and, therefore, only a very small anomalous magnetic 
moment is expected. In fact, the experimental value
is $\kappa_d=-0.143$ resulting in a GDH prediction of 
$I^{GDH}_d (\infty) = 0.65\,\mu$b,
which is more than two orders of magnitude smaller
than the nucleon values. On the other hand, considering the possible
absorptive processes, 
\noindent
(i) photodisintegration $\gamma + d \, \rightarrow  \,n + p$, 
\noindent
(ii) single pion production, coherent $\gamma + d \, \rightarrow  
\,d + \pi^0$ and incoherent $\gamma + d \,\rightarrow  \,N+N+\pi $,
\noindent
(iii) two pion production etc., 
\noindent
one notes first, that the incoherent pion production on 
the deuteron is dominated by the quasifree production on the nucleons bound
in the deuteron. Thus it is plausible to expect from these processes a
contribution to the GDH sum rule roughly given by the sum of the proton and
neutron GDH values, i.e., 438 $\mu$b. Additional contributions arise from
the coherent $\pi^0$ production channel. In order to obtain the small total
deuteron GDH sum rule, one, therefore, needs a large negative contribution of
about the same size for cancellation. Indeed, one has an additional channel 
not
present for the nucleon, namely the photodisintegration channel which is
the only photoabsorption process below the pion production threshold.
A closer look shows in fact that at very low energies near threshold a
sizable negative contribution arises from the $M1$ transition to the
resonant $^1S_0$ state, because this state can only be reached if the spins
of photon and deuteron are antiparallel, and is forbidden for the parallel
situation as has been pointed out, for example in \cite{BaD67}.

Recently, we have evaluated explicitly the GDH sum rule for the deuteron by
integrating the difference of the two total photoabsorption cross sections
with photon and deuteron spins parallel and antiparallel up to a photon 
energy of 550 MeV \cite{ArK97}. Three contributions have been included: (i) the
photodisintegration channel $\gamma d \rightarrow n p$, (ii) the coherent
pion production $\gamma d \rightarrow \pi^0 d$, and (iii) the
incoherent pion production $\gamma d \rightarrow \pi N N$.
As already mentioned, the upper integration limit of 550 MeV has been 
chosen because on the one hand one finds sufficient convergence for the 
photodisintegration channel, while on the other hand, only single pion 
photoproduction has been considered, thus limiting the
applicability of the present theoretical treatment to energies not too far
above the two pion production threshold as long as significant
contributions from multipion production cannot be expected. Indeed, the
recent evaluation of $I^{GDH}$ using experimental data for the nucleon
by Sandorfi et al.\ \cite{SaW94} indicates
that significant contributions from two-pion production start only above
this energy. I will now discuss the three contributions separately.
\\

\noindent
{\bf (i) GDH contribution from photodisintegration}

At low energies one has dominant contributions from the $E1$ and $M1$ 
multipoles. However, the $E1$ transitions cancel each other almost 
completely. Thus, at
low energies remain the $M1$ transitions, essentially to $^1S_0$ and
$^3S_1$ states. Of these, the $^1S_0$ contribution is dominant because
of the large isovector part of the $M1$ operator coming from the large
isovector anomalous magnetic moment of the nucleon. It is particularly
strong close to break-up threshold, where the
$^1S_0$ state is resonant, and can only be reached by the antiparallel spin
combination resulting in a strong negative contribution to the GDH sum 
rule. The results are summarized in Fig.\ \ref{figgdh_fig1}, where the cross 
section difference and the GDH integral is shown. The GDH values are listed 
in Tab.\ \ref{tabdis}. 
\begin{figure}[h]
\centerline {\psfig{file=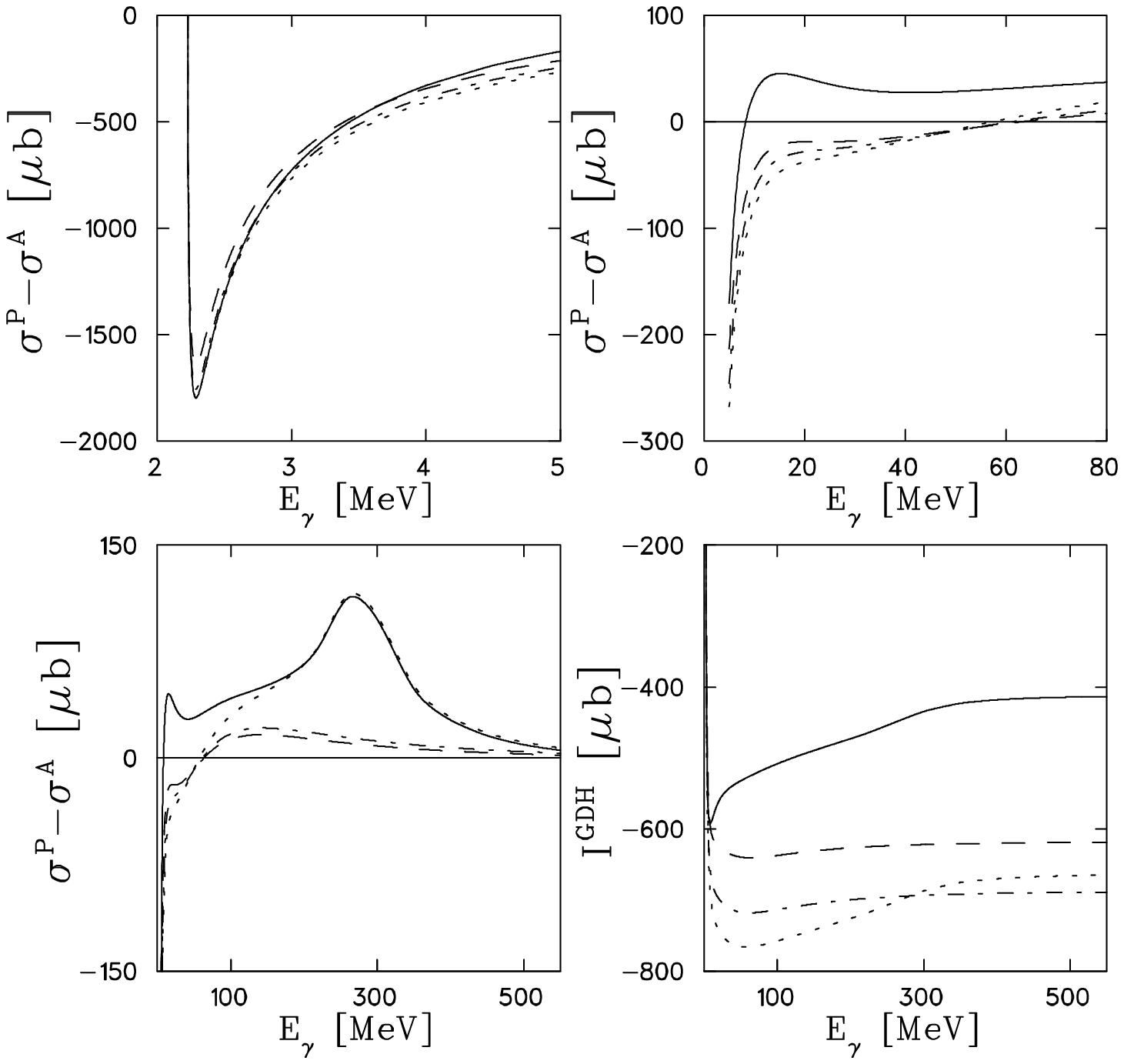,width=12cm,angle=0}}
\caption{figgdh_fig1}
{Contribution of the photodisintegration channel to the GDH sum rule for 
the
deuteron. Two upper and lower left panels: difference of the cross sections
in various energy regions; lower right panel:
$I^{GDH}_{\gamma d \to np}$ as function of the upper integration energy. 
Dashed curves: normal (N), dash-dot: N+MEC, dotted: N+MEC+IC, and 
full curves N+MEC+IC+RC (from \cite{ArK97}).}
\end{figure}
\begin{table}[h]
\fcaption{tabdis}
{Various contributions of the photodisintegration channel to the
GDH integral for the deuteron integrated up to 550 MeV in $\mu$b.}
\begin{center} 
\begin{tabular}{|ccc|c|}
\hline
  N & N+MEC & N+MEC+IC & N+MEC+IC+RC\\
\hline
 $-619$ & $-689$ & $-665$ & $-413$ \\
\hline
\end{tabular}
\end{center}
\end{table}

One readily notes the huge negative contribution from
the $^1S_0$ state at low energies (see the upper left panel of Fig.\
\ref{figgdh_fig1}). It is the result of the large isovector anomalous magnetic 
moment of the nucleon. Indeed, for a vanishing anomalous nucleon magnetic 
moment one finds 
$I^{GDH}_d(550\,\mbox{MeV})_{ \kappa_N=0}\,=\, 7.3\, \mu\mbox{b}$.
The effects from MEC
are relatively strong, resulting in an enhancement of the negative value by
about 15 percent. It corresponds to the well-known 10 percent enhancement
of the radiative capture of thermal neutrons on protons.
Isobar effects are significant in the region of
the $\Delta$ resonance, as expected. They give a positive contribution, but
considerably smaller in absolute size than MEC. The largest positive
contribution stems from relativistic effects (RC) 
in the energy region up to about 100 MeV
(see the upper right panel of Fig.\ \ref{figgdh_fig1})
reducing the GDH value in absolute size by more than 30 percent. This
strong influence of relativistic effects is not surprising in view of the
fact, that the correct form of the term linear in the photon momentum of
the low energy expansion of the forward Compton scattering amplitude
is only obtained if leading order relativistic contributions
are included \cite{Fri77}.
\\

\noindent
{\bf (ii) GDH contribution from coherent pion production}

As contributions from single pion production one has to consider 
coherent and incoherent processes. 
The theoretical model used to calculate the contribution of the
coherent pion production channel has been described above in Sect.\ 
\ref{sec6meson}. The reaction is clearly dominated by the
magnetic dipole excitation of the $\Delta$ resonance from which one expects
a strong positive $I^{GDH}_{\gamma d \to d\pi^0}$ contribution. The reason
for this is that the
$\Delta$ excitation is favoured if photon and nucleon spins are parallel
compared to the antiparallel situation. 
Fig.\ \ref{figgdh_fig2} shows the result of our calculation. One sees the strong
positive contribution from the $\Delta$ excitation giving a value of 
$I^{GDH}_{\gamma d \to d\pi^0}(550\,\mbox{MeV})=63\,\mu$b. The comparison
with the unpolarized cross section, also plotted in Fig.\ \ref{figgdh_fig2},
demonstrates the dominance of $\sigma^P$ over $\sigma^A$.
\begin{figure}[h]
\centerline {\psfig{file=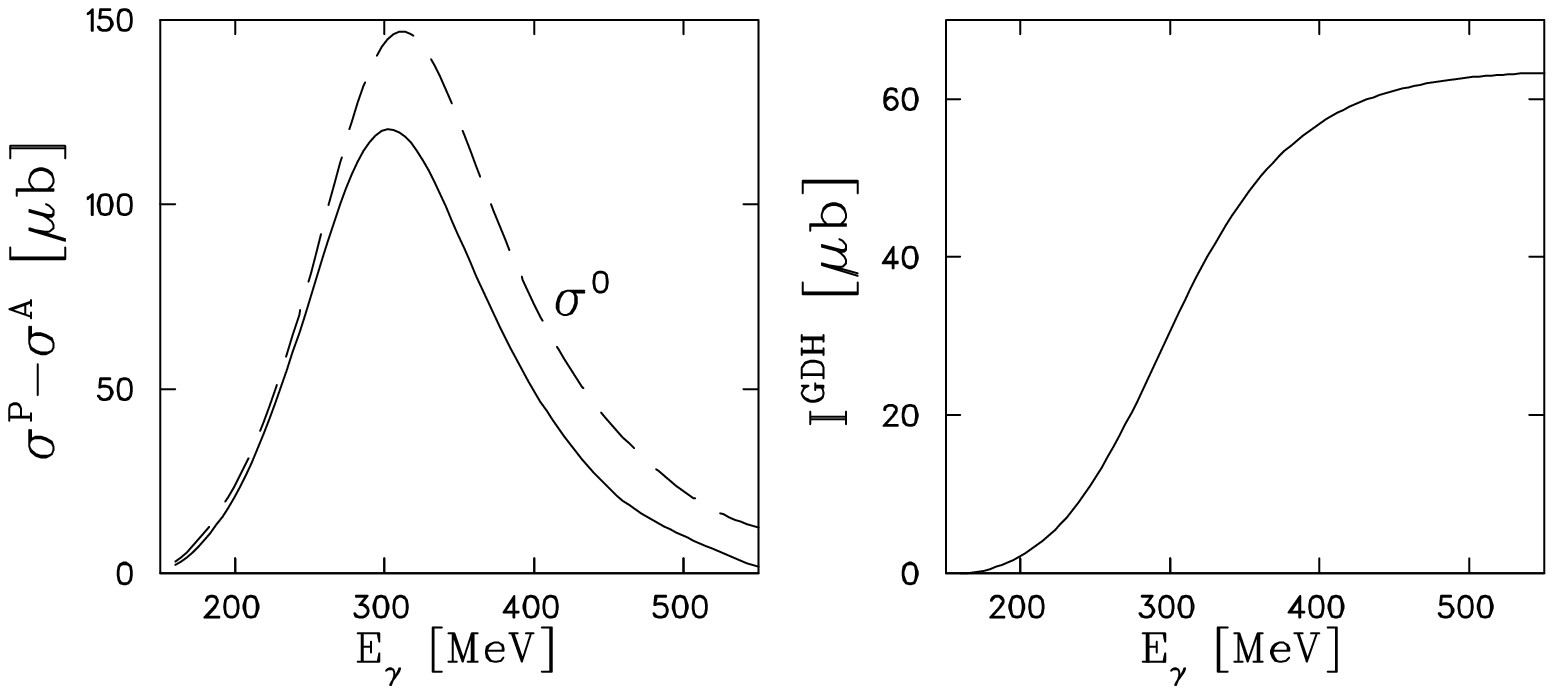,width=12cm,angle=0}}
\caption{figgdh_fig2}
{Contribution of coherent $\pi^0$ production to the GDH sum rule for the
deuteron. Left panel: difference of the cross sections (full curve), the
dashed curve shows the unpolarized cross section; Right panel:
$I^{GDH}_{\gamma d \to d\pi^0}$ as function of the upper integration 
energy (from \cite{ArK97}).}
\end{figure}
\\

\noindent
{\bf (iii) GDH contribution from incoherent pion production}

The calculation of the incoherent $\gamma d \rightarrow \pi NN$ 
contributions to the GDH integral is based on the spectator-nucleon 
approach of Schmidt et al.\ \cite{ScA96}. In this framework, the 
reaction proceeds
via the pion production on one nucleon while the other nucleon acts
merely as a spectator. Thus, the $\gamma d \rightarrow \pi NN$
operator is given as the sum of the elementary $\gamma N \rightarrow \pi N$
operators of the two nucleons. 
The results are collected in Fig.\ \ref{figgdh_fig3}. The upper part shows the
individual contributions from the different charge states of the pion
and their total sum to the spin asymmetry 
for pion photoproduction on both the deuteron
and for comparison on the nucleon. One notes qualitatively a similar
behaviour although the maxima and minima are smaller and also slightly
shifted towards higher energies for the deuteron. In the lower part of
Fig.\ \ref{figgdh_fig3} the corresponding GDH integrals are shown. A large 
positive
contribution comes from $\pi^0$ production whereas the charged pions give a
negative but - in absolute size - smaller contribution to the GDH value. Up 
to an energy of 550 MeV, one finds for the total contribution of the 
incoherent pion production channels a value
$I^{GDH}_{\gamma d \to NN\pi}(550\,\mbox{MeV})=167\,\mu$b, 
which is remarkably
close to the sum of the neutron and proton values for the given elementary
model $I^{GDH}_n(550\,\mbox{MeV})+I^{GDH}_p(550\,\mbox{MeV})=163\,\mu$b.
This fact is also indicated by the dashed curves in the lower part of
Fig.\ \ref{figgdh_fig3} which represent the appropriate differences of
$I^{GDH}_{\gamma d \to NN\pi}-I^{GDH}_p-I^{GDH}_n$.
It underlines again that the total cross
section is dominated by the quasifree process. However, as is evident 
from Fig.\ \ref{figgdh_fig3}, convergence is certainly not reached at this
energy. Furthermore, the elementary pion production operator used in 
\cite{ScA96} had been
constructed primarily to give a realistic description of the $\Delta$
resonance region. In fact, it underestimates the GDH inegral up to 550 MeV
by about a factor two compared to a corresponding evaluation based on a
multipole analysis of experimental pion photoproduction data. 
For this reason we cannot expect that this model gives also a good 
description of experimental data above 400 MeV. However, the important result 
is, that the total GDH contribution from the incoherent process is very 
close to the sum of the free proton and neutron GDH integrals, which we 
expect to remain valid for an improved elementary production operator.
\begin{figure}
\centerline {\psfig{file=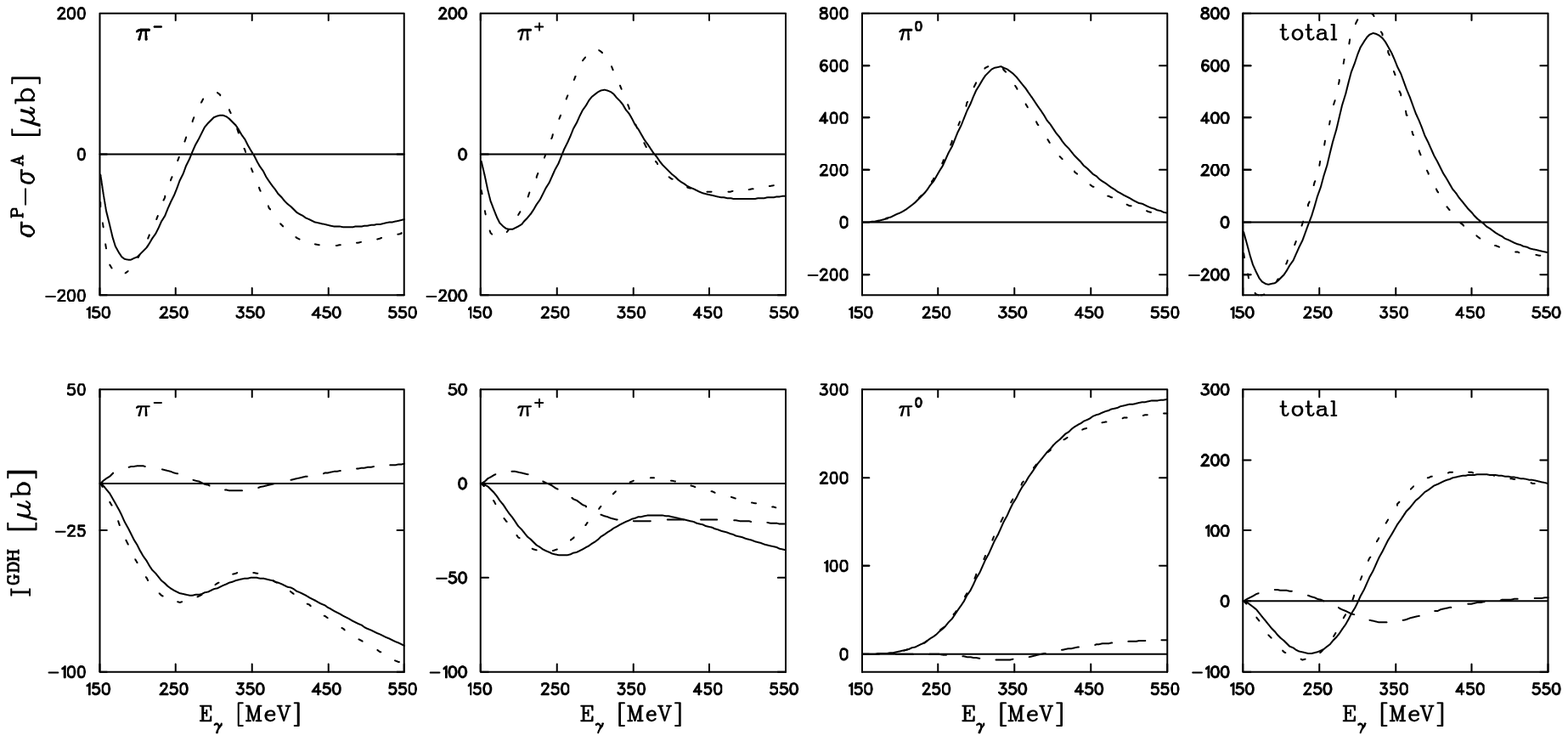,width=14cm,angle=0}}
\caption{figgdh_fig3}
{Contribution of the incoherent $\pi$ production to the GDH sum rule for 
the
deuteron and the nucleon. Upper part: difference of the cross sections;
lower part: $I^{GDH}_{\gamma d \to NN\pi}$ as function of the upper 
integration limit, full curves for the deuteron, dotted curves for 
the nucleon. In the case of $\pi^0$
production, the dotted curve shows the summed proton and neutron
contributions. The dashed curves show the appropriate differences
$I^{GDH}_{\gamma d \to NN\pi}-I^{GDH}_p-I^{GDH}_n$ (from \cite{ArK97}).}
\end{figure}
\\

\noindent
{\bf (iv) Total GDH contribution for the deuteron}

The contributions from all three channels and their sum are listed in Tab.\
\ref{tab1}. A very interesting and important result is the large negative
contribution from the photodisintegration channel near and not too far
above the break-up threshold with surprisingly large relativistic effects
below 100 MeV. Hopefully, this low energy feature of
the GDH sum rule could be checked experimentally in the near future.
For the total GDH value from explicit integration up to 550 MeV, we find a
negative value $I^{GDH}_d(550\,\mbox{MeV})=-183\,\mu$b. However, as we have
mentioned above, some larger uncertainty lies in the contribution of the 
incoherent
pion production channel because of shortcomings of the model of the
elementary production amplitude above the $\Delta$ resonance. If we use
instead of the model value $I^{GDH}_{\gamma d \to NN\pi}(550\,\mbox{MeV})=
167\,\mu$b (cf.\ previous section)
the sum of the GDH values of neutron and proton by integrating the
cross section difference, obtained from a multipole analysis of experimental
data (fit SM95 from \cite{SAID}), giving $I^{GDH}_n(550\,\mbox{MeV})
+I^{GDH}_p(550\,\mbox{MeV})=331\,\mu$b, we find for the deuteron
$I^{GDH}_d(550\,\mbox{MeV})=-19\,\mu$b, which we consider a more realistic
estimate. Since this value is still negative, a positive contribution of
about the same size should come from contributions at higher energies in
order to fulfil the small GDH sum rule for the deuteron, provided that the
sum rule is valid. These contributions should come from the incoherent
single pion production above 550 MeV because here convergence had not been
reached in contrast to the other two channels, photodisintegration and
coherent pion production, and in addition, from multipion production.
\begin{table}
\fcaption{tab1}
{Contributions of the different absorption channels to the
GDH integral for the deuteron integrated up to 550 MeV in $\mu$b.}
\begin{center}
\begin{tabular}{|ccccc|c|}
\hline
$\gamma d \to np$ & $\gamma  d \to d \pi^0$ & $\gamma d \to np\pi^0$
  &$\gamma d \to nn\pi^+$   & $\gamma d \to pp\pi^-$   &total\\
\hline
  $-413$ &   63 &  288 &  $-35$ &  $-86$ & $-183$ \\
\hline
\end{tabular}
\end{center} 
\end{table}

The foregoing results show that the spin asymmetry and the GDH sum rule
of the deuteron are very interesting observables 
because of a strong  anticorrelation  of
photodisintegration and pion production.
Further  improvements  are necessary with 
respect to the elementary single and double pion production and the 
treatment of photodisintegration above the $\Delta$ resonance.

\subsection{Parity Violation in Quasifree Electron Scattering}
\label{sec6parity}
%(G.\ K\"uster et al., Nucl.\ Phys.\ (in print), MKPH-T-97-9)

The recent interest in studying parity violation in electron scattering 
by electroweak interference is motivated by the possibility to investigate 
the so-called strangeness or better
$s\bar s$ content of the nucleon, a quantity of particular interest with
respect to the nucleon spin structure functions as measured
in deep inelastic scattering \cite{AsB93,AbA95,AdA94}. In fact,
several experiments to measure parity violation in electron scattering
off hydrogen and deuterium are presently underway
\cite{Pro90,FiS91,Har93,BeA96}.
In these experiments, deuterium serves
as a neutron target and for this reason quasifree kinematics is
preferred in order to minimize possible interaction effects.

Parity violation in inelastic electron deuteron scattering has been studied
theoretically by Hwang et al.\ \cite{HwH80,HwH81} in the low energy
and momentum transfer domain, by Hadjimichael et al.\ \cite{HaP92} for
a larger kinematical range, in particular for quasifree kinematics at high
momentum transfers, and very recently by Mosconi and Ricci \cite{MoR97}. 
Relativistic contributions have been considered by Poulis \cite{Pou96a,Pou96b}. 
While in \cite{HwH81} the parity violation by electroweak interference as well 
as through parity violating nuclear components has been considered, the latter
has been neglected throughout in the more recent evaluations
\cite{HaP92,Pou96a,Pou96b,MoR97}. Already in \cite{HaP92} it has been 
remarked that such effects should be investigated in order to see for which
kinematical situations they can indeed be neglected or whether they have 
to be included. This has been done in a recent work by K\"uster and myself 
\cite{KuA97}. 

The presence of a parity violating $NN$ potential $V^{pnc}$ has the 
consequence that the nuclear states are no longer states of good parity, 
i.e., besides a dominant state $\ \vert \,J^{\pi}\, \rangle$ of angular momentum
$J$ and parity $\pi$, there will be a small admixture of opposite parity 
$\ \vert \,J^{-\pi}\,\rangle$, with an amplitude ${\cal F}$ of
the order $10^{-6}$ \cite{McK69}. In the case of the deuteron, the
admixture of opposite parity (pnc) components must result in additional small 
$P$ waves. For the calculation of these $P$ wave components, the 
one-boson-exchange model of Desplanques, Donoghue and Holstein \cite{DeD80} 
has been used for $V^{pnc}$. 
Based on the quark model and $SU(6)_w$ symmetry, they made
predictions for all meson-nucleon couplings from both charged and neutral
current pieces of the weak Hamiltonian.

In order to calculate the opposite parity admixture of the deuteron wave
function, first-order perturbation theory is sufficient. The radial 
functions of the pnc components are shown in Fig.\ \ref{figdeut_wf}. For the
unperturbed wave function, the parametrization by Machleidt et al.\
\cite{MaH87,Mac89} for the Bonn OBEPQ model has been used. In momentum space,
the parity violating $P$ waves are much
more spread out than the normal $S$ and $D$ waves, indicating a much shorter
range in coordinate space. This is particularly pronounced in the
$^1P_1$-part, because the $\pi$ exchange part of $V^{pnc}$ does not
contribute here. The total $P$ state probability is
$P_P=2.7 \times 10^{-14}$, which corresponds to an admixture amplitude that 
is of the order of magnitude expected \cite{McK69} and in qualitative 
agreement with \cite{HwH81}.
\begin{figure}
\centerline {\psfig{file=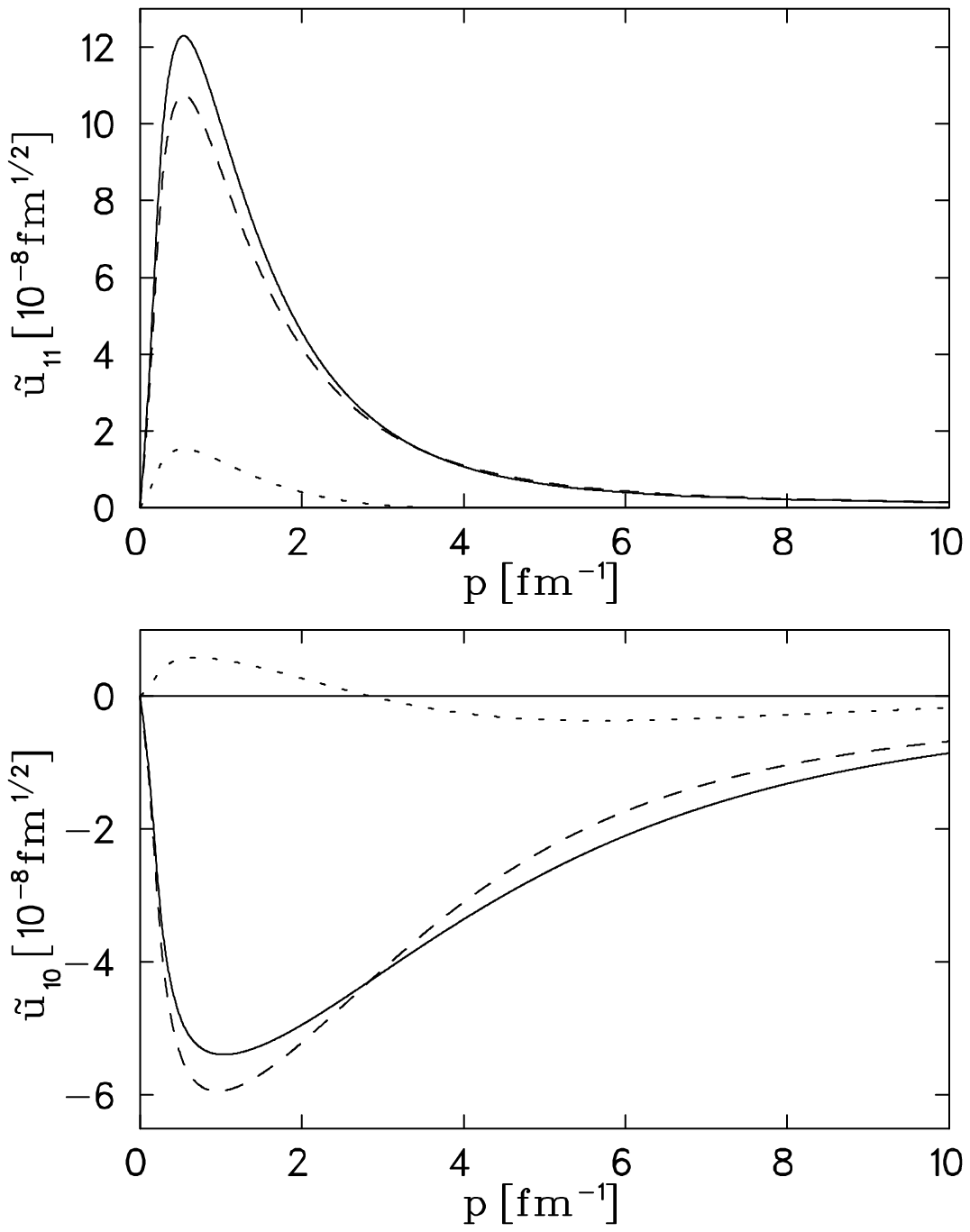,width=6cm,angle=0}}
\caption{figdeut_wf}
{Radial parts of pnc $P$ wave components  (full) of the deuteron 
in momentum space. Also shown are the contributions generated separately from 
$S$ (dashed) and $D$ (dotted) components (from \protect{\cite{KuA97}}).
}
\end{figure}
In principle, parity violating components will also appear in the  
final scattering states. However, one expects a negligible effect for 
quasifree kinematics. 
\begin{figure}[h]
\centerline {\psfig{file=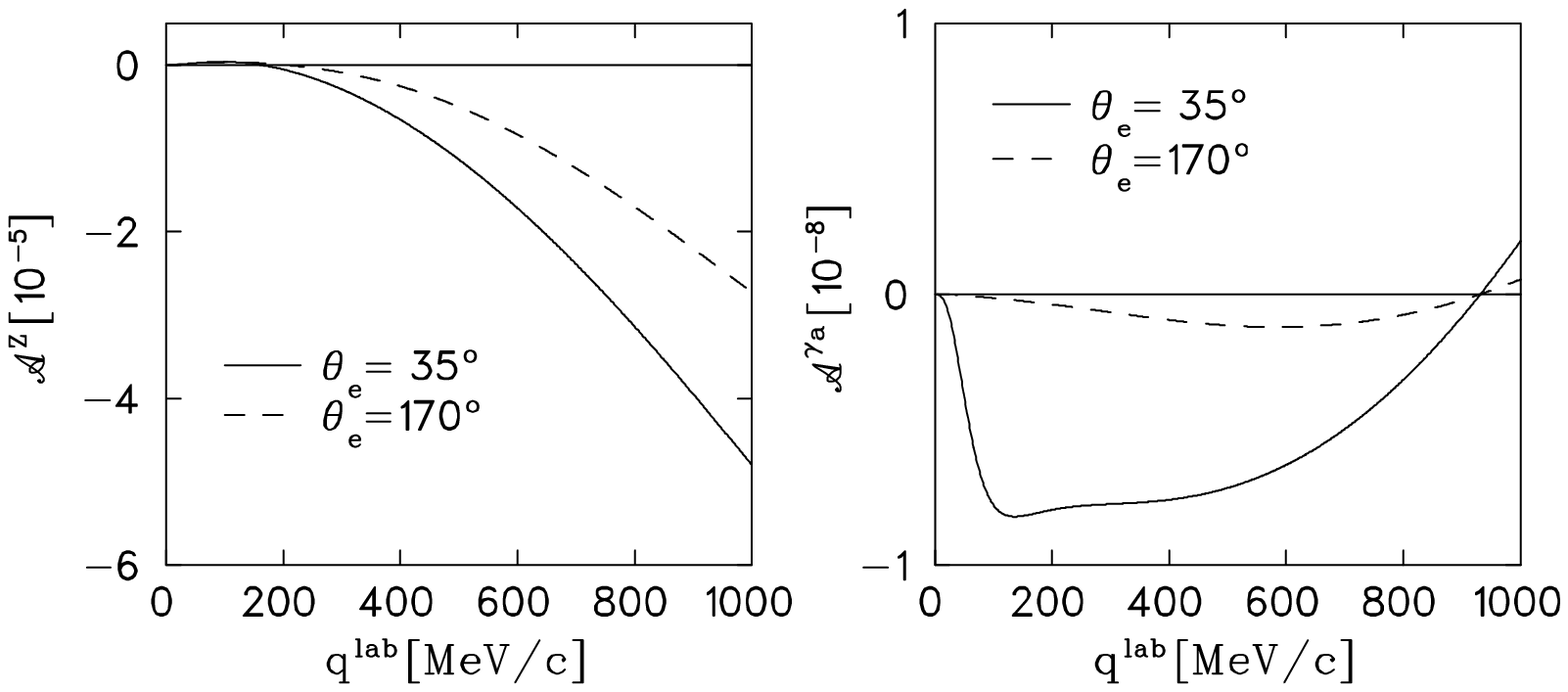,width=12cm,angle=0}}
\caption{figlongasy}
{Longitudinal asymmetry for polarized electrons along the
quasifree ridge from  electroweak interference (left panel) 
and from the  pnc deuteron components (right panel) 
for forward and backward electron
scattering ($\theta_e=35^\circ\,\mbox{and}\, 170^\circ$) in the laboratory
frame (from \cite{KuA97}).
}
\end{figure}

The resulting longitudinal beam asymmetries ${\cal A}^{\gamma_a}$ 
and ${\cal A}^{Z}$ (see (\ref{asymmg}) and (\ref{asymmz}) for their definition) 
are shown in
Fig.~\ref{figlongasy}. Since for a given momentum transfer $q^{lab}$
the asymmetry depends also on the electron kinematics through the lepton
density matrices, we have chosen two laboratory scattering angles, one at
a more forward direction ($\theta_e=35^\circ$) and one backward angle
($\theta_e=170^\circ$). One readily notes that the
beam asymmetry ${\cal A}^{\gamma_a}$ due to the $P$ wave in the deuteron
varies strongly with the scattering angle. Indeed, it is relatively more
suppressed for backward scattering than ${\cal A}^{Z}$. Furthermore, it is
apparent that ${\cal A}^{\gamma_a}$ is negligible compared to ${\cal A}^{Z}
$ over the whole range of momentum transfers considered here. 
For the beam asymmetry due to electroweak interference, the dominant
contribution comes from the term proportional to $a_aF_T^{Z_v}$, but the
dependence on the electron scattering angle is mainly a result of the term
proportional to $a_aF_L^{Z_v}$. In order to compare
our results on the asymmetry ${\cal A}^{Z}$ from electroweak interference
with those reported in \cite{HaP92}, we show in Fig.~\ref{figlongasylog} the
asymmetries on a logarithmic scale. For ${\cal A}^{Z}$ we find very good
agreement with their results for the quasifree case in plane wave Born
approximation.
\begin{figure}[h]
\centerline {\psfig{file=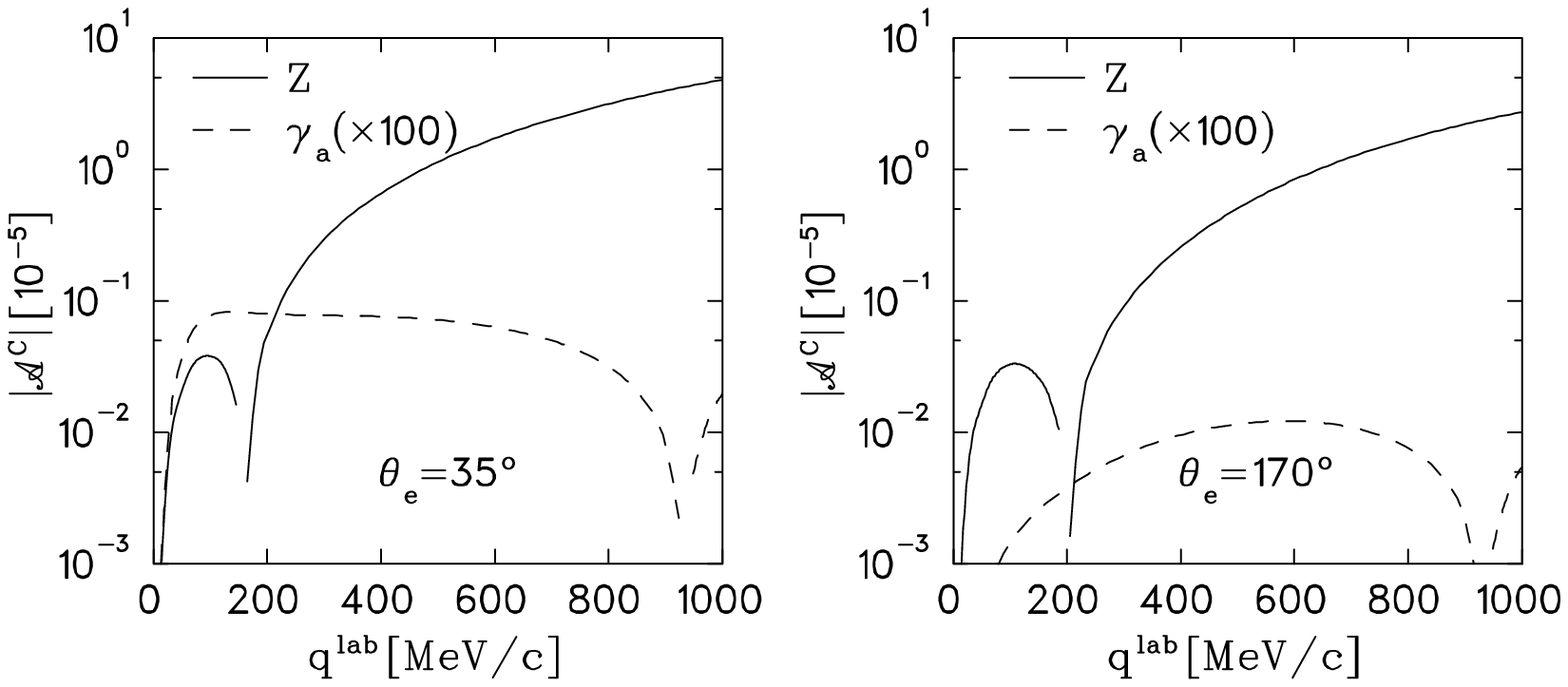,width=12cm,angle=0}}
\caption{figlongasylog}
{As Fig.\ \ref{figlongasy} but on a logarithmic scale (from \cite{KuA97}).
}
\end{figure}

As already mentioned, there is a great deal of interest in experiments to
determine strange quark contributions to hadronic matrix elements.
In connection
with this study, the SAMPLE experiment at MIT-Bates \cite{BeA96} is of
special interest, since it measures the strange magnetic form factor
$G_M^{(s)}$ at quite low momentum transfer as determined in parity 
violating
electron scattering off hydrogen and deuterium. Even here, we have not
found significant effects from parity violation in the hadronic sector.

In order to study parity violation which originates from the hadronic 
sector
in electromagnetic deuteron break-up, one has to go away from the quasifree
kinematics to lower momentum transfer where it may become more comparable
in size to the contribution from electroweak interference. But then one has
to consider also pnc components in the final state and the contribution 
from
meson exchange currents. In this respect, the present study should be 
considered as a starting point for further investigations, in
particular with respect to the role of final state interaction, meson 
exchange currents and isobar configurations.

    %

%\newpage
\setcounter{equation}{0}
\setcounter{figure}{0}
\setcounter{table}{0}

\section{Conclusions and Outlook}

I hope these lectures have demonstrated convincingly, that the electroweak 
probe is still an interesting, very important and versatile tool, allowing 
one to reveal the internal structure 
of hadrons in great detail. Although a lot of insight has been gained 
from the past experimental and theoretical work, there still exist 
largely unexplored regions waiting to be unveiled. In the more distant 
past, mostly inclusive reactions had been studied due to the low duty cycle 
of the at that time existing electron machines. Only during the last decade 
the new generation of high duty cycle machines became available and 
allows one now to exploit the full power of the electroweak probe. Thus it is 
clear, that the thrust of present and future experimental research will lie on 
the study of more and more exclusive reactions. 

In particular, the recent developments with respect to more intense and 
highly polarized beams and targets of higher density and polarization, and 
the availability of highly efficient polarimeters for the analysis of 
the final reaction products will open up a new era of studying polarization 
observables in much greater detail than could be done before, and thus will 
give us even more detailed information on the properties of hadrons. 
Because polarization observables in general constitute much more stringent 
tests for theoretical models, being more sensitive to small, but 
interesting amplitudes. To this end, we need further developments 
of even more intense polarized beams. Similarly, highly polarized targets 
are desirable, 
which can resist high intensity beams. Furthermore, the efficiency of 
polarimeters, for example, neutron polarimeters, needs to be 
pushed to higher values. 

On the theory side, the role of effective degrees of freedom in nuclei 
in terms of nucleon, meson and isobar degrees of freedom and their relation 
to the underlying more fundamental quark-gluon degrees of freedom of QCD 
should be further clarified. Specifically, it will be extremely important 
to understand, how far in energy and momentum transfer we can push this 
framework of effective nucleon, meson and isobar degrees of freedom and 
whether signatures of explicit quark-gluon effects will be manifest in nuclei 
under certain kinematic conditions. 
Very likely, we will not find a clear cut borderline between the 
perturbative and nonperturbative regimes of QCD and it will be difficult to 
pin down genuine quark-gluon effects in view of the partly phenomenological 
ingredients of present theoretical models on both sides, effective nuclear 
d.o.f.\ and quark-gluon descriptions. 
Similarly, with respect to the internal nucleon, or in general,  
hadron structure, the nature of the effective constituent quark degrees of 
freedom should be revealed.
In addition, in view of the increase of energy and momentum transfers 
involved in present and future experiments, the effects which 
arise from relativity should be considered with great care.

\section*{Acknowledgement} I would like to thank the organizers of this 
workshop, A.C.\ Fonseca, W.\ Plessas and P.\ Sauer, for giving me the 
opportunity to deliver these lectures. 
The collaboration of Ji\v{r}i Adam, Elena Breitmoser, Alexander 
Fix, Gunilla K\"uster, Frank Ritz, Jr., R\"udiger Schmidt, 
Michael Schwamb, Thomas Wilbois, and Paul Wilhelm in various parts of the 
work reported here is gratefully acknowledged. 
Furthermore, I would like to thank Mariana Kirchbach for various helpful 
discussions and a critical reading of the manuscript.

%\newpage

\end{document}